\documentclass[longauth]{aaEC}

\usepackage{txfonts}           

\usepackage{graphicx}
\usepackage[table]{xcolor}
\usepackage{natbib} 
\usepackage{scalerel}
\usepackage{lastpage}
\usepackage{comment}
\usepackage{makecell}
\usepackage{float}
\usepackage{algorithm}
\usepackage{csvsimple}
\usepackage{algorithmic}
\usepackage{csvsimple}
\usepackage{tabularx}
\usepackage{siunitx}
\usepackage{pdflscape}
\usepackage{xcolor}

\usepackage{euclid}

\usepackage[pdfencoding=auto,psdextra]{hyperref}
\usepackage{cleveref}
\hypersetup{
  colorlinks=true,
  linkcolor=blue,
  citecolor=blue,
  urlcolor=blue
}

\usepackage[utf8]{inputenc}

\usepackage[switch, modulo]{lineno}

\providecommand{\orcid}[1]{}

\begin{document}
%
%
   \title{Euclid Quick Data Release (Q1)}
   \subtitle{AgileLens: A scalable
CNN-based pipeline for strong gravitational lens identification}


\author{Euclid Collaboration: X.~Xu\orcid{0009-0002-0985-2309}\thanks{\email{xinmeix@uci.edu}}\inst{\ref{aff1},\ref{aff2}}
\and R.~Chen\orcid{0009-0005-7381-5416}\inst{\ref{aff1}}
\and T.~Li\orcid{0009-0003-5064-1914}\inst{\ref{aff1}}
\and A.~R.~Cooray\orcid{0000-0002-3892-0190}\inst{\ref{aff1}}
\and S.~Schuldt\orcid{0000-0003-2497-6334}\inst{\ref{aff3},\ref{aff4}}
\and J.~A.~Acevedo~Barroso\orcid{0000-0002-9654-1711}\inst{\ref{aff5}}
\and D.~Stern\orcid{0000-0003-2686-9241}\inst{\ref{aff5}}
\and D.~Scott\orcid{0000-0002-6878-9840}\inst{\ref{aff6}}
\and M.~Meneghetti\orcid{0000-0003-1225-7084}\inst{\ref{aff7},\ref{aff8}}
\and G.~Despali\orcid{0000-0001-6150-4112}\inst{\ref{aff9},\ref{aff7},\ref{aff8}}
\and J.~Chopra\orcid{0009-0001-6562-7846}\inst{\ref{aff1}}
\and Y.~Cao\orcid{0000-0001-6111-4157}\inst{\ref{aff1}}
\and M.~Cheng\inst{\ref{aff1}}
\and J.~Buda\inst{\ref{aff1}}
\and J.~Zhang\orcid{0009-0003-2801-2919}\inst{\ref{aff1}}
\and J.~Furumizo\orcid{0009-0006-1559-6750}\inst{\ref{aff1}}
\and R.~Valencia\inst{\ref{aff1}}
\and Z.~Jiang\inst{\ref{aff2}}
\and C.~Tortora\orcid{0000-0001-7958-6531}\inst{\ref{aff10}}
\and N.~E.~P.~Lines\orcid{0009-0004-7751-1914}\inst{\ref{aff11}}
\and T.~E.~Collett\orcid{0000-0001-5564-3140}\inst{\ref{aff11}}
\and S.~Fotopoulou\orcid{0000-0002-9686-254X}\inst{\ref{aff12}}
\and A.~Galan\orcid{0000-0003-2547-9815}\inst{\ref{aff13},\ref{aff14}}
\and A.~Manj\'on-Garc\'ia\orcid{0000-0002-7413-8825}\inst{\ref{aff15}}
\and R.~Gavazzi\orcid{0000-0002-5540-6935}\inst{\ref{aff16},\ref{aff17}}
\and L.~Iwamoto\orcid{0009-0006-4812-2033}\inst{\ref{aff18}}
\and S.~Kruk\orcid{0000-0001-8010-8879}\inst{\ref{aff19}}
\and M.~Millon\orcid{0000-0001-7051-497X}\inst{\ref{aff20}}
\and P.~Nugent\orcid{0000-0002-3389-0586}\inst{\ref{aff21}}
\and C.~Saulder\orcid{0000-0002-0408-5633}\inst{\ref{aff22},\ref{aff23}}
\and D.~Sluse\orcid{0000-0001-6116-2095}\inst{\ref{aff24}}
\and J.~Wilde\orcid{0000-0002-4460-7379}\inst{\ref{aff25}}
\and M.~Walmsley\orcid{0000-0002-6408-4181}\inst{\ref{aff26},\ref{aff27}}
\and F.~Courbin\orcid{0000-0003-0758-6510}\inst{\ref{aff25},\ref{aff28},\ref{aff29}}
\and R.~B.~Metcalf\orcid{0000-0003-3167-2574}\inst{\ref{aff9},\ref{aff7}}
\and B.~Altieri\orcid{0000-0003-3936-0284}\inst{\ref{aff19}}
\and A.~Amara\inst{\ref{aff30}}
\and S.~Andreon\orcid{0000-0002-2041-8784}\inst{\ref{aff31}}
\and N.~Auricchio\orcid{0000-0003-4444-8651}\inst{\ref{aff7}}
\and C.~Baccigalupi\orcid{0000-0002-8211-1630}\inst{\ref{aff32},\ref{aff33},\ref{aff34},\ref{aff35}}
\and M.~Baldi\orcid{0000-0003-4145-1943}\inst{\ref{aff36},\ref{aff7},\ref{aff8}}
\and A.~Balestra\orcid{0000-0002-6967-261X}\inst{\ref{aff37}}
\and S.~Bardelli\orcid{0000-0002-8900-0298}\inst{\ref{aff7}}
\and P.~Battaglia\orcid{0000-0002-7337-5909}\inst{\ref{aff7}}
\and R.~Bender\orcid{0000-0001-7179-0626}\inst{\ref{aff22},\ref{aff23}}
\and A.~Biviano\orcid{0000-0002-0857-0732}\inst{\ref{aff33},\ref{aff32}}
\and E.~Branchini\orcid{0000-0002-0808-6908}\inst{\ref{aff38},\ref{aff39},\ref{aff31}}
\and M.~Brescia\orcid{0000-0001-9506-5680}\inst{\ref{aff40},\ref{aff10}}
\and S.~Camera\orcid{0000-0003-3399-3574}\inst{\ref{aff41},\ref{aff42},\ref{aff43}}
\and V.~Capobianco\orcid{0000-0002-3309-7692}\inst{\ref{aff43}}
\and C.~Carbone\orcid{0000-0003-0125-3563}\inst{\ref{aff4}}
\and V.~F.~Cardone\inst{\ref{aff44},\ref{aff45}}
\and J.~Carretero\orcid{0000-0002-3130-0204}\inst{\ref{aff46},\ref{aff47}}
\and S.~Casas\orcid{0000-0002-4751-5138}\inst{\ref{aff48},\ref{aff49}}
\and M.~Castellano\orcid{0000-0001-9875-8263}\inst{\ref{aff44}}
\and G.~Castignani\orcid{0000-0001-6831-0687}\inst{\ref{aff7}}
\and S.~Cavuoti\orcid{0000-0002-3787-4196}\inst{\ref{aff10},\ref{aff50}}
\and A.~Cimatti\inst{\ref{aff51}}
\and C.~Colodro-Conde\inst{\ref{aff52}}
\and G.~Congedo\orcid{0000-0003-2508-0046}\inst{\ref{aff53}}
\and C.~J.~Conselice\orcid{0000-0003-1949-7638}\inst{\ref{aff27}}
\and L.~Conversi\orcid{0000-0002-6710-8476}\inst{\ref{aff54},\ref{aff19}}
\and Y.~Copin\orcid{0000-0002-5317-7518}\inst{\ref{aff55}}
\and H.~M.~Courtois\orcid{0000-0003-0509-1776}\inst{\ref{aff56}}
\and M.~Cropper\orcid{0000-0003-4571-9468}\inst{\ref{aff57}}
\and A.~Da~Silva\orcid{0000-0002-6385-1609}\inst{\ref{aff58},\ref{aff59}}
\and H.~Degaudenzi\orcid{0000-0002-5887-6799}\inst{\ref{aff60}}
\and G.~De~Lucia\orcid{0000-0002-6220-9104}\inst{\ref{aff33}}
\and C.~Dolding\orcid{0009-0003-7199-6108}\inst{\ref{aff57}}
\and H.~Dole\orcid{0000-0002-9767-3839}\inst{\ref{aff61}}
\and F.~Dubath\orcid{0000-0002-6533-2810}\inst{\ref{aff60}}
\and X.~Dupac\inst{\ref{aff19}}
\and S.~Dusini\orcid{0000-0002-1128-0664}\inst{\ref{aff62}}
\and S.~Escoffier\orcid{0000-0002-2847-7498}\inst{\ref{aff63}}
\and M.~Farina\orcid{0000-0002-3089-7846}\inst{\ref{aff64}}
\and R.~Farinelli\inst{\ref{aff7}}
\and S.~Farrens\orcid{0000-0002-9594-9387}\inst{\ref{aff65}}
\and S.~Ferriol\inst{\ref{aff55}}
\and F.~Finelli\orcid{0000-0002-6694-3269}\inst{\ref{aff7},\ref{aff66}}
\and P.~Fosalba\orcid{0000-0002-1510-5214}\inst{\ref{aff67},\ref{aff68}}
\and M.~Frailis\orcid{0000-0002-7400-2135}\inst{\ref{aff33}}
\and E.~Franceschi\orcid{0000-0002-0585-6591}\inst{\ref{aff7}}
\and M.~Fumana\orcid{0000-0001-6787-5950}\inst{\ref{aff4}}
\and S.~Galeotta\orcid{0000-0002-3748-5115}\inst{\ref{aff33}}
\and K.~George\orcid{0000-0002-1734-8455}\inst{\ref{aff69}}
\and W.~Gillard\orcid{0000-0003-4744-9748}\inst{\ref{aff63}}
\and B.~Gillis\orcid{0000-0002-4478-1270}\inst{\ref{aff53}}
\and C.~Giocoli\orcid{0000-0002-9590-7961}\inst{\ref{aff7},\ref{aff8}}
\and P.~G\'omez-Alvarez\orcid{0000-0002-8594-5358}\inst{\ref{aff70},\ref{aff19}}
\and J.~Gracia-Carpio\inst{\ref{aff22}}
\and A.~Grazian\orcid{0000-0002-5688-0663}\inst{\ref{aff37}}
\and F.~Grupp\inst{\ref{aff22},\ref{aff23}}
\and S.~V.~H.~Haugan\orcid{0000-0001-9648-7260}\inst{\ref{aff71}}
\and W.~Holmes\inst{\ref{aff5}}
\and F.~Hormuth\inst{\ref{aff72}}
\and A.~Hornstrup\orcid{0000-0002-3363-0936}\inst{\ref{aff73},\ref{aff74}}
\and K.~Jahnke\orcid{0000-0003-3804-2137}\inst{\ref{aff75}}
\and M.~Jhabvala\inst{\ref{aff76}}
\and B.~Joachimi\orcid{0000-0001-7494-1303}\inst{\ref{aff77}}
\and S.~Kermiche\orcid{0000-0002-0302-5735}\inst{\ref{aff63}}
\and A.~Kiessling\orcid{0000-0002-2590-1273}\inst{\ref{aff5}}
\and B.~Kubik\orcid{0009-0006-5823-4880}\inst{\ref{aff55}}
\and M.~K\"ummel\orcid{0000-0003-2791-2117}\inst{\ref{aff23}}
\and M.~Kunz\orcid{0000-0002-3052-7394}\inst{\ref{aff20}}
\and H.~Kurki-Suonio\orcid{0000-0002-4618-3063}\inst{\ref{aff78},\ref{aff79}}
\and A.~M.~C.~Le~Brun\orcid{0000-0002-0936-4594}\inst{\ref{aff80}}
\and S.~Ligori\orcid{0000-0003-4172-4606}\inst{\ref{aff43}}
\and P.~B.~Lilje\orcid{0000-0003-4324-7794}\inst{\ref{aff71}}
\and V.~Lindholm\orcid{0000-0003-2317-5471}\inst{\ref{aff78},\ref{aff79}}
\and I.~Lloro\orcid{0000-0001-5966-1434}\inst{\ref{aff81}}
\and G.~Mainetti\orcid{0000-0003-2384-2377}\inst{\ref{aff82}}
\and E.~Maiorano\orcid{0000-0003-2593-4355}\inst{\ref{aff7}}
\and O.~Mansutti\orcid{0000-0001-5758-4658}\inst{\ref{aff33}}
\and S.~Marcin\inst{\ref{aff83}}
\and O.~Marggraf\orcid{0000-0001-7242-3852}\inst{\ref{aff84}}
\and M.~Martinelli\orcid{0000-0002-6943-7732}\inst{\ref{aff44},\ref{aff45}}
\and N.~Martinet\orcid{0000-0003-2786-7790}\inst{\ref{aff16}}
\and F.~Marulli\orcid{0000-0002-8850-0303}\inst{\ref{aff9},\ref{aff7},\ref{aff8}}
\and R.~J.~Massey\orcid{0000-0002-6085-3780}\inst{\ref{aff85}}
\and E.~Medinaceli\orcid{0000-0002-4040-7783}\inst{\ref{aff7}}
\and S.~Mei\orcid{0000-0002-2849-559X}\inst{\ref{aff86},\ref{aff87}}
\and M.~Melchior\inst{\ref{aff88}}
\and E.~Merlin\orcid{0000-0001-6870-8900}\inst{\ref{aff44}}
\and G.~Meylan\inst{\ref{aff89}}
\and A.~Mora\orcid{0000-0002-1922-8529}\inst{\ref{aff90}}
\and M.~Moresco\orcid{0000-0002-7616-7136}\inst{\ref{aff9},\ref{aff7}}
\and L.~Moscardini\orcid{0000-0002-3473-6716}\inst{\ref{aff9},\ref{aff7},\ref{aff8}}
\and R.~Nakajima\orcid{0009-0009-1213-7040}\inst{\ref{aff84}}
\and C.~Neissner\orcid{0000-0001-8524-4968}\inst{\ref{aff91},\ref{aff47}}
\and R.~C.~Nichol\orcid{0000-0003-0939-6518}\inst{\ref{aff30}}
\and S.-M.~Niemi\orcid{0009-0005-0247-0086}\inst{\ref{aff92}}
\and J.~W.~Nightingale\orcid{0000-0002-8987-7401}\inst{\ref{aff93}}
\and C.~Padilla\orcid{0000-0001-7951-0166}\inst{\ref{aff91}}
\and S.~Paltani\orcid{0000-0002-8108-9179}\inst{\ref{aff60}}
\and F.~Pasian\orcid{0000-0002-4869-3227}\inst{\ref{aff33}}
\and K.~Pedersen\inst{\ref{aff94}}
\and W.~J.~Percival\orcid{0000-0002-0644-5727}\inst{\ref{aff95},\ref{aff96},\ref{aff97}}
\and V.~Pettorino\orcid{0000-0002-4203-9320}\inst{\ref{aff92}}
\and G.~Polenta\orcid{0000-0003-4067-9196}\inst{\ref{aff98}}
\and M.~Poncet\inst{\ref{aff99}}
\and L.~A.~Popa\inst{\ref{aff100}}
\and F.~Raison\orcid{0000-0002-7819-6918}\inst{\ref{aff22}}
\and A.~Renzi\orcid{0000-0001-9856-1970}\inst{\ref{aff101},\ref{aff62}}
\and J.~Rhodes\orcid{0000-0002-4485-8549}\inst{\ref{aff5}}
\and G.~Riccio\inst{\ref{aff10}}
\and E.~Romelli\orcid{0000-0003-3069-9222}\inst{\ref{aff33}}
\and M.~Roncarelli\orcid{0000-0001-9587-7822}\inst{\ref{aff7}}
\and R.~Saglia\orcid{0000-0003-0378-7032}\inst{\ref{aff23},\ref{aff22}}
\and Z.~Sakr\orcid{0000-0002-4823-3757}\inst{\ref{aff102},\ref{aff103},\ref{aff104}}
\and D.~Sapone\orcid{0000-0001-7089-4503}\inst{\ref{aff105}}
\and M.~Schirmer\orcid{0000-0003-2568-9994}\inst{\ref{aff75}}
\and P.~Schneider\orcid{0000-0001-8561-2679}\inst{\ref{aff84}}
\and T.~Schrabback\orcid{0000-0002-6987-7834}\inst{\ref{aff106}}
\and A.~Secroun\orcid{0000-0003-0505-3710}\inst{\ref{aff63}}
\and G.~Seidel\orcid{0000-0003-2907-353X}\inst{\ref{aff75}}
\and E.~Sihvola\orcid{0000-0003-1804-7715}\inst{\ref{aff107}}
\and P.~Simon\inst{\ref{aff84}}
\and C.~Sirignano\orcid{0000-0002-0995-7146}\inst{\ref{aff101},\ref{aff62}}
\and G.~Sirri\orcid{0000-0003-2626-2853}\inst{\ref{aff8}}
\and L.~Stanco\orcid{0000-0002-9706-5104}\inst{\ref{aff62}}
\and P.~Tallada-Cresp\'{i}\orcid{0000-0002-1336-8328}\inst{\ref{aff46},\ref{aff47}}
\and A.~N.~Taylor\inst{\ref{aff53}}
\and I.~Tereno\orcid{0000-0002-4537-6218}\inst{\ref{aff58},\ref{aff108}}
\and N.~Tessore\orcid{0000-0002-9696-7931}\inst{\ref{aff57}}
\and S.~Toft\orcid{0000-0003-3631-7176}\inst{\ref{aff109},\ref{aff110}}
\and R.~Toledo-Moreo\orcid{0000-0002-2997-4859}\inst{\ref{aff111}}
\and F.~Torradeflot\orcid{0000-0003-1160-1517}\inst{\ref{aff47},\ref{aff46}}
\and I.~Tutusaus\orcid{0000-0002-3199-0399}\inst{\ref{aff68},\ref{aff67},\ref{aff103}}
\and L.~Valenziano\orcid{0000-0002-1170-0104}\inst{\ref{aff7},\ref{aff66}}
\and J.~Valiviita\orcid{0000-0001-6225-3693}\inst{\ref{aff78},\ref{aff79}}
\and T.~Vassallo\orcid{0000-0001-6512-6358}\inst{\ref{aff33},\ref{aff69}}
\and G.~Verdoes~Kleijn\orcid{0000-0001-5803-2580}\inst{\ref{aff112}}
\and A.~Veropalumbo\orcid{0000-0003-2387-1194}\inst{\ref{aff31},\ref{aff39},\ref{aff38}}
\and Y.~Wang\orcid{0000-0002-4749-2984}\inst{\ref{aff113}}
\and J.~Weller\orcid{0000-0002-8282-2010}\inst{\ref{aff23},\ref{aff22}}
\and A.~Zacchei\orcid{0000-0003-0396-1192}\inst{\ref{aff33},\ref{aff32}}
\and G.~Zamorani\orcid{0000-0002-2318-301X}\inst{\ref{aff7}}
\and F.~M.~Zerbi\orcid{0000-0002-9996-973X}\inst{\ref{aff31}}
\and E.~Zucca\orcid{0000-0002-5845-8132}\inst{\ref{aff7}}
\and M.~Ballardini\orcid{0000-0003-4481-3559}\inst{\ref{aff114},\ref{aff115},\ref{aff7}}
\and M.~Bolzonella\orcid{0000-0003-3278-4607}\inst{\ref{aff7}}
\and C.~Burigana\orcid{0000-0002-3005-5796}\inst{\ref{aff116},\ref{aff66}}
\and R.~Cabanac\orcid{0000-0001-6679-2600}\inst{\ref{aff103}}
\and M.~Calabrese\orcid{0000-0002-2637-2422}\inst{\ref{aff117},\ref{aff4}}
\and A.~Cappi\inst{\ref{aff118},\ref{aff7}}
\and T.~Castro\orcid{0000-0002-6292-3228}\inst{\ref{aff33},\ref{aff34},\ref{aff32},\ref{aff119}}
\and J.~A.~Escartin~Vigo\inst{\ref{aff22}}
\and L.~Gabarra\orcid{0000-0002-8486-8856}\inst{\ref{aff120}}
\and S.~Hemmati\orcid{0000-0003-2226-5395}\inst{\ref{aff113}}
\and J.~Macias-Perez\orcid{0000-0002-5385-2763}\inst{\ref{aff121}}
\and R.~Maoli\orcid{0000-0002-6065-3025}\inst{\ref{aff122},\ref{aff44}}
\and J.~Mart\'{i}n-Fleitas\orcid{0000-0002-8594-569X}\inst{\ref{aff123}}
\and N.~Mauri\orcid{0000-0001-8196-1548}\inst{\ref{aff51},\ref{aff8}}
\and P.~Monaco\orcid{0000-0003-2083-7564}\inst{\ref{aff124},\ref{aff33},\ref{aff34},\ref{aff32}}
\and A.~A.~Nucita\inst{\ref{aff125},\ref{aff126},\ref{aff127}}
\and A.~Pezzotta\orcid{0000-0003-0726-2268}\inst{\ref{aff31}}
\and M.~P\"ontinen\orcid{0000-0001-5442-2530}\inst{\ref{aff78}}
\and I.~Risso\orcid{0000-0003-2525-7761}\inst{\ref{aff31},\ref{aff39}}
\and V.~Scottez\orcid{0009-0008-3864-940X}\inst{\ref{aff128},\ref{aff129}}
\and M.~Sereno\orcid{0000-0003-0302-0325}\inst{\ref{aff7},\ref{aff8}}
\and M.~Tenti\orcid{0000-0002-4254-5901}\inst{\ref{aff8}}
\and M.~Tucci\inst{\ref{aff60}}
\and M.~Viel\orcid{0000-0002-2642-5707}\inst{\ref{aff32},\ref{aff33},\ref{aff35},\ref{aff34},\ref{aff119}}
\and M.~Wiesmann\orcid{0009-0000-8199-5860}\inst{\ref{aff71}}
\and Y.~Akrami\orcid{0000-0002-2407-7956}\inst{\ref{aff102},\ref{aff130}}
\and I.~T.~Andika\orcid{0000-0001-6102-9526}\inst{\ref{aff69}}
\and G.~Angora\orcid{0000-0002-0316-6562}\inst{\ref{aff10},\ref{aff114}}
\and S.~Anselmi\orcid{0000-0002-3579-9583}\inst{\ref{aff62},\ref{aff101},\ref{aff131}}
\and M.~Archidiacono\orcid{0000-0003-4952-9012}\inst{\ref{aff3},\ref{aff132}}
\and F.~Atrio-Barandela\orcid{0000-0002-2130-2513}\inst{\ref{aff133}}
\and L.~Bazzanini\orcid{0000-0003-0727-0137}\inst{\ref{aff114},\ref{aff7}}
\and P.~Bergamini\orcid{0000-0003-1383-9414}\inst{\ref{aff7}}
\and D.~Bertacca\orcid{0000-0002-2490-7139}\inst{\ref{aff101},\ref{aff37},\ref{aff62}}
\and M.~Bethermin\orcid{0000-0002-3915-2015}\inst{\ref{aff134}}
\and F.~Beutler\orcid{0000-0003-0467-5438}\inst{\ref{aff53}}
\and L.~Blot\orcid{0000-0002-9622-7167}\inst{\ref{aff135},\ref{aff80}}
\and S.~Borgani\orcid{0000-0001-6151-6439}\inst{\ref{aff124},\ref{aff32},\ref{aff33},\ref{aff34},\ref{aff119}}
\and M.~L.~Brown\orcid{0000-0002-0370-8077}\inst{\ref{aff27}}
\and S.~Bruton\orcid{0000-0002-6503-5218}\inst{\ref{aff136}}
\and A.~Calabro\orcid{0000-0003-2536-1614}\inst{\ref{aff44}}
\and B.~Camacho~Quevedo\orcid{0000-0002-8789-4232}\inst{\ref{aff32},\ref{aff35},\ref{aff33}}
\and F.~Caro\inst{\ref{aff44}}
\and C.~S.~Carvalho\inst{\ref{aff108}}
\and F.~Cogato\orcid{0000-0003-4632-6113}\inst{\ref{aff9},\ref{aff7}}
\and S.~Conseil\orcid{0000-0002-3657-4191}\inst{\ref{aff55}}
\and O.~Cucciati\orcid{0000-0002-9336-7551}\inst{\ref{aff7}}
\and S.~Davini\orcid{0000-0003-3269-1718}\inst{\ref{aff39}}
\and G.~Desprez\orcid{0000-0001-8325-1742}\inst{\ref{aff112}}
\and A.~D\'iaz-S\'anchez\orcid{0000-0003-0748-4768}\inst{\ref{aff15}}
\and S.~Di~Domizio\orcid{0000-0003-2863-5895}\inst{\ref{aff38},\ref{aff39}}
\and J.~M.~Diego\orcid{0000-0001-9065-3926}\inst{\ref{aff137}}
\and P.-A.~Duc\orcid{0000-0003-3343-6284}\inst{\ref{aff134}}
\and V.~Duret\orcid{0009-0009-0383-4960}\inst{\ref{aff63}}
\and M.~Y.~Elkhashab\orcid{0000-0001-9306-2603}\inst{\ref{aff124},\ref{aff33},\ref{aff34},\ref{aff32}}
\and A.~Enia\orcid{0000-0002-0200-2857}\inst{\ref{aff7}}
\and Y.~Fang\orcid{0000-0002-0334-6950}\inst{\ref{aff23}}
\and A.~Finoguenov\orcid{0000-0002-4606-5403}\inst{\ref{aff78}}
\and A.~Franco\orcid{0000-0002-4761-366X}\inst{\ref{aff126},\ref{aff125},\ref{aff127}}
\and K.~Ganga\orcid{0000-0001-8159-8208}\inst{\ref{aff86}}
\and T.~Gasparetto\orcid{0000-0002-7913-4866}\inst{\ref{aff44}}
\and E.~Gaztanaga\orcid{0000-0001-9632-0815}\inst{\ref{aff68},\ref{aff67},\ref{aff11}}
\and F.~Giacomini\orcid{0000-0002-3129-2814}\inst{\ref{aff8}}
\and F.~Gianotti\orcid{0000-0003-4666-119X}\inst{\ref{aff7}}
\and G.~Gozaliasl\orcid{0000-0002-0236-919X}\inst{\ref{aff138},\ref{aff78}}
\and M.~Guidi\orcid{0000-0001-9408-1101}\inst{\ref{aff36},\ref{aff7}}
\and C.~M.~Gutierrez\orcid{0000-0001-7854-783X}\inst{\ref{aff52},\ref{aff139}}
\and A.~Hall\orcid{0000-0002-3139-8651}\inst{\ref{aff53}}
\and C.~Hern\'andez-Monteagudo\orcid{0000-0001-5471-9166}\inst{\ref{aff139},\ref{aff52}}
\and H.~Hildebrandt\orcid{0000-0002-9814-3338}\inst{\ref{aff140}}
\and J.~Hjorth\orcid{0000-0002-4571-2306}\inst{\ref{aff94}}
\and J.~J.~E.~Kajava\orcid{0000-0002-3010-8333}\inst{\ref{aff141},\ref{aff142},\ref{aff143}}
\and Y.~Kang\orcid{0009-0000-8588-7250}\inst{\ref{aff60}}
\and V.~Kansal\orcid{0000-0002-4008-6078}\inst{\ref{aff144},\ref{aff145}}
\and D.~Karagiannis\orcid{0000-0002-4927-0816}\inst{\ref{aff114},\ref{aff146}}
\and K.~Kiiveri\inst{\ref{aff107}}
\and J.~Kim\orcid{0000-0003-2776-2761}\inst{\ref{aff120}}
\and C.~C.~Kirkpatrick\inst{\ref{aff107}}
\and F.~Lepori\orcid{0009-0000-5061-7138}\inst{\ref{aff147}}
\and G.~Leroy\orcid{0009-0004-2523-4425}\inst{\ref{aff148},\ref{aff85}}
\and G.~F.~Lesci\orcid{0000-0002-4607-2830}\inst{\ref{aff9},\ref{aff7}}
\and J.~Lesgourgues\orcid{0000-0001-7627-353X}\inst{\ref{aff48}}
\and T.~I.~Liaudat\orcid{0000-0002-9104-314X}\inst{\ref{aff149}}
\and S.~J.~Liu\orcid{0000-0001-7680-2139}\inst{\ref{aff64}}
\and M.~Magliocchetti\orcid{0000-0001-9158-4838}\inst{\ref{aff64}}
\and E.~A.~Magnier\orcid{0000-0002-7965-2815}\inst{\ref{aff150}}
\and F.~Mannucci\orcid{0000-0002-4803-2381}\inst{\ref{aff151}}
\and C.~J.~A.~P.~Martins\orcid{0000-0002-4886-9261}\inst{\ref{aff152},\ref{aff153}}
\and L.~Maurin\orcid{0000-0002-8406-0857}\inst{\ref{aff61}}
\and M.~Miluzio\inst{\ref{aff19},\ref{aff154}}
\and C.~Moretti\orcid{0000-0003-3314-8936}\inst{\ref{aff33},\ref{aff32},\ref{aff34}}
\and G.~Morgante\inst{\ref{aff7}}
\and K.~Naidoo\orcid{0000-0002-9182-1802}\inst{\ref{aff11},\ref{aff75}}
\and A.~Navarro-Alsina\orcid{0000-0002-3173-2592}\inst{\ref{aff84}}
\and S.~Nesseris\orcid{0000-0002-0567-0324}\inst{\ref{aff102}}
\and D.~Paoletti\orcid{0000-0003-4761-6147}\inst{\ref{aff7},\ref{aff66}}
\and F.~Passalacqua\orcid{0000-0002-8606-4093}\inst{\ref{aff101},\ref{aff62}}
\and K.~Paterson\orcid{0000-0001-8340-3486}\inst{\ref{aff75}}
\and L.~Patrizii\inst{\ref{aff8}}
\and A.~Pisani\orcid{0000-0002-6146-4437}\inst{\ref{aff63}}
\and D.~Potter\orcid{0000-0002-0757-5195}\inst{\ref{aff155}}
\and G.~W.~Pratt\inst{\ref{aff65}}
\and S.~Quai\orcid{0000-0002-0449-8163}\inst{\ref{aff9},\ref{aff7}}
\and M.~Radovich\orcid{0000-0002-3585-866X}\inst{\ref{aff37}}
\and K.~Rojas\orcid{0000-0003-1391-6854}\inst{\ref{aff83}}
\and W.~Roster\orcid{0000-0002-9149-6528}\inst{\ref{aff22}}
\and S.~Sacquegna\orcid{0000-0002-8433-6630}\inst{\ref{aff156}}
\and M.~Sahl\'en\orcid{0000-0003-0973-4804}\inst{\ref{aff157}}
\and D.~B.~Sanders\orcid{0000-0002-1233-9998}\inst{\ref{aff150}}
\and E.~Sarpa\orcid{0000-0002-1256-655X}\inst{\ref{aff35},\ref{aff119},\ref{aff33}}
\and C.~Scarlata\orcid{0000-0002-9136-8876}\inst{\ref{aff158}}
\and A.~Schneider\orcid{0000-0001-7055-8104}\inst{\ref{aff155}}
\and M.~Schultheis\inst{\ref{aff118}}
\and D.~Sciotti\orcid{0009-0008-4519-2620}\inst{\ref{aff44},\ref{aff45}}
\and E.~Sellentin\inst{\ref{aff159},\ref{aff160}}
\and L.~C.~Smith\orcid{0000-0002-3259-2771}\inst{\ref{aff161}}
\and K.~Tanidis\orcid{0000-0001-9843-5130}\inst{\ref{aff162}}
\and C.~Tao\orcid{0000-0001-7961-8177}\inst{\ref{aff63}}
\and F.~Tarsitano\orcid{0000-0002-5919-0238}\inst{\ref{aff163},\ref{aff60}}
\and G.~Testera\inst{\ref{aff39}}
\and R.~Teyssier\orcid{0000-0001-7689-0933}\inst{\ref{aff164}}
\and S.~Tosi\orcid{0000-0002-7275-9193}\inst{\ref{aff38},\ref{aff31},\ref{aff39}}
\and A.~Troja\orcid{0000-0003-0239-4595}\inst{\ref{aff101},\ref{aff62}}
\and A.~Venhola\orcid{0000-0001-6071-4564}\inst{\ref{aff165}}
\and D.~Vergani\orcid{0000-0003-0898-2216}\inst{\ref{aff7}}
\and G.~Vernardos\orcid{0000-0001-8554-7248}\inst{\ref{aff166},\ref{aff167}}
\and G.~Verza\orcid{0000-0002-1886-8348}\inst{\ref{aff168},\ref{aff169}}
\and S.~Vinciguerra\orcid{0009-0005-4018-3184}\inst{\ref{aff16}}
\and N.~A.~Walton\orcid{0000-0003-3983-8778}\inst{\ref{aff161}}
\and A.~H.~Wright\orcid{0000-0001-7363-7932}\inst{\ref{aff140}}
\and H.~W.~Yeung\orcid{0000-0002-4993-9014}\inst{\ref{aff53}}}
										   
\institute{Department of Physics \& Astronomy, University of California Irvine, Irvine CA 92697, USA\label{aff1}
\and
University of Southern California, 3551 Trousdale Parkway, Los Angeles, CA 90089, USA\label{aff2}
\and
Dipartimento di Fisica "Aldo Pontremoli", Universit\`a degli Studi di Milano, Via Celoria 16, 20133 Milano, Italy\label{aff3}
\and
INAF-IASF Milano, Via Alfonso Corti 12, 20133 Milano, Italy\label{aff4}
\and
Jet Propulsion Laboratory, California Institute of Technology, 4800 Oak Grove Drive, Pasadena, CA, 91109, USA\label{aff5}
\and
Department of Physics and Astronomy, University of British Columbia, Vancouver, BC V6T 1Z1, Canada\label{aff6}
\and
INAF-Osservatorio di Astrofisica e Scienza dello Spazio di Bologna, Via Piero Gobetti 93/3, 40129 Bologna, Italy\label{aff7}
\and
INFN-Sezione di Bologna, Viale Berti Pichat 6/2, 40127 Bologna, Italy\label{aff8}
\and
Dipartimento di Fisica e Astronomia "Augusto Righi" - Alma Mater Studiorum Universit\`a di Bologna, via Piero Gobetti 93/2, 40129 Bologna, Italy\label{aff9}
\and
INAF-Osservatorio Astronomico di Capodimonte, Via Moiariello 16, 80131 Napoli, Italy\label{aff10}
\and
Institute of Cosmology and Gravitation, University of Portsmouth, Portsmouth PO1 3FX, UK\label{aff11}
\and
School of Physics, HH Wills Physics Laboratory, University of Bristol, Tyndall Avenue, Bristol, BS8 1TL, UK\label{aff12}
\and
Max-Planck-Institut f\"ur Astrophysik, Karl-Schwarzschild-Str.~1, 85748 Garching, Germany\label{aff13}
\and
Technical University of Munich, TUM School of Natural Sciences, Physics Department, James-Franck-Str.~1, 85748 Garching, Germany\label{aff14}
\and
Departamento F\'isica Aplicada, Universidad Polit\'ecnica de Cartagena, Campus Muralla del Mar, 30202 Cartagena, Murcia, Spain\label{aff15}
\and
Aix-Marseille Universit\'e, CNRS, CNES, LAM, Marseille, France\label{aff16}
\and
Institut d'Astrophysique de Paris, UMR 7095, CNRS, and Sorbonne Universit\'e, 98 bis boulevard Arago, 75014 Paris, France\label{aff17}
\and
Center for Astrophysics | Harvard \& Smithsonian, 60 Garden St., Cambridge, MA 02138, USA\label{aff18}
\and
ESAC/ESA, Camino Bajo del Castillo, s/n., Urb. Villafranca del Castillo, 28692 Villanueva de la Ca\~nada, Madrid, Spain\label{aff19}
\and
Universit\'e de Gen\`eve, D\'epartement de Physique Th\'eorique and Centre for Astroparticle Physics, 24 quai Ernest-Ansermet, CH-1211 Gen\`eve 4, Switzerland\label{aff20}
\and
Lawrence Berkeley National Laboratory, One Cyclotron Road, Berkeley, CA 94720, USA\label{aff21}
\and
Max Planck Institute for Extraterrestrial Physics, Giessenbachstr. 1, 85748 Garching, Germany\label{aff22}
\and
Universit\"ats-Sternwarte M\"unchen, Fakult\"at f\"ur Physik, Ludwig-Maximilians-Universit\"at M\"unchen, Scheinerstr.~1, 81679 M\"unchen, Germany\label{aff23}
\and
STAR Institute, University of Li{\`e}ge, Quartier Agora, All\'ee du six Ao\^ut 19c, 4000 Li\`ege, Belgium\label{aff24}
\and
Institut de Ci\`{e}ncies del Cosmos (ICCUB), Universitat de Barcelona (IEEC-UB), Mart\'{i} i Franqu\`{e}s 1, 08028 Barcelona, Spain\label{aff25}
\and
David A. Dunlap Department of Astronomy \& Astrophysics, University of Toronto, 50 St George Street, Toronto, Ontario M5S 3H4, Canada\label{aff26}
\and
Jodrell Bank Centre for Astrophysics, Department of Physics and Astronomy, University of Manchester, Oxford Road, Manchester M13 9PL, UK\label{aff27}
\and
Instituci\'o Catalana de Recerca i Estudis Avan\c{c}ats (ICREA), Passeig de Llu\'{\i}s Companys 23, 08010 Barcelona, Spain\label{aff28}
\and
Institut de Ciencies de l'Espai (IEEC-CSIC), Campus UAB, Carrer de Can Magrans, s/n Cerdanyola del Vall\'es, 08193 Barcelona, Spain\label{aff29}
\and
School of Mathematics and Physics, University of Surrey, Guildford, Surrey, GU2 7XH, UK\label{aff30}
\and
INAF-Osservatorio Astronomico di Brera, Via Brera 28, 20122 Milano, Italy\label{aff31}
\and
IFPU, Institute for Fundamental Physics of the Universe, via Beirut 2, 34151 Trieste, Italy\label{aff32}
\and
INAF-Osservatorio Astronomico di Trieste, Via G. B. Tiepolo 11, 34143 Trieste, Italy\label{aff33}
\and
INFN, Sezione di Trieste, Via Valerio 2, 34127 Trieste TS, Italy\label{aff34}
\and
SISSA, International School for Advanced Studies, Via Bonomea 265, 34136 Trieste TS, Italy\label{aff35}
\and
Dipartimento di Fisica e Astronomia, Universit\`a di Bologna, Via Gobetti 93/2, 40129 Bologna, Italy\label{aff36}
\and
INAF-Osservatorio Astronomico di Padova, Via dell'Osservatorio 5, 35122 Padova, Italy\label{aff37}
\and
Dipartimento di Fisica, Universit\`a di Genova, Via Dodecaneso 33, 16146, Genova, Italy\label{aff38}
\and
INFN-Sezione di Genova, Via Dodecaneso 33, 16146, Genova, Italy\label{aff39}
\and
Department of Physics "E. Pancini", University Federico II, Via Cinthia 6, 80126, Napoli, Italy\label{aff40}
\and
Dipartimento di Fisica, Universit\`a degli Studi di Torino, Via P. Giuria 1, 10125 Torino, Italy\label{aff41}
\and
INFN-Sezione di Torino, Via P. Giuria 1, 10125 Torino, Italy\label{aff42}
\and
INAF-Osservatorio Astrofisico di Torino, Via Osservatorio 20, 10025 Pino Torinese (TO), Italy\label{aff43}
\and
INAF-Osservatorio Astronomico di Roma, Via Frascati 33, 00078 Monteporzio Catone, Italy\label{aff44}
\and
INFN-Sezione di Roma, Piazzale Aldo Moro, 2 - c/o Dipartimento di Fisica, Edificio G. Marconi, 00185 Roma, Italy\label{aff45}
\and
Centro de Investigaciones Energ\'eticas, Medioambientales y Tecnol\'ogicas (CIEMAT), Avenida Complutense 40, 28040 Madrid, Spain\label{aff46}
\and
Port d'Informaci\'{o} Cient\'{i}fica, Campus UAB, C. Albareda s/n, 08193 Bellaterra (Barcelona), Spain\label{aff47}
\and
Institute for Theoretical Particle Physics and Cosmology (TTK), RWTH Aachen University, 52056 Aachen, Germany\label{aff48}
\and
Deutsches Zentrum f\"ur Luft- und Raumfahrt e. V. (DLR), Linder H\"ohe, 51147 K\"oln, Germany\label{aff49}
\and
INFN section of Naples, Via Cinthia 6, 80126, Napoli, Italy\label{aff50}
\and
Dipartimento di Fisica e Astronomia "Augusto Righi" - Alma Mater Studiorum Universit\`a di Bologna, Viale Berti Pichat 6/2, 40127 Bologna, Italy\label{aff51}
\and
Instituto de Astrof\'{\i}sica de Canarias, E-38205 La Laguna, Tenerife, Spain\label{aff52}
\and
Institute for Astronomy, University of Edinburgh, Royal Observatory, Blackford Hill, Edinburgh EH9 3HJ, UK\label{aff53}
\and
European Space Agency/ESRIN, Largo Galileo Galilei 1, 00044 Frascati, Roma, Italy\label{aff54}
\and
Universit\'e Claude Bernard Lyon 1, CNRS/IN2P3, IP2I Lyon, UMR 5822, Villeurbanne, F-69100, France\label{aff55}
\and
UCB Lyon 1, CNRS/IN2P3, IUF, IP2I Lyon, 4 rue Enrico Fermi, 69622 Villeurbanne, France\label{aff56}
\and
Mullard Space Science Laboratory, University College London, Holmbury St Mary, Dorking, Surrey RH5 6NT, UK\label{aff57}
\and
Departamento de F\'isica, Faculdade de Ci\^encias, Universidade de Lisboa, Edif\'icio C8, Campo Grande, PT1749-016 Lisboa, Portugal\label{aff58}
\and
Instituto de Astrof\'isica e Ci\^encias do Espa\c{c}o, Faculdade de Ci\^encias, Universidade de Lisboa, Campo Grande, 1749-016 Lisboa, Portugal\label{aff59}
\and
Department of Astronomy, University of Geneva, ch. d'Ecogia 16, 1290 Versoix, Switzerland\label{aff60}
\and
Universit\'e Paris-Saclay, CNRS, Institut d'astrophysique spatiale, 91405, Orsay, France\label{aff61}
\and
INFN-Padova, Via Marzolo 8, 35131 Padova, Italy\label{aff62}
\and
Aix-Marseille Universit\'e, CNRS/IN2P3, CPPM, Marseille, France\label{aff63}
\and
INAF-Istituto di Astrofisica e Planetologia Spaziali, via del Fosso del Cavaliere, 100, 00100 Roma, Italy\label{aff64}
\and
Universit\'e Paris-Saclay, Universit\'e Paris Cit\'e, CEA, CNRS, AIM, 91191, Gif-sur-Yvette, France\label{aff65}
\and
INFN-Bologna, Via Irnerio 46, 40126 Bologna, Italy\label{aff66}
\and
Institut d'Estudis Espacials de Catalunya (IEEC),  Edifici RDIT, Campus UPC, 08860 Castelldefels, Barcelona, Spain\label{aff67}
\and
Institute of Space Sciences (ICE, CSIC), Campus UAB, Carrer de Can Magrans, s/n, 08193 Barcelona, Spain\label{aff68}
\and
University Observatory, LMU Faculty of Physics, Scheinerstr.~1, 81679 Munich, Germany\label{aff69}
\and
FRACTAL S.L.N.E., calle Tulip\'an 2, Portal 13 1A, 28231, Las Rozas de Madrid, Spain\label{aff70}
\and
Institute of Theoretical Astrophysics, University of Oslo, P.O. Box 1029 Blindern, 0315 Oslo, Norway\label{aff71}
\and
Felix Hormuth Engineering, Goethestr. 17, 69181 Leimen, Germany\label{aff72}
\and
Technical University of Denmark, Elektrovej 327, 2800 Kgs. Lyngby, Denmark\label{aff73}
\and
Cosmic Dawn Center (DAWN), Denmark\label{aff74}
\and
Max-Planck-Institut f\"ur Astronomie, K\"onigstuhl 17, 69117 Heidelberg, Germany\label{aff75}
\and
NASA Goddard Space Flight Center, Greenbelt, MD 20771, USA\label{aff76}
\and
Department of Physics and Astronomy, University College London, Gower Street, London WC1E 6BT, UK\label{aff77}
\and
Department of Physics, P.O. Box 64, University of Helsinki, 00014 Helsinki, Finland\label{aff78}
\and
Helsinki Institute of Physics, Gustaf H{\"a}llstr{\"o}min katu 2, University of Helsinki, 00014 Helsinki, Finland\label{aff79}
\and
Laboratoire d'etude de l'Univers et des phenomenes eXtremes, Observatoire de Paris, Universit\'e PSL, Sorbonne Universit\'e, CNRS, 92190 Meudon, France\label{aff80}
\and
SKAO, Jodrell Bank, Lower Withington, Macclesfield SK11 9FT, UK\label{aff81}
\and
Centre de Calcul de l'IN2P3/CNRS, 21 avenue Pierre de Coubertin 69627 Villeurbanne Cedex, France\label{aff82}
\and
University of Applied Sciences and Arts of Northwestern Switzerland, School of Computer Science, 5210 Windisch, Switzerland\label{aff83}
\and
Universit\"at Bonn, Argelander-Institut f\"ur Astronomie, Auf dem H\"ugel 71, 53121 Bonn, Germany\label{aff84}
\and
Department of Physics, Institute for Computational Cosmology, Durham University, South Road, Durham, DH1 3LE, UK\label{aff85}
\and
Universit\'e Paris Cit\'e, CNRS, Astroparticule et Cosmologie, 75013 Paris, France\label{aff86}
\and
CNRS-UCB International Research Laboratory, Centre Pierre Bin\'etruy, IRL2007, CPB-IN2P3, Berkeley, USA\label{aff87}
\and
University of Applied Sciences and Arts of Northwestern Switzerland, School of Engineering, 5210 Windisch, Switzerland\label{aff88}
\and
Institute of Physics, Laboratory of Astrophysics, Ecole Polytechnique F\'ed\'erale de Lausanne (EPFL), Observatoire de Sauverny, 1290 Versoix, Switzerland\label{aff89}
\and
Telespazio UK S.L. for European Space Agency (ESA), Camino bajo del Castillo, s/n, Urbanizacion Villafranca del Castillo, Villanueva de la Ca\~nada, 28692 Madrid, Spain\label{aff90}
\and
Institut de F\'{i}sica d'Altes Energies (IFAE), The Barcelona Institute of Science and Technology, Campus UAB, 08193 Bellaterra (Barcelona), Spain\label{aff91}
\and
European Space Agency/ESTEC, Keplerlaan 1, 2201 AZ Noordwijk, The Netherlands\label{aff92}
\and
School of Mathematics, Statistics and Physics, Newcastle University, Herschel Building, Newcastle-upon-Tyne, NE1 7RU, UK\label{aff93}
\and
DARK, Niels Bohr Institute, University of Copenhagen, Jagtvej 155, 2200 Copenhagen, Denmark\label{aff94}
\and
Waterloo Centre for Astrophysics, University of Waterloo, Waterloo, Ontario N2L 3G1, Canada\label{aff95}
\and
Department of Physics and Astronomy, University of Waterloo, Waterloo, Ontario N2L 3G1, Canada\label{aff96}
\and
Perimeter Institute for Theoretical Physics, Waterloo, Ontario N2L 2Y5, Canada\label{aff97}
\and
Space Science Data Center, Italian Space Agency, via del Politecnico snc, 00133 Roma, Italy\label{aff98}
\and
Centre National d'Etudes Spatiales -- Centre spatial de Toulouse, 18 avenue Edouard Belin, 31401 Toulouse Cedex 9, France\label{aff99}
\and
Institute of Space Science, Str. Atomistilor, nr. 409 M\u{a}gurele, Ilfov, 077125, Romania\label{aff100}
\and
Dipartimento di Fisica e Astronomia "G. Galilei", Universit\`a di Padova, Via Marzolo 8, 35131 Padova, Italy\label{aff101}
\and
Instituto de F\'isica Te\'orica UAM-CSIC, Campus de Cantoblanco, 28049 Madrid, Spain\label{aff102}
\and
Institut de Recherche en Astrophysique et Plan\'etologie (IRAP), Universit\'e de Toulouse, CNRS, UPS, CNES, 14 Av. Edouard Belin, 31400 Toulouse, France\label{aff103}
\and
Universit\'e St Joseph; Faculty of Sciences, Beirut, Lebanon\label{aff104}
\and
Departamento de F\'isica, FCFM, Universidad de Chile, Blanco Encalada 2008, Santiago, Chile\label{aff105}
\and
Universit\"at Innsbruck, Institut f\"ur Astro- und Teilchenphysik, Technikerstr. 25/8, 6020 Innsbruck, Austria\label{aff106}
\and
Department of Physics and Helsinki Institute of Physics, Gustaf H\"allstr\"omin katu 2, University of Helsinki, 00014 Helsinki, Finland\label{aff107}
\and
Instituto de Astrof\'isica e Ci\^encias do Espa\c{c}o, Faculdade de Ci\^encias, Universidade de Lisboa, Tapada da Ajuda, 1349-018 Lisboa, Portugal\label{aff108}
\and
Cosmic Dawn Center (DAWN)\label{aff109}
\and
Niels Bohr Institute, University of Copenhagen, Jagtvej 128, 2200 Copenhagen, Denmark\label{aff110}
\and
Universidad Polit\'ecnica de Cartagena, Departamento de Electr\'onica y Tecnolog\'ia de Computadoras,  Plaza del Hospital 1, 30202 Cartagena, Spain\label{aff111}
\and
Kapteyn Astronomical Institute, University of Groningen, PO Box 800, 9700 AV Groningen, The Netherlands\label{aff112}
\and
Caltech/IPAC, 1200 E. California Blvd., Pasadena, CA 91125, USA\label{aff113}
\and
Dipartimento di Fisica e Scienze della Terra, Universit\`a degli Studi di Ferrara, Via Giuseppe Saragat 1, 44122 Ferrara, Italy\label{aff114}
\and
Istituto Nazionale di Fisica Nucleare, Sezione di Ferrara, Via Giuseppe Saragat 1, 44122 Ferrara, Italy\label{aff115}
\and
INAF, Istituto di Radioastronomia, Via Piero Gobetti 101, 40129 Bologna, Italy\label{aff116}
\and
Astronomical Observatory of the Autonomous Region of the Aosta Valley (OAVdA), Loc. Lignan 39, I-11020, Nus (Aosta Valley), Italy\label{aff117}
\and
Universit\'e C\^{o}te d'Azur, Observatoire de la C\^{o}te d'Azur, CNRS, Laboratoire Lagrange, Bd de l'Observatoire, CS 34229, 06304 Nice cedex 4, France\label{aff118}
\and
ICSC - Centro Nazionale di Ricerca in High Performance Computing, Big Data e Quantum Computing, Via Magnanelli 2, Bologna, Italy\label{aff119}
\and
Department of Physics, Oxford University, Keble Road, Oxford OX1 3RH, UK\label{aff120}
\and
Univ. Grenoble Alpes, CNRS, Grenoble INP, LPSC-IN2P3, 53, Avenue des Martyrs, 38000, Grenoble, France\label{aff121}
\and
Dipartimento di Fisica, Sapienza Universit\`a di Roma, Piazzale Aldo Moro 2, 00185 Roma, Italy\label{aff122}
\and
Aurora Technology for European Space Agency (ESA), Camino bajo del Castillo, s/n, Urbanizacion Villafranca del Castillo, Villanueva de la Ca\~nada, 28692 Madrid, Spain\label{aff123}
\and
Dipartimento di Fisica - Sezione di Astronomia, Universit\`a di Trieste, Via Tiepolo 11, 34131 Trieste, Italy\label{aff124}
\and
Department of Mathematics and Physics E. De Giorgi, University of Salento, Via per Arnesano, CP-I93, 73100, Lecce, Italy\label{aff125}
\and
INFN, Sezione di Lecce, Via per Arnesano, CP-193, 73100, Lecce, Italy\label{aff126}
\and
INAF-Sezione di Lecce, c/o Dipartimento Matematica e Fisica, Via per Arnesano, 73100, Lecce, Italy\label{aff127}
\and
Institut d'Astrophysique de Paris, 98bis Boulevard Arago, 75014, Paris, France\label{aff128}
\and
ICL, Junia, Universit\'e Catholique de Lille, LITL, 59000 Lille, France\label{aff129}
\and
CERCA/ISO, Department of Physics, Case Western Reserve University, 10900 Euclid Avenue, Cleveland, OH 44106, USA\label{aff130}
\and
Laboratoire Univers et Th\'eorie, Observatoire de Paris, Universit\'e PSL, Universit\'e Paris Cit\'e, CNRS, 92190 Meudon, France\label{aff131}
\and
INFN-Sezione di Milano, Via Celoria 16, 20133 Milano, Italy\label{aff132}
\and
Departamento de F{\'\i}sica Fundamental. Universidad de Salamanca. Plaza de la Merced s/n. 37008 Salamanca, Spain\label{aff133}
\and
Universit\'e de Strasbourg, CNRS, Observatoire astronomique de Strasbourg, UMR 7550, 67000 Strasbourg, France\label{aff134}
\and
Center for Data-Driven Discovery, Kavli IPMU (WPI), UTIAS, The University of Tokyo, Kashiwa, Chiba 277-8583, Japan\label{aff135}
\and
California Institute of Technology, 1200 E California Blvd, Pasadena, CA 91125, USA\label{aff136}
\and
Instituto de F\'isica de Cantabria, Edificio Juan Jord\'a, Avenida de los Castros, 39005 Santander, Spain\label{aff137}
\and
Department of Computer Science, Aalto University, PO Box 15400, Espoo, FI-00 076, Finland\label{aff138}
\and
Universidad de La Laguna, Dpto. Astrof\'\i sica, E-38206 La Laguna, Tenerife, Spain\label{aff139}
\and
Ruhr University Bochum, Faculty of Physics and Astronomy, Astronomical Institute (AIRUB), German Centre for Cosmological Lensing (GCCL), 44780 Bochum, Germany\label{aff140}
\and
Department of Physics and Astronomy, Vesilinnantie 5, University of Turku, 20014 Turku, Finland\label{aff141}
\and
Finnish Centre for Astronomy with ESO (FINCA), Quantum, Vesilinnantie 5, University of Turku, 20014 Turku, Finland\label{aff142}
\and
Serco for European Space Agency (ESA), Camino bajo del Castillo, s/n, Urbanizacion Villafranca del Castillo, Villanueva de la Ca\~nada, 28692 Madrid, Spain\label{aff143}
\and
ARC Centre of Excellence for Dark Matter Particle Physics, Melbourne, Australia\label{aff144}
\and
Centre for Astrophysics \& Supercomputing, Swinburne University of Technology,  Hawthorn, Victoria 3122, Australia\label{aff145}
\and
Department of Physics and Astronomy, University of the Western Cape, Bellville, Cape Town, 7535, South Africa\label{aff146}
\and
Departement of Theoretical Physics, University of Geneva, Switzerland\label{aff147}
\and
Department of Physics, Centre for Extragalactic Astronomy, Durham University, South Road, Durham, DH1 3LE, UK\label{aff148}
\and
IRFU, CEA, Universit\'e Paris-Saclay 91191 Gif-sur-Yvette Cedex, France\label{aff149}
\and
Institute for Astronomy, University of Hawaii, 2680 Woodlawn Drive, Honolulu, HI 96822, USA\label{aff150}
\and
INAF-Osservatorio Astrofisico di Arcetri, Largo E. Fermi 5, 50125, Firenze, Italy\label{aff151}
\and
Centro de Astrof\'{\i}sica da Universidade do Porto, Rua das Estrelas, 4150-762 Porto, Portugal\label{aff152}
\and
Instituto de Astrof\'isica e Ci\^encias do Espa\c{c}o, Universidade do Porto, CAUP, Rua das Estrelas, PT4150-762 Porto, Portugal\label{aff153}
\and
HE Space for European Space Agency (ESA), Camino bajo del Castillo, s/n, Urbanizacion Villafranca del Castillo, Villanueva de la Ca\~nada, 28692 Madrid, Spain\label{aff154}
\and
Department of Astrophysics, University of Zurich, Winterthurerstrasse 190, 8057 Zurich, Switzerland\label{aff155}
\and
INAF - Osservatorio Astronomico d'Abruzzo, Via Maggini, 64100, Teramo, Italy\label{aff156}
\and
Theoretical astrophysics, Department of Physics and Astronomy, Uppsala University, Box 516, 751 37 Uppsala, Sweden\label{aff157}
\and
Minnesota Institute for Astrophysics, University of Minnesota, 116 Church St SE, Minneapolis, MN 55455, USA\label{aff158}
\and
Mathematical Institute, University of Leiden, Einsteinweg 55, 2333 CA Leiden, The Netherlands\label{aff159}
\and
Leiden Observatory, Leiden University, Einsteinweg 55, 2333 CC Leiden, The Netherlands\label{aff160}
\and
Institute of Astronomy, University of Cambridge, Madingley Road, Cambridge CB3 0HA, UK\label{aff161}
\and
Center for Astrophysics and Cosmology, University of Nova Gorica, Nova Gorica, Slovenia\label{aff162}
\and
Institute for Particle Physics and Astrophysics, Dept. of Physics, ETH Zurich, Wolfgang-Pauli-Strasse 27, 8093 Zurich, Switzerland\label{aff163}
\and
Department of Astrophysical Sciences, Peyton Hall, Princeton University, Princeton, NJ 08544, USA\label{aff164}
\and
Space physics and astronomy research unit, University of Oulu, Pentti Kaiteran katu 1, FI-90014 Oulu, Finland\label{aff165}
\and
Department of Physics and Astronomy, Lehman College of the CUNY, Bronx, NY 10468, USA\label{aff166}
\and
American Museum of Natural History, Department of Astrophysics, New York, NY 10024, USA\label{aff167}
\and
International Centre for Theoretical Physics (ICTP), Strada Costiera 11, 34151 Trieste, Italy\label{aff168}
\and
Center for Computational Astrophysics, Flatiron Institute, 162 5th Avenue, 10010, New York, NY, USA\label{aff169}}



%
%

\abstract{
We present an end-to-end, iterative pipeline for efficient identification of strong galaxy--galaxy lensing systems, applied to the \emph{Euclid} Q1 imaging data. Starting from VIS catalogues, we reject point sources, apply a magnitude cut ($\IE\leq 24$) on deflectors, and run a pixel-level artefact/noise filter to build $96\times 96$\,pix cutouts; VIS+NISP colour composites are constructed with a VIS-anchored luminance scheme that preserves VIS morphology and NISP colour contrast. A VIS-only seed classifier supplies clear positives and typical impostors, from which we curate a morphology-balanced negative set and augment scarce positives. Among the six compact CNNs studied initially, a modified VGG16 (GlobalAveragePooling + 256/128 dense layers with the last nine layers trainable) performs best; the training set grows from 27 seed lenses (augmented $67\times$ to 1809) plus 2000 negatives to a colour dataset of 30\,686 images. After three rounds of iterative fine-tuning, human grading of the top 4000 candidates ranked by the final model yields 441 Grade A/B candidate lensing systems, including 311 overlapping with the existing Q1 strong-lens catalogue, and 130 additional A/B candidates (9 As and 121 Bs) not previously reported. Independently, the model recovers 740 out of 905 (81.8\%) candidate Q1 lenses within its top 20\,000 predictions, considering off-centred samples. Candidates span $\IE\simeq 17$--24\,AB\,mag (median 21.3\,AB\,mag) and are redder in $\YE-\HE$ than the parent population, consistent with massive early-type deflectors. Each training iteration required about a week for a small team, and the approach easily scales to future wide-area \emph{Euclid} releases; future work will calibrate the selection function via lens injection, extend recall through uncertainty-aware active learning, and explore multi-scale or attention-based neural networks with fast post-hoc vetters that incorporate lens models into the classification.
}

%
%
%
%
\titlerunning{\Euclid Q1: A Scalable CNN-based Pipeline for Strong Gravitational Lens
Identification}
\authorrunning{Euclid Collaboration: X. Xu et al.}

 \maketitle
%
%
%
%
   
\section{\label{sc:Intro}Introduction}
Strong gravitational lensing, a phenomenon predicted by Einstein's theory of general relativity, occurs when the gravitational potential of a massive foreground astronomical object, such as a galaxy or galaxy cluster, deflects and magnifies light from a background source. This deflection can lead to the formation of multiple, distorted images or, in cases of precise alignment, characteristic structures such as Einstein rings. The observational properties of strong lensing systems enable wide astrophysical and cosmological applications. Given a relative alignment, the lensing features of the background depend entirely on the gravitational potential of the foreground, making it a useful probe of luminous and dark masses within the foreground \citep[e.g.,][]{Koopman_review,Treu_2010,Vegetti_review}.

The lensed images are sensitive to small perturbations in the lensing potential, therefore it can be used to probe dark matter substructures in the foreground \citep[e.g.,][]{Dalal_2002,Koopmans_2005,Keeton_2009,Vegetti_2010,Vegetti_2014,Hezaveh_2016,Gannon_review}. The determination of the total foreground mass from lensing in combination with stellar mass analysis from photometry yields valuable constraints on the initial mass function (IMF) of galaxies outside the Local Group \citep[e.g.,][]{Treu_2010b,Sonnenfeld_2013,Sonnenfeld_2015}. Since multiple image copies of the same lensed background traverse different paths to Earth, there is often a time delay between different images. This difference is measurable if the background source is time variable, and can be used to measure cosmological parameters such as the Hubble constant $H_0$ \citep[e.g.,][]{Wong_2020,Rodney_2021}. Statistics of a large population of lensing systems can be used to constrain cosmological evolution and galaxy mass structures \citep[e.g.,][]{Cen_1994,Chae_2003,Sonnenfeld_2021a,Sonnenfeld_2021b}, and well-defined selection criteria can further strengthen these constraints \citep{Sonnenfeld_2022}. Strong lensing can also be used to aid the observation of very high-redshift galaxies as it significantly magnifies and amplifies the light from the background source \citep[e.g.,][]{Chiriv_2020,Fudamoto_2025}. For all of its scientific value, the discovery of more strong lensing systems is highly desirable.

The \Euclid Mission \citep{EuclidSkyOverview}, developed by the European Space Agency with contributions from NASA, is projected to survey most of the extragalactic sky and will observe billions of galaxies \citep{Scaramella-EP1} with the space-based telescope. Initial estimates from simulations suggest that $\sim$~170\,000 strong lenses should be observed in the \Euclid dataset \citep{Collett_2015}. Extrapolations from strong lenses identified in the preliminary results also show broadly consistent expectations \citep{AcevedoBarroso24,Q1-SP048}. Though visual inspection has historically been utilised for finding lensing systems \citep[e.g.,][]{Jackson_2008,Faure_2008}, it is infeasible for a survey of \Euclid's scale. Consequently, the development of efficient and reliable automated detection algorithms is imperative to systematically identify and catalogue strong lensing systems, thereby maximising the scientific return of the \Euclid survey. 

The detection of strong lenses in large-scale astronomical surveys presents several inherent challenges. Statistically, strong lensing events are exceptionally rare, occurring approximately once per $\mathcal{O}(10^4)$ observed galaxies; nevertheless, \Euclid is projected to observe a larger sample of lenses than any previous surveys \citep[see Fig. 1 in][]{Q1-SP048}. Furthermore, lensing features exhibit considerable morphological diversity and are frequently confused with other astrophysical objects, such as interacting and ring galaxies. Consequently, it is essential to deploy classification algorithms with high precision and recall to minimise false positives while ensuring high completeness of the true lens sample.

Previous strong lensing surveys and associated detection efforts have employed a range of techniques. The initial discoveries were often serendipitous. Subsequent systematic searches, such as the Cosmic Lens All-Sky Survey \citep{Myers_2003} and the Sloan Lens ACS Survey \citep{Bolton_2006}, utilised targeted observational strategies combined with visual inspection or semi-automated algorithms based on predefined morphological or photometric criteria. Although these surveys yielded significant discoveries, their methodologies possess inherent limitations with respect to scalability and potential selection biases. More recent wide-field surveys, including the Kilo-Degree Survey \citep{Jong_2013}, the Dark Energy Survey \citep{Abbott_2016}, and the Hyper Suprime-Cam SSP Survey \citep{Aihara_2018}, have further underscored the need for advanced automated detection techniques commensurate with increasing data volumes.

The application of machine learning techniques in astronomy has seen considerable growth, offering powerful tools for data analysis and interpretation. Specifically, deep learning architectures, such as Convolutional Neural Networks (CNN; \citealt{LeCunBengio1995}), have proven effective in tasks involving image classification from astronomical datasets \citep{Carleo_2019}, such as morphological classification of SDSS galaxies \citep{Zhu_2019}. Several studies have successfully deployed CNNs for the identification of strong lens candidates in observational survey data, frequently achieving performance metrics that surpass those of traditional algorithms in terms of accuracy and computational efficiency \citep{Petrillo2017_KiDS_lenses,Jacobs_2017,Pourrahmani2018,Jacobs2019_DES_catalog,Cheng_2020,Li_2020,Huang2021_DESI_lenses,Storfer_2024}.

Recently, much effort has been made to find strong lenses in the new \Euclid data releases. Initial experiments have been conducted on the \Euclid ERO data to yield preliminary validation of the lens finding pipelines \citep{Pearce-Casey24,AcevedoBarroso24,Nagam25}. A much larger search was then conducted in the \Euclid Q1 data, notably by the strong lens discovery engine (SLDE) series. SLDE A \citep{Q1-SP048} encompasses an overview of SLDE research in \Euclid Q1. SLDE B \citep{Q1-SP052} reports early discoveries based on spectroscopically confirmed high-velocity-dispersion galaxies, forming training samples for machine learning models in SLDE C \citep{Q1-SP053}, which also evaluates their performance using real \Euclid imaging data. SLDE D \citep{Q1-SP054} focuses on the double source plane lenses identified in Q1 and provides preliminary modelling for a forecast of cosmological parameters. SLDE E \citep{Q1-SP059} introduces a Bayesian ensemble approach that combines multiple lens classifiers to further optimise lens discovery in the DR1 discovery engine. SLDE F \citep{Q1-SP099} presents additional bright and low redshift lens samples found in Q1 by looking into the \textit{Gaia} cross-matched sources previously excluded in SLDE. In addition to CNN-based pipelines, vision transformers have recently been explored for strong lens detection, motivated by their capacity to model long-range spatial correlations \citep{astrovink}, though their relative advantages over CNNs for \Euclid data are not yet firmly established.

These prior investigations provide a robust foundation and compelling evidence for the utility of deep learning in the domain of strong lens detection. This paper aims to extend these efforts by developing and evaluating a CNN-based methodology specifically optimised for the characteristics of the \Euclid survey data, addressing the critical need for high-fidelity lens identification in the era of large-scale astronomical surveys.

This paper is organised as follows. Section~\ref{sc:Data} gives an overview of the \Euclid Q1 dataset, then describes our expectation and how this work complements existing lens search efforts on Q1 dataset. Section~\ref{sc:Blunders} presents the framework of the automated strong-lens detection pipeline. It then details the specific data generation procedure to procure the full dataset used for training and inference. These include: catalogue based pre-filtering, cutout generation and subsequent enhancements, pixel-level artefact/noise cleaning, balanced hard-negative curation, colour-set data composition, and the iterative fine-tuning of compact CNN approaches, with an emphasis on the modified VGG16 model that is our primary classifier. It ends with an explanation of the procedure applied for visual grading of the top-ranked candidates. Section~\ref{results} reports performance and validation: recovery of lenses from the Q1 imaging data, and basic photometric trends of the discovered systems. Section~\ref{sec:strandlim} discusses strengths and limitations, selection bias of the detection pipeline, comparisons with simulation-heavy approaches, and scalability to future \Euclid releases. We conclude with a brief summary and outlook in Sect.~\ref{conclusion}.

\section{\label{sc:Data}\Euclid imaging data}
\subsection{Overview of \Euclid VIS and NISP instruments}

The \Euclid \textit{Mission} \citep{EuclidSkyOverview} is equipped with two instruments for measuring electromagnetic radiation: VIS (Visible Imager; \citealt{EuclidSkyVIS}) and NISP (Near-Infrared Spectrometer and Photometer; \citealt{EuclidSkyNISP}). When used together, these instruments deliver high-quality imaging data in the visible and near-infrared bands.

Of the two, the VIS instrument is optimally suited for resolving the morphology of astronomical sources due to its superior angular resolution. It observes in a single wide passband, approximately corresponding to the combined $r + i + z$ filters, and covers the \SIrange[round-mode=places,round-precision=0,range-units=single]{550}{900}{\nano\metre} wavelength range. The VIS focal plane comprises 36 charge-coupled devices (CCDs), each $4132 \times 4096$ pixels ($\approx$ 16\,megapixels per CCD), yielding an aggregate image size of approximately 609\,megapixels. With a pixel scale of $0\farcs1$, the total field of view of VIS is approximately 0.54\,deg$^2$. Its combination of wide field, broadband filter, and long exposure time confers high sensitivity to faint sources across a broad spectral range, achieving a signal-to-noise ratio of $\geq 10$ for sources with $\IE$ $\leq$ 24.5 \citep{EuclidSkyVIS}. In this work, all magnitudes are in AB. 

Conversely, the NISP instrument is designed to provide detailed spectroscopic and photometric data. It images in three separate bands, \YE, \JE, and \HE, collectively spanning the 950--2020\,nm range \citep{Schirmer-EP18}. NISP employs 16 detectors, each with $2040 \times 2040$ pixels (approximately 4.16\,megapixels). With a field of view comparable to VIS at 0.57\,deg$^2$, NISP offers about one-third the angular resolution, corresponding to a pixel scale of $0\farcs3$. It is also less sensitive than VIS, detecting $\IE$ $\leq$ 24.5 sources at a signal-to-noise ratio of $\geq 5$ in each band \citep{EuclidSkyNISP}. 

\subsection{Characteristics of Q1 release imaging data}

The full \Euclid survey aims to observe approximately one-third of the celestial sphere using the VIS and NISP instruments over its eight-year mission duration. This work utilises a small portion of the full survey: the Quick Release~1 (Q1) dataset \citep{Q1cite}. Q1 comprises data from three deep survey fields \citep{Q1-TP001}, covering a total area of 63.2\,deg$^2$, which represents approximately 0.45\% of the mission’s total planned sky coverage of 14\,000\,deg$^2$. Based on Q1 dataset, a broad range of scientific investigations were conducted in anticipation of the future data releases covering wider areas on the sky (e.g., \citealt{EP-Enia,Q1-SP044,Q1-SP059}).

\subsection{\label{expected-data-lens}Expected data volume and lens distribution}

The survey fields for \Euclid were selected to minimise contamination from Galactic stars and dust, thereby maximising the detection depth for extragalactic sources. As the most sensitive and accurate wide-field cosmological survey to date, \Euclid is expected to detect roughly 6\,billion galaxies \citep{EuclidSkyOverview}. According to the estimates of \citet{Collett_2015}, this dataset is projected to include up to 170\,000 strong gravitational lensing events. Scaling this estimate to the Q1 dataset suggests that $\approx 760$ strong lenses may be present, of which $\approx 670$ are expected after applying the \IE$<24$ magnitude cut. Given that Q1 contains approximately 30 million sources \citep{Q1-TP001}, the data volume renders manual identification impractical. Prior studies such as SLDE A employed a combination of citizen science, simulations, ML models, modelling, and expert inspection to identify lenses in Q1, yielding a catalogue of 500 expert-validated grade A/B lens candidates \citep{Q1-SP048}. SLDE C demonstrated the efficacy of automated methods for lens discovery in \Euclid data, where the best performing model, \texttt{Zoobot}, detected 163 grade A/B lenses in its top 1000 ranked objects. The other four models used in SLDE C were trained on a combination of mostly simulated lenses and real lenses. 

This work builds upon and complements previous works for three reasons. The \texttt{GAIA\_CROSSMATCH} filter used to remove stars in SLDE A
was later found by SLDE F to filter out galaxies, including numerous obvious gravitational lenses associated with mostly low-redshift, bright galaxies. Our study does not apply \texttt{GAIA\_CROSSMATCH} for object filtering, ensuring that all possible lenses are kept. Secondly, SLDE A searched down to a magnitude of 22.5 by discarding 25\% of possible foreground lensing galaxies with $\IE$ $>$ 22.5. In this work, our data range up to $\IE$ = 24.0 for a more comprehensive investigation. Throughout this paper, magnitudes are expressed in the AB system unless explicitly stated otherwise. Finally, we implement a very different model-building approach compared with SLDE C where the strongest model \texttt{Zoobot} was trained from a general-purpose galaxy classification model using a transfer learning approach. In this work, we fine-tuned a CNN model from scratch with an iterative pipeline (Sect.~\ref{sc:pipeline_overview}). 

One notable advantage of our approach lies in the construction of a fully independent pipeline, developed and executed from start to finish. This independence enhances confidence that the final combined catalogue -- integrating results from multiple lens searches -- is more comprehensive and less prone to systematic biases inherent to any single method. Such robustness is especially important for Q1, as it represents the first \Euclid lens search and will form the foundation for subsequent catalogues and analyses.

\section{\label{sc:Blunders}Iterative fine-tuning pipeline}
\subsection{\label{sc:pipeline_overview}Pipeline architecture overview}

An iterative, human-in-the-loop classification pipeline based on deep-learning methods is employed for efficient strong-lens detection. The aim is to perform binary classification of galaxy cutouts into "lens'' and "non-lens'' using the model progressively refined by the pipeline. At its core is a CNN, widely recognised for extracting complex, hierarchical features from images and therefore well suited to identifying the morphological signatures of gravitational lensing -- such as arcs and Einstein rings.

For training, real \Euclid imaging data are preferable to simulated images, ensuring the model learns features consistent with the observational characteristics of future data. While negative examples ("non-lens'') are plentiful, the positive class (confirmed lens candidates) is intrinsically sparse and costly to assemble, owing to the rarity of lensing events and the need for expert annotation. A common approach in the literature is to rely on simulated lens images \citep{Delchambre2019_GaiaGraLIII,Q1-SP048, Metcalf_2019, Schuldt2022_HOLISMOKES_SimTrain, Thuruthipilly2021_SelfAttention, Sheng2022_UnsupervisedSims}. However, such simulations risk visual divergence from genuine \Euclid data and can impair model generalisability -- as demonstrated by \citet{PearceCasey2024_Euclid_CNN}, where models trained on simulations achieved high accuracy yet performed poorly on real \Euclid observations. It is demonstrated in \citet{EuclidQ1LinesLensML} that, by using real \Euclid lens images and non-lens samples as a training set, the number of expected lenses improved 25--30\%, in which 20\% came from the use of real data compared to simulated methods. 

To enhance model performance, we adopt an agile, iterative training procedure that uses simulated images only at the outset. First, a model pre-trained on simulated lenses assigns lensing probabilities to real objects in the Q1 dataset. From the highest-probability cases, experts select confident lens candidates (true positives) to assemble a positive training set of real strong lenses. Objects misidentified as lenses (false positives) and those classified as "non-lens'' inform the construction of the negative training set. This initial, all-real training set is then used to fine-tune a chosen CNN architecture end-to-end. Thereafter, the training set is updated iteratively from the model’s outputs and used to fine-tune the model in the next round. Here, "fine-tuning'' denotes tuning selected layers of the network on progressively refined datasets so that it adapts to the nuances of real \Euclid data. Since fine-tuning is one form of training, when the term "training'' is used in this paper such as "training dataset'', it still should refer to the fine-tuning process where not all layers are allowed to vary, instead of the narrower definition of "training'' where all layers can vary.

\begin{figure}[htbp]
    \centering
    \includegraphics[width=\linewidth]{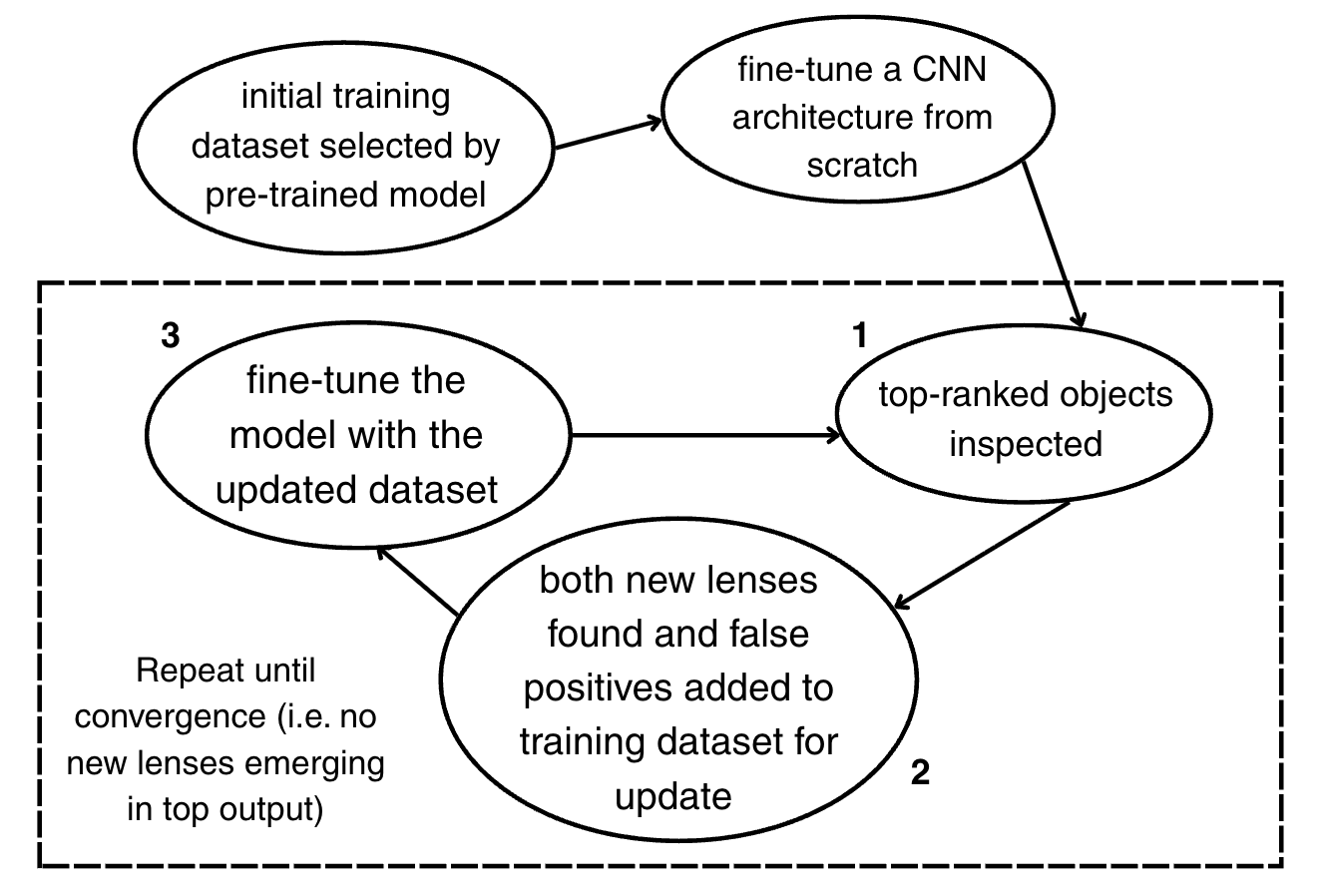}
    \caption{
    Schematic representation of the iterative training pipeline. Each iteration comprises three steps: Step~1: inspection of the top-ranked candidate objects in the model output; Step~2: update of the training dataset through the inclusion of newly identified gravitational lens candidates and false positives; and Step~3: fine-tuning of the model with the updated dataset. The process is repeated until convergence, i.e., when no additional lenses are identified among the top-ranked candidates.}
    \label{fig:pipeline_workflow}
\end{figure}

The pipeline’s dynamic design affords high flexibility in curating the training data. In each round, newly identified lenses are added to the positive set, while emerging false positives are incorporated into the negative set, which is balanced across different negative types (see Sect.~\ref{sc:select TP and FP}). Building a comprehensive negative set is as important as the quality of the positive images, since it increases the purity of positively classified objects. This workflow is repeated to convergence, yielding a robust classifier optimised for the identification of strong gravitational-lens systems (Fig.~\ref{fig:pipeline_workflow}).

The early round seeds were chosen for their morphological clarity and high signal-to-noise, to provide a clean foundation for iterative CNN training. Although a larger number of lenses were previously published, the independent seed set enables detection of strong lenses with morphologies distinct from prior catalogues, increasing the diversity and completeness of the final candidate list. Moreover, this work demonstrates the feasibility of expanding the number of strong lenses found through an iterative approach given limited training samples. The seed set was expanded through augmentation and regularization was applied throughout to limit overfitting and shortcut learning.

Most current studies neither train their models over multiple rounds nor train on fully real \Euclid data. Instead, they apply transfer learning from models trained for broader galaxy-classification tasks that are not purpose-built for lens identification. For example, \texttt{Zoobot} is a Bayesian CNN originally trained for general galaxy morphology and later adapted for lens detection by fine-tuning only the final layers, using a mixture of simulated and real objects in the training set \citep{Q1-SP053}. The chief advantage of fine-tuning a related model is speed, motivated by tight time constraints: it can be faster than fine-tuning a new model across several rounds. However, we propose that training a dedicated model on \Euclid data from the outset, for the sole purpose of lens-galaxy identification, allows all parameters to be fully utilised and optimised for this single task, improving both accuracy and runtime efficiency.

A well-performing single-channel model previously trained for lens identification \citep{Li2023} was used to select \Euclid VIS band images to form the initial training dataset. The details of the pre-trained model are elaborated in Sect.~\ref{sc:pre-trained-model}. Before running the model on \Euclid $\IE$-band images, the \Euclid Q1 data underwent several necessary processing and cleaning stages, as outlined and reasoned in Sect.~\ref{sec:imgprocandclean}. 

\subsubsection{\label{sec:imgprocandclean}Image processing and cleaning}

The Q1 data used in this work are composed of 344 MER tile mosaics and their corresponding catalogues \citep{Q1-TP004}. We apply two filters based on the catalogue entries. The first, \texttt{POINT\_LIKE\_FLAG} $\neq$ 1, excludes point-like sources, especially stars. The second criterion is \IE$\leq 24.0$ (converted from \texttt{FLUX\_DETECTION\_TOTAL} in $\mu\mathrm{Jy}$), since we have found that objects fainter than $\IE=24$ likely would not yield detectable strong lenses. 

In SLDE A, as a reference for comparison, the source selection included a cut at \IE$ \leq 22.5$, with the 5th percentile corresponding to a \texttt{SEGMENTATION\_AREA} of 259 pixels. Additional criteria include a selection limited to sources with \texttt{VIS\_DET} = 1 (i.e., detected in VIS), exclusion of sources with a \textit{Gaia} cross-match identification, and a requirement that sources satisfy \texttt{MUMAX\_MINUS\_MAG} $\geq -2.6$ and \texttt{MU\_MAX} $\geq 15.0$ -- these latter two criteria remove stars and saturated objects, respectively. Furthermore, a threshold of \texttt{SPURIOUS\_PROB} $<$ 0.05 was imposed to filter out likely artefacts. This filtering resulted in a sample of 1\,086\,556 sources in the earlier study of selecting lenses in \Euclid Q1 imaging data.

Each selected object was cropped into a 96~$\times$~96 pixel cutout, corresponding to an angular size of $9\farcs6 \times 9\farcs6$. We chose a \num{96}~\(\times\)~\num{96} cutout based on both lens morphology and network downsampling constraints. \citep{Collett_2015} show that essentially all \Euclid-discoverable strong lenses are expected to have Einstein radii \(\theta_\mathrm{E} \lesssim 3\arcsecond\). With the \Euclid VIS pixel scale of approximately 0\farcs1 per pixel, this corresponds to a radius of \(\sim\)\num{30} pixels, implying that a cutout must be at least \(\sim\)\num{60} pixels across to fully contain the lensed features. We additionally include a margin to accommodate centring uncertainty, PSF wings, and surrounding context.

From a modelling perspective, \num{96} provides a convenient downsampling path: one factor-of-three reduction (e.g. a single \(3\times 3\) pooling with stride 3) maps \num{96} \(\rightarrow\) \num{32}, after which standard \(2\times 2\) pooling or strided operations reduce the dimensionality as \num{32} \(\rightarrow\) \num{16} \(\rightarrow\) \num{8} \(\rightarrow\) \num{4} \(\rightarrow\) \num{2} \(\rightarrow\) \num{1}, cleanly. This keeps the cutout large enough to capture lens structure while avoiding the computational cost of \num{128}\(\times\)\num{128} and the tight framing associated with \num{64}\(\times\)\num{64}.

The relaxed filtering used in this study yielded a total of 5\,461\,976 sources with \IE$\leq 24.0$, subdivided into 2\,594\,180 with \IE$\leq 23.0$ and 2\,867\,796 with $23.0 <$ \IE$\leq 24.0$ (Table~\ref{tab:processing_stats}). In our filtered cutouts, to enhance visual contrast for inspection, a pixel value clipping threshold at the 0.5\% level was applied to each image.

\begin{figure}[htbp]
    \centering
    \includegraphics[width=\linewidth]{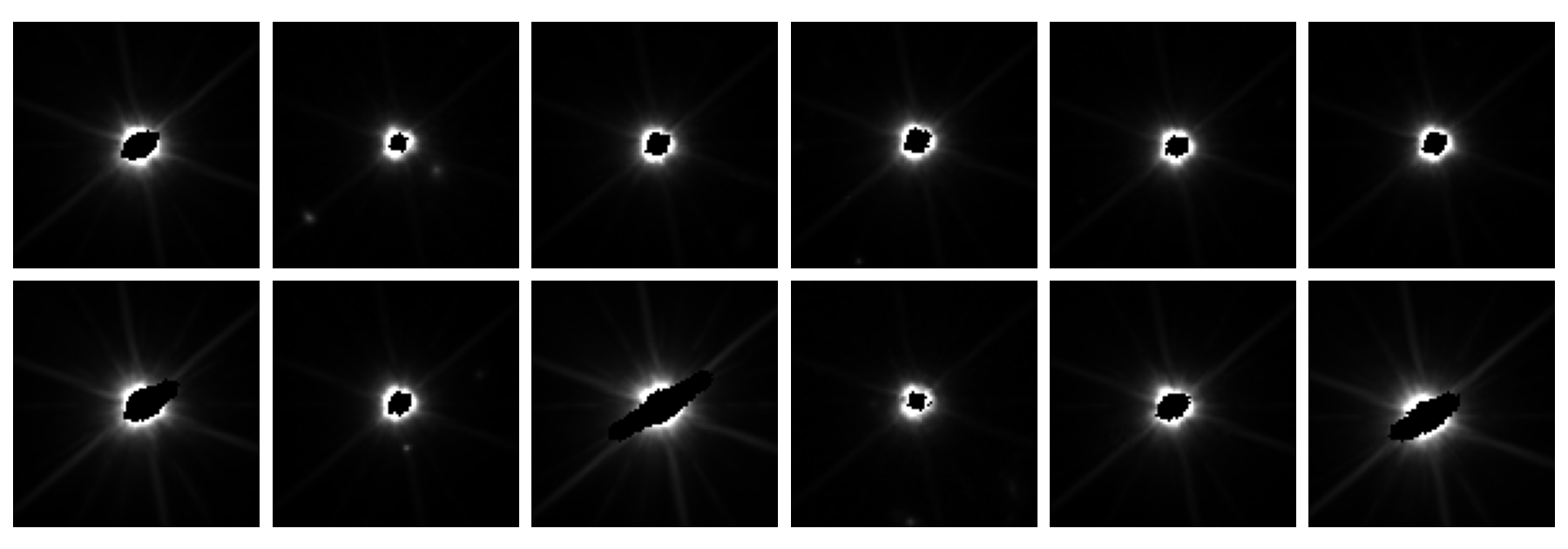}
    \caption{Examples of randomly selected artefacts (displayed in $\IE$-band images) not removed by the \texttt{SPURIOUS\_PROB}~$<0.05$ filter. In almost all cases, these are bright stars with centrally saturated pixels that are captured by L2 image flags.}
    \label{fig:spurious_examples}
\end{figure}

Due to relaxed filtering steps, numerous instrumental artefacts persisted in our cutouts. These artefacts often mimic lens-like morphology, leading to a high false-positive rates in CNN models, as confirmed by preliminary tests. Although application of a \texttt{SPURIOUS\_PROB}~$<~0.05$ filter, used in prior studies such as SLDE A, mitigates this issue partially, many artefacts remain and are falsely flagged as candidates (see Fig.~\ref{fig:spurious_examples}).

To address this, we implemented an image-level artefact-removal algorithm to the VIS band images. Artefacts are broadly classified into two types: large and small defects. Large-defect removal is based on a 6~$\times$~6 pix sliding window with a stride of 1. If all pixels in the window exhibit identical intensity, the cutout is discarded. Small-defect removal employs a 2~$\times$~2 pix detector over the central 10~$\times$~10 pix region, a frequent location for compact artefacts.

In addition, noisy images were identified as a recurrent source of misclassification. To mitigate this, a final noise-based filter assesses the outer 
region of each image, excluding the central sector to preserve bright core features. Images with excessive background brightness in the outer region are removed.

\begin{figure}[htbp]
    \centering
    \includegraphics[width=\linewidth]{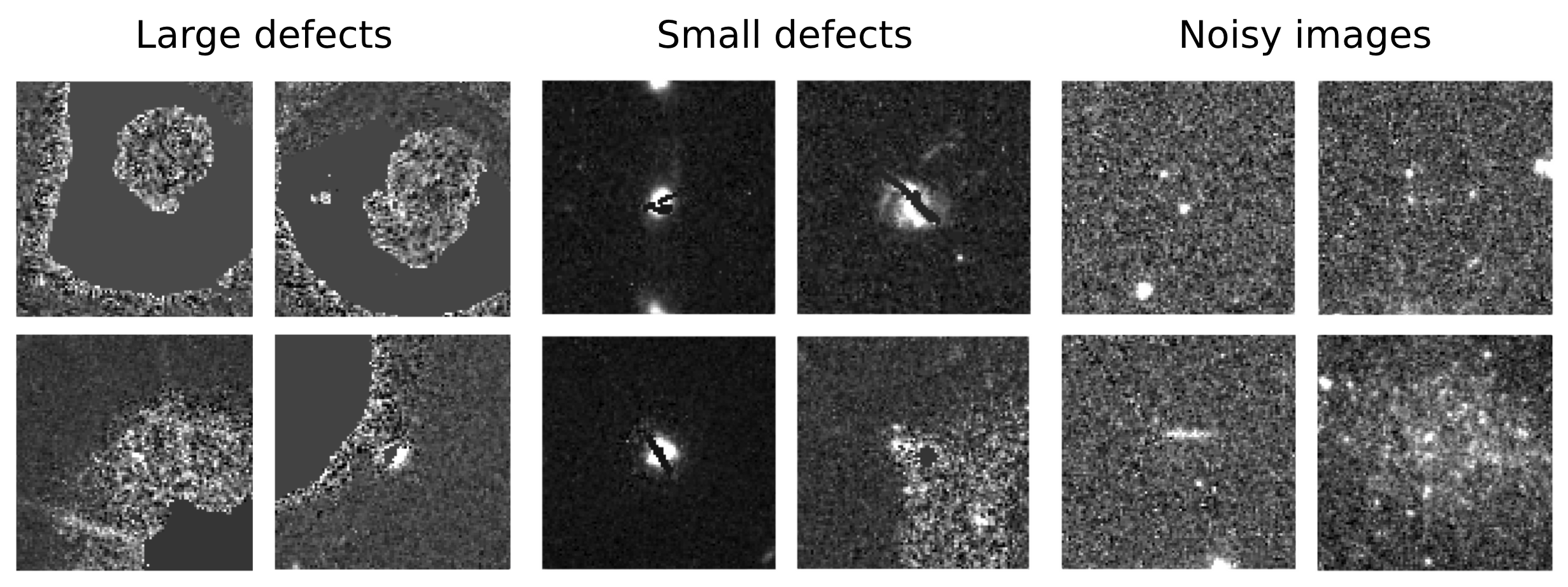}
    \caption{Examples of VIS band images images containing defects made of large and small pixel areas, and noisy images that are removed by our artefact removal algorithm.}
    \label{fig:artifremoveexmpls}
\end{figure}

The complete cleaning pipeline results in a well-vetted dataset suitable~for strong lens identification. The statistics are outlined in Table~\ref{tab:processing_stats}, where we also compare the cleaning effect of \texttt{SPURIOUS\_PROB} $<0.05$ and our artefact cleaner separately. The final \Euclid Q1 dataset with minimal contamination includes 2\,170\,933 objects with $\mathrm{mag} \leq 23.0$ and 1\,265\,417 objects with $23.0 < \mathrm{mag} \leq 24.0$. A total of 3\,436\,350 detected sources were retained.

\btable
\caption{Number of detected sources per magnitude bin after successive quality filters.}
\label{tab:processing_stats}
\centering
\renewcommand{\arraystretch}{1.3}
\begin{tabular}{lccc}
\hline\hline
\rule{0pt}{3.0ex}
$\IE$  & Point. & Point.+Spur. & Point.+Artefact \\
\hline
0--23  & 2\,594\,180 & 2\,361\,825 & 2\,170\,933 \\
23--24 & 2\,867\,796 & 2\,025\,908 & 1\,265\,417 \\
\hline
\end{tabular}

\medskip
{\footnotesize
\textbf{Notes.} Column (2) contains sources with \texttt{POINT\_LIKE\_FLAG $\neq 1$}. 
Column (3) additionally applies the cut \texttt{SPURIOUS\_PROB $<0.05$}. 
Column (4) further includes the artefact and noise cleaning procedure.
}

\etable

\subsubsection{\label{sc:pre-trained-model}Pre-trained single-channel model}

From the processed \Euclid imaging cutouts, we selected an initial training sample using the pre-trained single-channel CNN described by \citet{Li2023}. This model had been trained on 60\,000 semi-simulated cutouts to distinguish lenses from non-lenses using $\IE$-band data. The positive class was constructed by pairing real galaxies from the \textit{Hubble}/CANDELS survey \citep{Grogin2011,Koekemoer2011} at various redshifts into foreground-background pairs. A Singular Isothermal Ellipsoid (SIE) lensing potential was applied to distort the background galaxy, with its orientation and ellipticity aligned to the foreground light profile and a randomly sampled Einstein radius. The resulting lensed background was overlaid onto the foreground cutout to produce realistic strong-lensing configurations at \textit{Hubble} resolution. These images were then convolved, resampled, and injected with noise to match \Euclid quality. Two CNNs were trained independently on simulated VIS and NISP images, yielding AUC scores of 0.987 and 0.797, respectively. Owing to its superior performance, the $\IE$-band model was selected to perform binary classifications on the Q1 VIS cutouts.
\subsubsection{\label{sc:select TP and FP}Selection of positive and negative training samples}

There are 73\,825 $\IE$-band images positively classified (with an assigned prediction value $>0$ with the sigmoid final layer) as lens candidates by the pre-trained single-channel model. We took quick observations of the top 10\,000 candidates by looking at slices at different cut-offs, and we set 1000 as a reasonable cut-off for internal expert team inspection. Among the top 1000 images given the highest lens probability, we examined and selected true positives and false positives by giving human ratings. Confidently real lenses as true positives were selected to build the initial positive training set, and false positives for the negative training set. 

\begin{figure}[htbp]
    \centering
    \includegraphics[width=\linewidth]{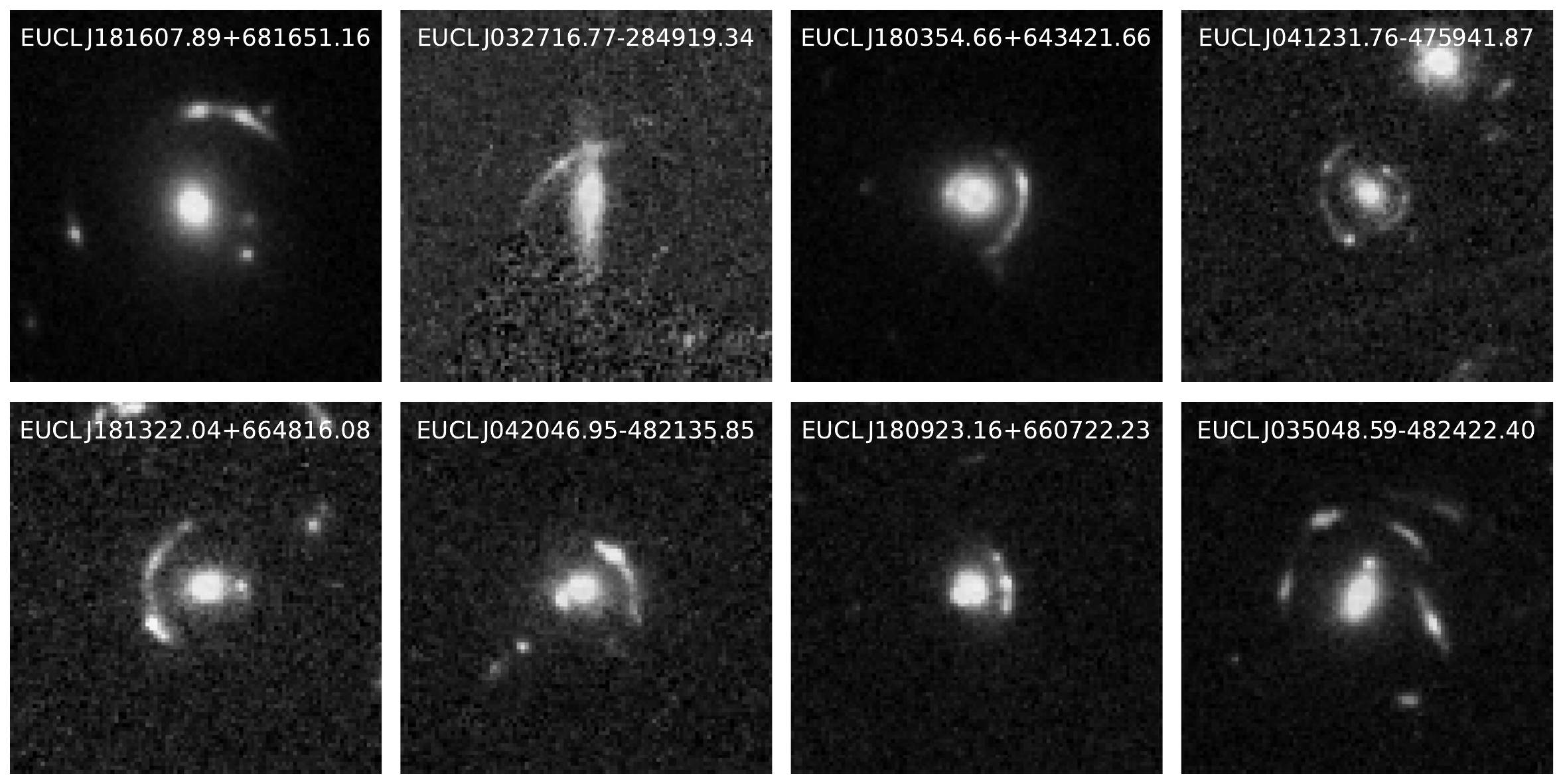}
    \caption{Examples of confident strong lens images used for building the initial \Euclid positive training set based out of VIS-only detection scheme used at Step~1. We augmented these initial detections from 27 candidate lenses to a total of $\sim2000$ true positive cases using a combination of rotation, translation, and colour transformations.}
    \label{fig:8chosenlenses}
\end{figure}

\begin{figure}[htbp]
    \centering
    \includegraphics[width=\linewidth]{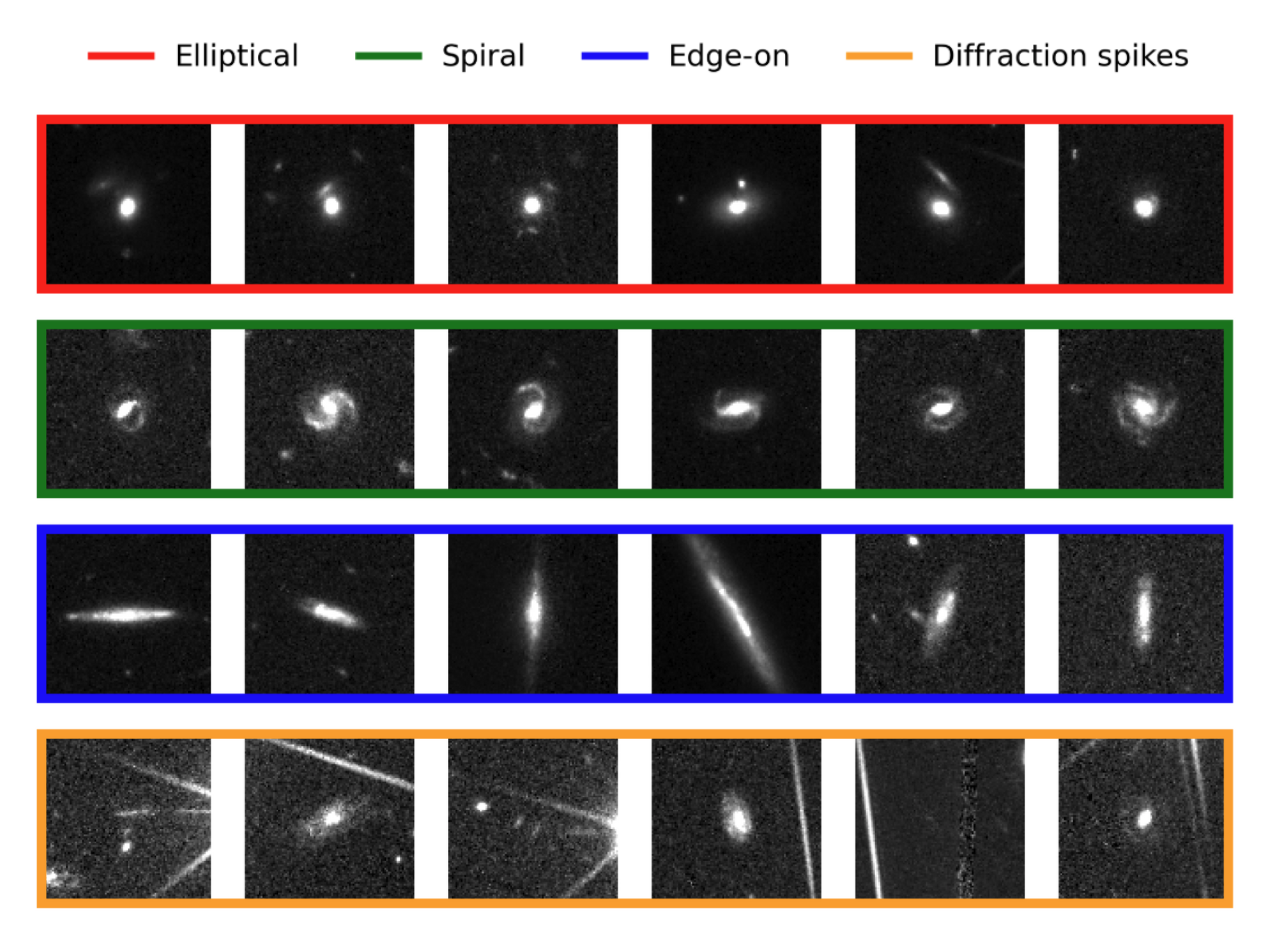}
    \caption{Examples of false positives identified from the output of pre-trained model. These false positives consist of four typical types: elliptical galaxies (top row), spiral galaxies (second row), edge-on spiral galaxies (third row), and images containing diffraction spikes of a bright star close to a galaxy or another source (buttom row).}
    \label{fig:1stfalsepositives}
\end{figure}

In the initial training dataset selection stage of the pipeline, each object received ten expert votes between lens and non-lens. Twenty-seven clearly true positives were selected, with a random subset shown in Fig.~\ref{fig:8chosenlenses}. The false positives resemble the morphological patterns of lenses, with the resemblance decreasing at lower assigned probability values. The $\sim$~70\,000 false-positive images mostly consist of four typical types, with unequal distribution: spiral galaxies, elliptical galaxies, edge-on galaxies, and diffraction spikes, separately shown in Fig.~\ref{fig:1stfalsepositives}.

In building the positive training set, the 27 true positives in \IE-band were converted to colour version (see Sect.~\ref{sc:colorset}) by merging with \YE-band and \HE-band imaging data. Then, they were augmented to around 2000 images using the methods outlined in Sect.~\ref{sc:aug}.

For building the negative training set, although abundant data are naturally available, it is essential to limit its size to that of the positive set in the same round so that the model is not affected by potential bias in the training data. A strongly imbalanced ratio (e.g., 1:10 positive-to-negative) would cause the loss function to be dominated by negative examples, shifting the decision boundary towards predicting the majority class and suppressing recall of true lenses. Moreover, the predicted probabilities would reflect the artificial training prior rather than the astrophysical prior, leading to systematic underestimation of lens likelihoods. It is also crucial to maintain the diversity and balance of the negative training set to prevent systematic bias and improve model robustness. In fact, it has been demonstrated by \cite{Canameras2024} that a balanced negative training dataset for lens classification tasks leads to better model performance. Thus, although objects such as spiral arms dominate the false positives, we would need a balanced negative set for different types of false positives.

We fine-tuned an EfficientNetB0-based CNN model \citep{tan2020efficientnet} to help select false positive images. For each category (spiral arms, elliptical, edge-on galaxies, and diffraction spikes), we manually selected around 100 samples to build a balanced training set. The trained EfficientNetB0 model achieved a categorical accuracy of 90\%, classifying the $\sim$~70\,000 false positives into four different groups. This classification does not intend to accurately identify specific galaxy types, but rather divides the false positives into reasonably different groups regarding their distinct morphologies in order to build a balanced negative dataset. 500 images were then selected from each group, followed by quick visual confirmation, to build a negative training set of size 2000 and were also converted to colour images. The full statistics of the initial training set is presented in Table~\ref{tab:trainingsetv1}.

\btable
\caption{Components and sizes of the initial \Euclid imaging training set later used to fine-tune models and evaluate their performance.}
\label{tab:trainingsetv1}
\centering
\footnotesize
\begin{tabular}{@{}lp{5.5cm}r@{}}
\hline\hline
\rule{0pt}{2.2ex}%
Initial set & Component & Size \\
\hline
\rule{0pt}{2.ex}%
Positive & 27 lens candidates, augmented 67-fold & 1809 \\
Negative & \parbox[t]{5.5cm}{
Balanced sample of:\newline
$\bullet$ spiral-arm galaxies\newline
$\bullet$ elliptical galaxies\newline
$\bullet$ edge-on galaxies\newline
$\bullet$ diffraction spikes
} & 2000 \\
\hline
\end{tabular}
\etable

\subsection{Data preparation}
\subsubsection{\label{sc:colorset}Construction of colour datasets}

To adapt to applying a three-channel model in the iterative pipeline, and for the purpose of improving model performance from data quality, we created colour versions for selected Q1 images and all training data. The colour images were generated by incorporating the \HE ($\sim$$1.5$--$1.8\,\micron$) and \YE ($\sim$$0.97$--$1.07\,\micron$) bands from the \Euclid NISP instrument. To construct colour composites of the original \IE dataset, we produced cutouts at identical sky positions in the \HE and \YE mosaics using the source catalogues, and passed these cutouts to an algorithm to combine them into a single RGB image.

This algorithm comprised two main components: a colour-combination function and a luminance adjustment function. In the colour-combination function, each of the \HE, \YE, and \IE cutouts were first normalised to the range $[0,1]$, and then divided by their median pixel values prior to ensured a uniform black background, which improved visual contrast between the lens and the source galaxy. In accordance with the ordering of wavelengths for each of the bands, we composite the processed \HE/\YE/\IE cutouts into the R/G/B channels to form an intermediate colour cutout.

Due to the inherently lower spatial resolution of the NISP data compared to \IE, we applied an additional luminance correction step to project the morphological features of the \IE image onto the final colour image. A similar approach was also adopted in the work of \cite{Q1-SP057}. Specifically, we extracted the luminance image of the intermediate colour cutout using the ITU-R Recommendation BT.709 standard\footnote{\url{https://www.itu.int/rec/R-REC-BT.709-6-201506-I}}, computed per pixel as $L=0.2126\,R + 0.7152\,G + 0.0722\,B$. We then divided the \IE image by this extracted luminance image and multiplied the resulting pixel-wise normalisation factor to the intermediate colour cutout. This procedure adjusted the contrast of the colour image to match that of the higher-resolution \IE data, yielding the final set of colour cutouts, stored as $96\times 96$ pixel TIFF images with \texttt{float32} precision.

\subsubsection{\label{colorcleaning}Colour image cleaning}

Following the creation of these colour version images, certain artefacts became apparent, particularly objects not covered by the corresponding \HE or \YE bands. To maintain a consistent standard of data quality, we excluded all images lacking the complete set of three bands from the dataset. Up to this stage of filtering, we have retained 3\,423\,516 objects from the Q1 data, cleaned, in colour versions.

\subsubsection{\label{sc:aug}Colour image augmentation}

In the data preparation stage of the first round of the iterative process, we applied image augmentation to the 27 true positives. We also applied augmentation to the increasing number of true positives in each subsequent round. This is a crucial step in the early stages of development, when true positives were exceedingly scarce. The augmentation procedures included horizontal and vertical flipping, affine transformations (e.g., zooming), and systematic variations in contrast, brightness, and saturation. Additional augmentations combined these operations, such as flipping with concurrent contrast and brightness adjustment, or simultaneous modifications of brightness, saturation, and contrast. These augmentations were implemented using \texttt{imgaug} sequences \citep{imgaug}, generating a diverse set of training examples with parameter variations typically in the range of $\pm 10\%$--$20\%$. Using this protocol, we increased the number of positive images in each dataset by a factor of 67 (see Table~\ref{tab:trainingsetv1}). A full scheme of augmentations used is outlined in Appendix~\ref{apd:imgaug} for reproducibility. 

\subsection{\label{sc:modeltraining}CNN model training and fine-tuning}

\subsubsection{Selection of pre-trained architectures \label{sc:selectingpretrained}}

To fine-tune a model from scratch, we selected six lightweight CNN architectures (see Table~\ref{tab:model_candidates}) to best match the complexity and dimensions of \Euclid galaxy images. Each architecture was fine-tuned with the dataset in Table~\ref{tab:trainingsetv1} using its optimal set of frozen layers, and then used to make predictions on our selected Q1 data. Their best performance was evaluated by counting the confident lenses flagged by experts from their top 128 predictions (as we found few lenses after this number), as summarised in Table~\ref{tab:model_candidates}. Among the six architectures, the VGG16 architecture \citep{simonyan2015vgg16} achieved the best performance, and was thus selected as the model architecture for subsequent fine-tuning rounds in the iterative process.

\btable[hbtp]
\caption{Performance of candidate CNN architectures fine-tuned on the initial training set. }
\label{tab:model_candidates}
\centering
\renewcommand{\arraystretch}{1.3}

\resizebox{\linewidth}{!}{
\begin{tabular}{lcc}
\hline\hline
\rule{0pt}{1.6ex}%
Model name & \rule{0pt}{2.4ex}Lenses found & Lens fraction \\
\hline
VGG16 \citep{simonyan2015vgg16}            & 34 & 0.27 \\
Xception \citep{chollet2017xception}      & 27 & 0.21 \\
ResNet50 \citep{he2015resnet}             & 24 & 0.19 \\
InceptionV3 \citep{szegedy2014inception}  & 23 & 0.18 \\
MobileNetV2 \citep{howard2017mobilenets}  & 8  & 0.06 \\
EfficientNetB0 \citep{tan2020efficientnet} & 6  & 0.05 \\
\hline
\end{tabular}
}

\medskip
\begin{minipage}{\linewidth}
\footnotesize
\raggedright
\textbf{Notes.} “Lenses found” gives the number of expert-graded, high-confidence strong-lens candidates 
among the top 128 highest-ranked objects predicted by each model. 
The “Lens fraction” is the corresponding fraction relative to the 128 inspected candidates.
\end{minipage}

\etable

\subsubsection{Fine-tuning strategy (VGG16)}

As can be seen from Table~\ref{tab:model_candidates}, there is still much room to increase the number of lenses found in top-ranked objects for VGG16. Although some progress has been achieved (34 lenses identified) with the initial training set, its diversity and volume can be further improved. The iterative fine-tuning process is designed to gradually enhance the performance of the model with a manually updated \Euclid imaging dataset and careful adjustment of model parameters in each round.

\begin{figure*}[htb!]
\centering
\includegraphics[width=0.9\textwidth]{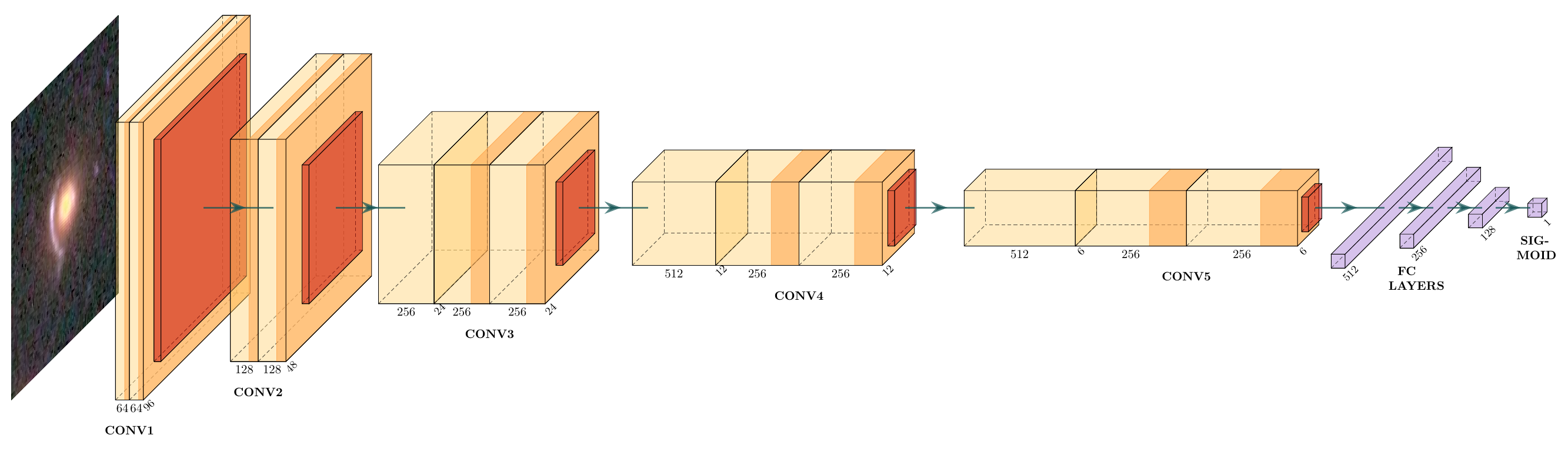}
\caption{Architecture of the modified VGG16 convolutional neural network used for lens classification. The original fully connected layers were replaced with a customised final structure, labelled in purple. This includes a \texttt{GlobalAveragePooling2D} layer to reduce spatial dimensions, followed by two dense layers with 256 and 128 units to capture high-level features, and a final sigmoid layer for binary classification.}
\label{fig:vgg16}
\end{figure*}

To further increase the fraction of true positives in the top predictions, we specifically modified the original VGG16 architecture to better adapt to the complexity and dimensions of galaxy images. The original final layers of VGG16 were replaced with a customised layer set, including a \texttt{GlobalAveragePooling2D} layer (to reduce spatial dimensions), a 256-unit fully connected layer and a 128-unit fully connected layer (to learn high-level feature representations), and a sigmoid binary classification layer (which outputs classification results between 0 and 1). The modified model architecture is visualised in Fig.~\ref{fig:vgg16}.

During each fine-tuning iteration, the number of trainable layers is determined based on both the fraction of lenses among the highest-ranked candidates and the model’s classification accuracy on a balanced test set. In the stage involving fine-tuning of the VGG16 architecture from scratch (see top-right panel of Fig.~\ref{fig:pipeline_workflow}), the optimal performance is achieved when the last nine layers are kept trainable. This is done by allowing the parameters of these layers to be updated while all preceding layers remain frozen. This iterative fine-tuning process is continued until a sufficiently large number of strong lens candidates appear among the top-ranked predictions.

\subsubsection{Evaluation metrics and validation}

In conventional supervised learning, model performance is typically assessed using held-out validation or test sets that contain labels unseen during training. Such evaluation is effective only when the withheld data are sufficiently numerous, representative, and disjoint from the training pool. For strong gravitational lensing, however, this paradigm is not directly applicable. \Euclid Q1 contains very few confirmed strong lenses, and many of the high-quality systems have been used -- iteratively and without persistent identifiers -- to refine the model itself. As a result, it is not currently possible to construct a purely independent test set of real \Euclid lenses large enough to support statistically meaningful precision or recall measurements. Even though a small number of \Euclid-confirmed lens candidates exist outside our training pool, they are far too few to form a reliable evaluation set.

Because of this limitation, classical supervised-learning metrics such as accuracy, precision, and recall cannot be reported in their standard form. Instead, evaluation must reflect the operational context of the \Euclid lens-search problem: identifying a small number of genuine lenses within millions of galaxy cutouts. Rather than relying on artificial balanced test sets -- whose statistical properties do not resemble the \Euclid population -- we therefore assess performance by examining how well the model prioritises likely lenses in its ranked candidate list.



Our primary evaluation metric is thus the recovery of known or visually confirmed lens candidates within the top-K ranked candidates, compared to the expected population of approximately $\sim$ 760 strong lenses in Q1 (670 with $\IE<24$). This approach provides a practical measure of the model’s capability: a method that successfully concentrates true lenses towards the top of the ranking directly reduces the human inspection effort required for scientific follow-up. Moreover, because this lens-search task is fundamentally dominated by extreme class imbalance, prioritisation-based evaluation more accurately captures real-world model behaviour than hypothetical accuracy-based metrics.

Another useful metric would be the measure of the continual improvement of the model throughout training: how successful iterations consistently increase the concentration of lens-like systems among the highest-ranked candidates. This progressive enhancement in ranking quality provides strong qualitative and quantitative evidence that the model learns more effective decision boundaries over time (elaborated in Sect.~\ref{sec:retraining}), even in the absence of a clean, fully independent \Euclid test set.

\subsubsection{\label{sec:retraining}Retraining and evaluation}

Human vetting is performed in two modes. During iterative training, we prioritise speed and use internal expert inspection of the top-ranked candidates, typically reviewing the top 4000 objects per iteration; this produces a labelled update to the training set on week timescales. For the final evaluation, we complement expert inspection with a larger-scale grading campaign using established citizen-science style tooling, in which each object receives multiple independent votes. This two-stage design separates rapid iteration (to improve the ranking model) from broader community-level validation (to quantify grading consistency and provide higher-confidence candidate labels).

In the first iteration, we started by inspecting the top-ranked objects (Step~1) in Fig.~\ref{fig:pipeline_workflow}. In Step~2, there were 123 strong lens candidates selected and augmented to 8241 to optimise the positive training set. Around 900 new false positives were picked and added to the previous balanced negative training set -- this formed an updated negative training set of size approximately 8900.

\begin{figure}
    \centering
    \includegraphics[width=\linewidth]{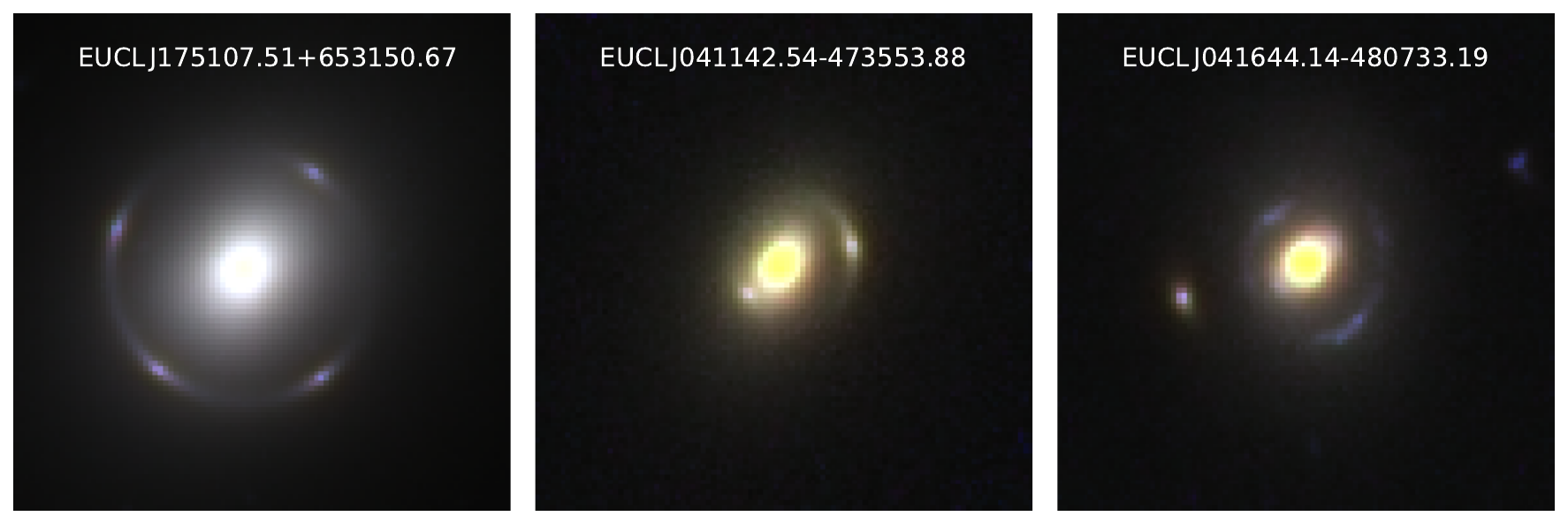}
    \caption{Three newly identified (relative to the previous iterative round) Grade A strong lens candidates from the fine-tuned model in the second round that exhibit atypical "lemon'' and white colouring compared to most other lensing systems that show a colour difference between the foreground lens and the background source. The first system (left) is NGC\,6505, a complete Einstein ring, discussed by~\citet{ORiordan25} based on the visual identification of the ring.}
    \label{fig:3-white-lenses}
\end{figure}

To further enhance the diversity of the positive training set, we created 100 hand-drawn lenses using the software Procreate\footnote{Procreate®, Savage Interactive Pty Ltd, \url{https://procreate.com}} (see Fig.~\ref{fig:app-art-lens}). The artistic lenses are purely experimental and are not derived from physical lens models or mass distributions. They were designed to test classifier sensitivity to extreme or atypical morphologies as a qualitative diversity regulariser. The artistic inspiration came from observing the dynamics and traits of the existing strong lens candidates found so far. Careful variations were designed to increase data diversity and robustness. These art lenses were added to the positive training set for fine-tuning the model. Among the approximately 1000 top-ranked objects of the fine-tuned model, by manual inspection, there were 229 strong lens candidates which were augmented and added to the positive training set. Among the 229 candidates, three new special lenses (relative to the previous round; Fig.~\ref{fig:3-white-lenses}) should be noted. These lenses display traits different from most other strong lenses, appearing more lemon-coloured and white based on our colour processing method. Specifically, the first one in Fig.~\ref{fig:3-white-lenses} is NGC\,6505, a complete Einstein ring, first identified by \citet{ORiordan25}.

In the second iteration, the three lenses were augmented to 201, with a higher weight in the positive set to help the model grasp the traits of diverse lenses. There were dot-like objects, appearing in green with our colour grading method, among the false positives of the second iteration. We added them to the negative training set to allow more true positives -- we pooled all the lenses we found up to that point and augmented them, creating a dataset of 30\,686 images, with 15\,343 each for positive and negative images. This dataset was used to produce the final iteration of the model.

\begin{figure*}[htbp]
    \centering
    \includegraphics[width=\textwidth]{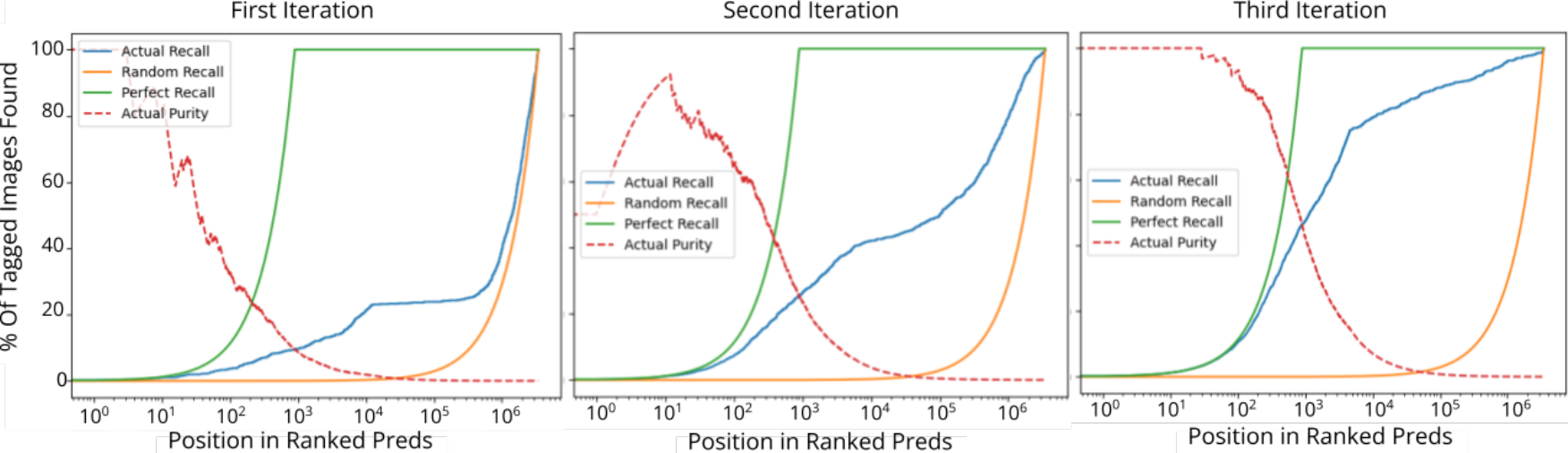}
    \caption{Three rounds of the iterative model retraining process cross-validated with 905 sources (containing off-centred samples) from this work and SLDE A/B/F. The left-hand plot denotes the first iteration trained on the initial training dataset selected by the pre-trained model. The blue curve shows recall, that is, where in the ranked predictions we could find each of the confirmed lens candidates. The green curve denotes the recall of a perfect classifier, while the orange line denotes a completely random classifier. The red curve denotes the precision of the current model, that is, how well the model moves non-lenses to lower rankings. Throughout development, we found that the actual recall curve mostly overlaps with the perfect classifier in the top 100 predicted candidates.}
    \label{fig:evalmetprogress}
\end{figure*}

In total, we fine-tuned the model three times, each resulting in noticeable improvements in performance (see the improving recall curve in Fig.~\ref{fig:evalmetprogress}; Sect.~\ref{model-performance} elaborates on how we generated a list of lenses as ground truth in producing Fig.~\ref{fig:evalmetprogress}). For all training iterations, optimisation was performed using the Adam optimiser \citep{adamoptimizer} with a learning rate of $5\times10^{-5}$, a batch size of 64, and a total of 15 epochs. During training, the model retained the final sigmoid activation so that outputs could be interpreted as probabilities, allowing the binary cross-entropy loss function to be applied effectively. 

\subsection{Inference on Q1 dataset}
\subsubsection{Linear output as raw scores \label{rawscores}}

To evaluate the output of the neural network for predictive purposes, we found it to be more beneficial to remove the final sigmoid layer and use the raw, unbounded output of the neural network's last dense layer; we will denote it as the "raw score'' in the rest of the paper. This is due to the fact that the sigmoid function compresses all values to a narrow range between 0 and 1, thereby losing granular information about the model's confidence for predictions that fall within the region where the sigmoid function plateaus. This issue is particularly critical in the context of this paper due to covariate shift, where the statistical properties of the real-world images can differ significantly from the manually selected training images. 

For instance, variations in instrumental signatures or unexpected artefacts can cause the distribution of the model's output scores to drift. When this occurs, the sigmoid function maps these shifted scores into its saturation regions, making it impossible to distinguish between varying levels of model confidence. 
In contrast, the raw output retains the full magnitude of the model's conviction and provides a wider distribution of scores. This makes the "raw score'' a more sensitive input for ranking-based evaluation and allows for a more nuanced selection of a decision threshold to better control the trade-off between precision and recall.

\subsubsection{Human evaluation and original grading results}

We selected the top 4000 candidates, without duplicates, ranked by "raw score'' as evaluated from our final model, as lens presence declines drastically thereafter (see Fig.~\ref{fig:lens_frac_dist}). The choice of top 4000 was quantitatively validated using classification results, as explained in Sect.~\ref{class-results}. 

The top 4000 images were visually inspected and graded by lensing expert volunteers from the \Euclid Collaboration. During the classification process, each participant was shown a randomly selected image from the top 4000 and prompted to assign it to one of the following categories: A+ (lens of individual scientific value), A (confidently a lens), B (probable lens, further information needed), C (lens-like, other explanations possible), or X (not a lens).
The displayed images were in PNG format due to the platform's compatibility. While the model input images were originally stored as TIFF files with \texttt{float32} precision, we manually verified that the PNG versions used for classification exhibited no perceptible visual differences from the original TIFF files. Each image was evaluated by ten independent assessors, resulting in a total of 40\,000 ratings.

For reference, in SLDE A, each candidate object likewise received ten expert votes, and calibration was performed by comparing each expert’s average grades with those of their peers on the same galaxies. If, for instance, an expert tended to assign systematically higher scores than the consensus (e.g., often giving candidate lens systems an A when the group average was B), their grades were adjusted downward to offset this optimism. However, this calibration method does not apply to this work, as there were 27 experts participating in the classification of our 4000 objects, each contributing to a different extent, from a ten classifications to all 4000. It is impractical to find groups with sufficiently large sample sizes that also received votes from the same sufficiently large number of experts. For instance, given an object that received 10 votes from 10 different experts, the grade of one expert is more optimistic than the grades of the other nine experts; the calibration method in SLDE A would then consider that expert as optimistic overall and discount their grade to all galaxies they classified. However, the conclusion of that expert being generally optimistic can only be made when there is a sufficient set of objects which received votes from those exact same ten experts for comparison. Due to the vast difference in the number of classifications contributed by each expert, we applied a different calibration method in Sect.~\ref{grade-process}.

\subsubsection{\label{grade-process}Grade aggregation}

Based on the original classification results, on a quick inspection we found that many strong lenses that also received human ratings in SLDE A underwent a "downgrade'' with a mode-first rating processing method. For instance, if an object receives two~A, three~B, one~C, and four non-lens votes, the mode-first method would assign a non-lens classification, even though lens-like features are present, as indicated by the A/B/C votes. To avoid such a collective downgrade and to ensure lens candidates remain visible, we implemented the following post-processing threshold:

\begin{algorithm}[H]
\caption{Grade aggregation threshold rule}
\begin{algorithmic}[1]
\IF {count(A$+$) is a majority}
    \STATE classify as grade A$+$
\ELSIF {count(A$+$) + count(A) is a majority}
    \STATE classify as grade A
\ELSIF {count(A$+$) + count(A) + count(B) is a majority}
    \STATE classify as grade B
\ELSIF {count(A$+$) + count(A) + count(B) + count(C) is a majority}
    \STATE classify as grade C
\ELSE
    \STATE classify as "not a lens''
\ENDIF
\end{algorithmic}
\end{algorithm}

\section{Results\label{results}}

\subsection{\label{model-results}Lens candidates identifications}

\subsubsection{\label{class-results}Human evaluation results}

The visual inspection was conducted within the Galaxy Judges project on the Zooniverse platform. The results of the processed ratings are shown in Table~\ref{tab:processed_classification}, which mitigates the downgrade effect of the mode-first processing method to some extent, as numerous objects have moved to higher grades. Using the grade calibration procedure in Algorithm~1, we have further mitigated the downgrade effect to a large extent.

\begin{table}[ht]
\caption{Distribution of post-processed classification outcomes from \Euclid expert ratings for the model's top 4000 predicted objects.}
\label{tab:processed_classification}

\renewcommand{\arraystretch}{1.3}
\begin{tabular}{lcc}
\hline\hline
\rule{0pt}{3.0ex}%
Processed classification & Vote No. & Fraction (\%) \\
\hline
Grade X: not a lens                  & 3095 & 77.4 \\
Grade C: lens-like feature  & 464 &  11.6 \\
Grade B: probable lens     & 314 &  7.9 \\
Grade A: confidently a lens                     &  111 &  2.8 \\
Grade A+: lens of science value      &  16 &  0.4 \\
\hline
Total                                           & 4000 & 100.0 \\
\hline
\end{tabular}
\end{table}

\begin{figure}
    \centering
    \includegraphics[width=\linewidth]{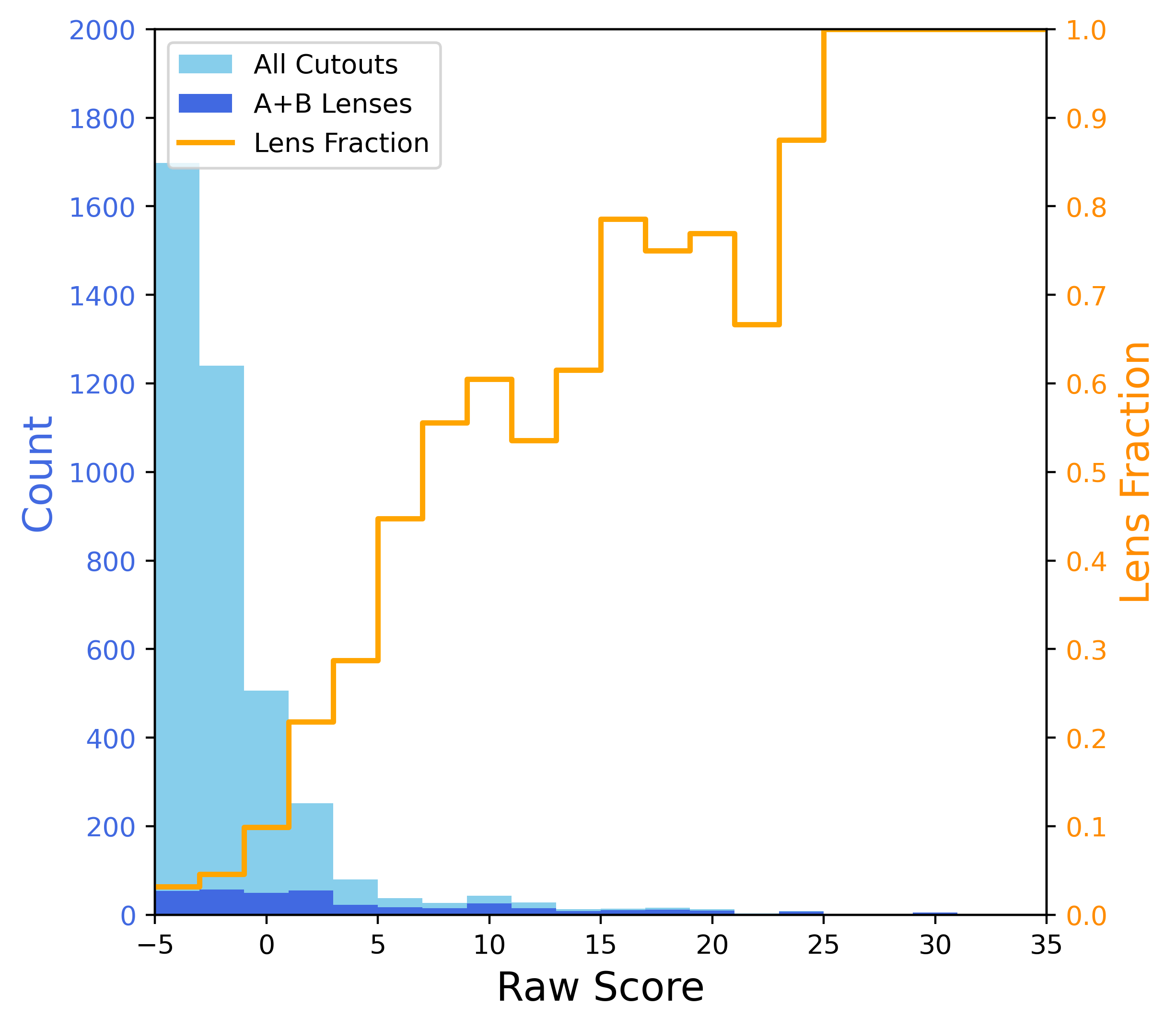}
    \caption{Light blue bars shows the "raw score'' of the top 4000 cutouts ranked by the model, and dark blue bars represent the identified A/B lens candidates within the 4000 cutouts. Orange line shows the fraction of lens candidates, computed as the ratio of dark blue over light blue in each bin. On the lower "raw score'' end, the fraction of lenses declines to less than 0.05. This justifies the threshold for visual inspections as lenses beyond top 4000 are too rare for effective human inspections.}
    \label{fig:lens_frac_dist}
\end{figure}

For clarity of communication, lenses graded as A+ have been merged into Grade~A. For further analysis, we will mostly consider grade A and grade B lens candidates since these candidates represent the more confident lens detections. The raw score distribution of these lens candidates are visualised in Fig.\ref{fig:lens_frac_dist}, which justifies the selection of only top 4000 ranked cutouts for visual inspections. 

\subsubsection{Variance in human evaluation results}

\begin{figure}[htbp]
    \centering
    \includegraphics[width=0.8\linewidth]{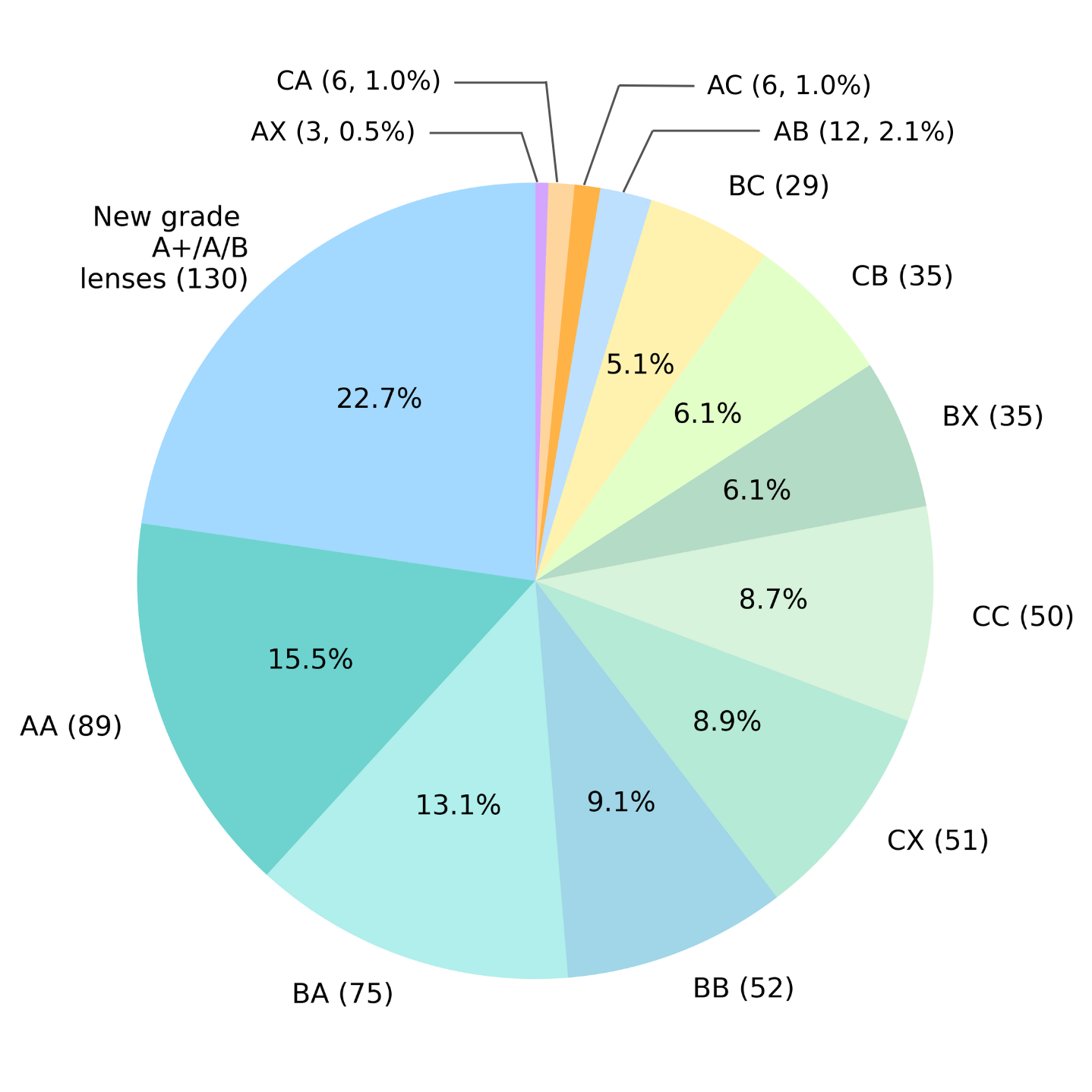}
    \caption{Pie chart showing distribution of 130 new lenses identified and lenses previously discovered by SLDE A in the top 4000 predictions. Each grade is a concatenation of grade in this work and in SLDE A. For instance, the BA slice indicates that 75 (13.1\%) lenses received grade B in this work but grade A in SLDE A.}
    \label{fig:lenspiechart}
\end{figure}

For lenses recovered in this work and in SLDE A, we concatenated our ratings and their ratings to produce a summary of results in Fig.~\ref{fig:lenspiechart}. Among the combined grades, there are 116 lenses that receive downgrade and 136 lenses that receive upgrade, resulting in a rough balance between grade changes. However, these discrepancies also indicate an inherent variance in human ratings, as demonstrated in \cite{Rojas2023}. To investigate possible reasons for such variance, we show examples of downgrade in Fig.~\ref{fig:app-a-to-b} and upgrade in Fig.~\ref{fig:app-mismatch-walmsley-lower} as separate panels.

We investigated a number of reasons for the differences in human ratings between this work and the previous study. One cause of this discrepancy could be in image visualisation and processing. The use of VIS as the luminance component in the colour composites may have enhanced the perceived resolution of lens features. This could contribute to improved human grading accuracy and CNN performance, and highlights the importance of image construction methodology in lens identification pipelines. Additionally, the images in this work were visualised with percentile clipping, which can cause certain features to be fainter than that with a non-linear stretching function such as arcsinh. However, many of the images that dropped from Grade A to Grade B do not appear to be due to the presence of faint features (see Fig.~\ref{fig:app-a-to-b}). Since one image is displayed for each object, a panel consisting differently scaled images may influence the grades received. Moreover, two different groups of people contributed to the grading process of SLDE A and this work, sharing different expectations and experience. The large number of downgraded candidates demonstrates the essential role in accounting for rating variance when using human labels as a recognised benchmark for model performance. Recognising these differences should also help evaluate the true performance of a machine-learning model and potentially establish more robust metrics for the comparison of different machine-learning models.

\begin{figure}[htbp]
    \centering
    \includegraphics[width=0.8\linewidth]{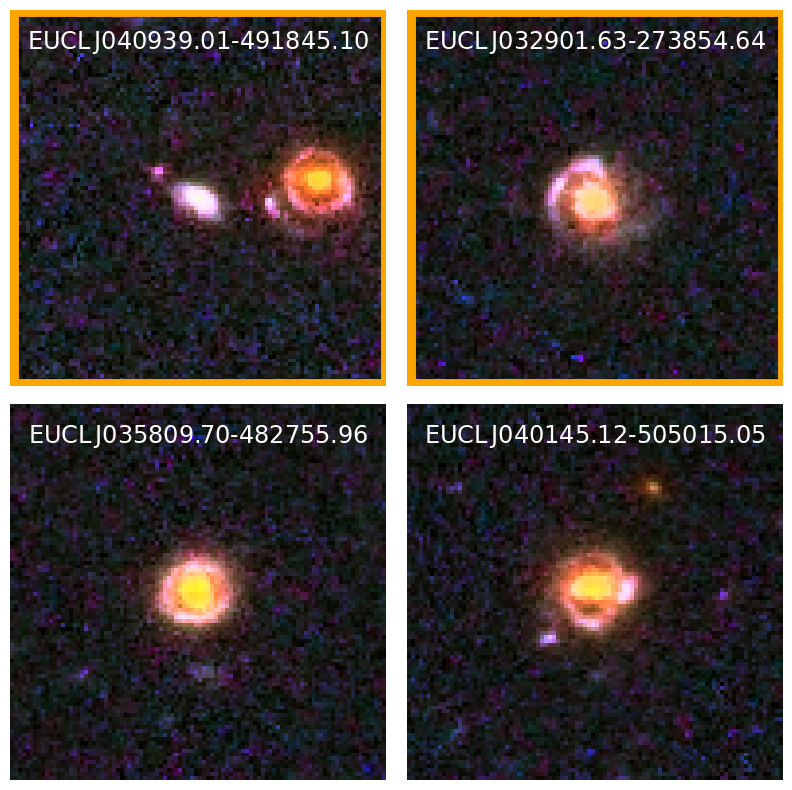}
    \caption{Four new lenses within the flux magnitude thresholds used in SLDE A but not identified as lens candidates in that work. Objects outlined in orange were identified as lenses by previous models in SLDE C but rejected by citizen scientists in Space Warps, a classification project hosted on Zooniverse platform. The other two were poorly ranked by SLDE C machine learning models.}
    \label{fig:prev_in_lines}
\end{figure}

\subsubsection{\label{model-performance}Final model (VGG-16) performance}

While the development of this model progressed independently of SLDE A, upon the release of their lens catalogue, we found it useful to validate our results. As model development progressed, we observed consistent improvements to both the recovery of known lenses and to the purity of the highest-ranked predictions (Fig.~\ref{fig:evalmetprogress}). The collection of lens samples used for validation is discussed next. 


Following the classification procedure described earlier, we identified a total of 441 Grade A and B lenses among the top 4000 ranked objects. These 441 objects include 302 lenses previously classified as Grade A/B in SLDE A, and 9 lenses identified in SLDE F. This results in a total of 130 new lenses not previously identified in \Euclid Q1 data. 

To evaluate the final model in a manner suited to the extreme class imbalance of the \Euclid lens-search problem, we focus on how effectively the classifier concentrates genuine lenses within its highest-ranked predictions. Because \Euclid Q1 does not contain a sufficiently large or independent set of real, unused lenses suitable for traditional supervised-learning metrics, we instead assess performance using a curated reference sample of 905 high-confidence, expert-graded lens candidates (including 130 newly identified lens candidates in this work, the NGC 6505 lens, and 500/3/9 Grade A or B systems from SLDE A/B/F with associated duplicates and off-centred cutouts). Among the model’s top 1000 ranked candidates, 427 of these 905 known lens candidate images appear, meaning nearly half of all high-confidence lens candidate systems in Q1 lie within the top 0.03\% of the 3.4 million cutouts. This demonstrates the model's capability to remove non-lenses, offering substantial reduction in human inspection effort.

Broadening the threshold illustrates the model’s ability to recover an increasingly large fraction of true lenses while maintaining meaningful prioritisation. The top 4000 candidates cover 644 known lens images (including duplicates), capturing progressively more of the 905 lens reference sample. Removing the duplicates yields 441 distinct candidates in the top 4000 candidates, which is 65.8\% of the ~670 lenses expected for $\IE$ $\leq$ 24. This indicates that the model assigns elevated scores to a wide range of lens morphologies and continues to retrieve true systems even at lower confidence ranks. Together with the consistent improvement trends observed across training iterations (Fig.~\ref{fig:evalmetprogress}), this concentration-based evaluation shows that the final model provides a robust, practical, and effective tool for surfacing strong gravitational lenses in large-scale survey data.

\begin{figure}
    \centering
    \includegraphics[width=\linewidth]{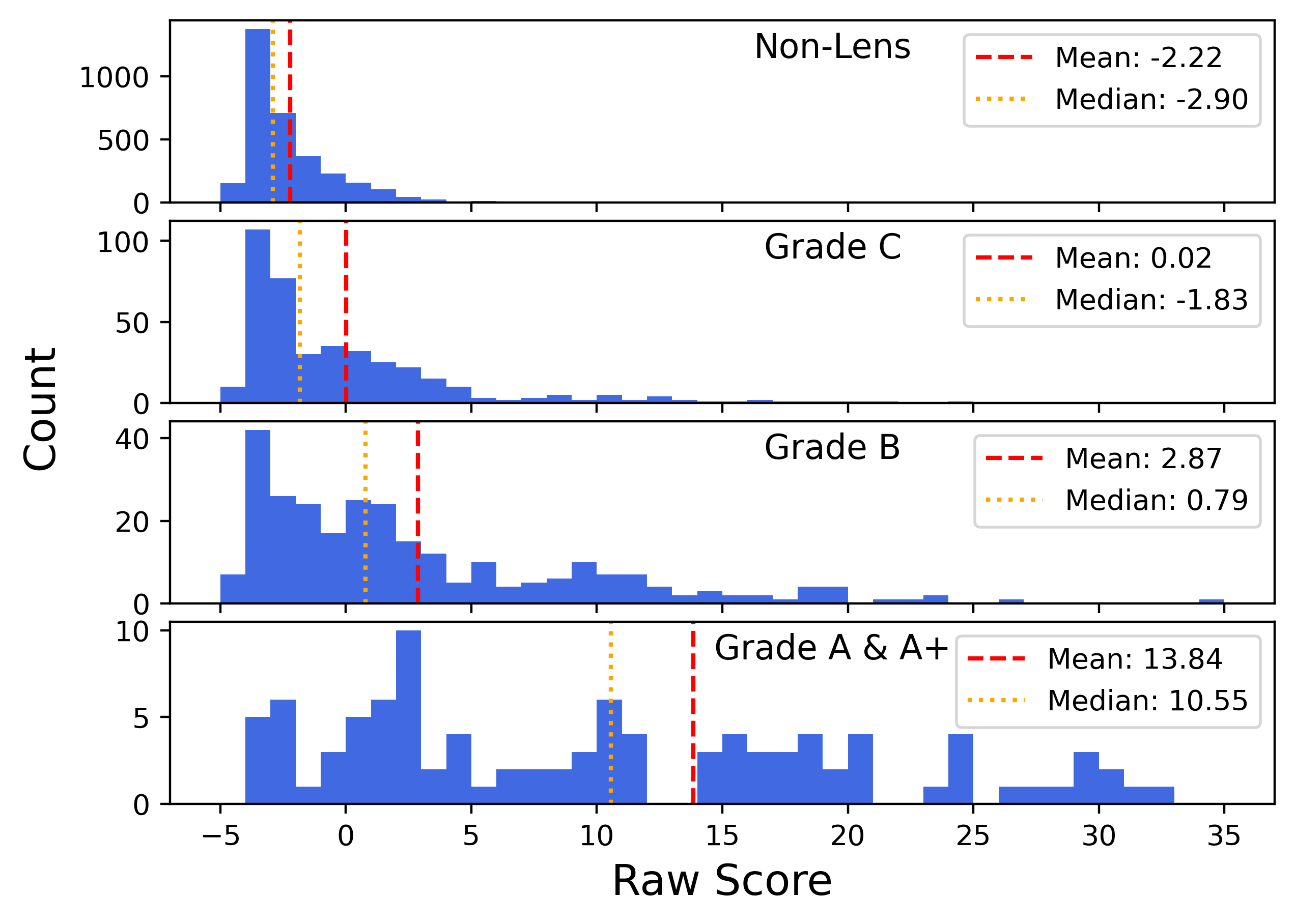}
    \caption{Distribution of model prediction "raw scores'' for candidates grouped by human-assigned grade: ``non-lens'', C, B, and A. Each subplot shows a histogram of prediction values, with red dashed lines indicating the mean and yellow dotted lines indicating the median for that grade. As the grade increases, the mean and median of the prediction values shift towards the right, indicating a higher prediction value. Thus, we can infer that the prediction value from our model correlates with a higher grade.}
    \label{fig:predictions_for_grades_bins}
\end{figure}

Furthermore, by analysing the model prediction scores and the human-assigned grades for the top 4000 candidates, we observe a clear correlation between the predicted probabilities and the human classifications (see Fig.~\ref{fig:predictions_for_grades_bins}). Higher model scores are associated with a greater likelihood, as determined by experts, that a given object is a gravitational lens. This is consistent with the grading rubric used in the visual inspection process, where higher grades reflect more prominent and unambiguous lensing features. The observed correlation suggests that the model has successfully learnt to identify and rely on visual features similar to those used by human classifiers to assess lensing confidence -- such as the presence of extended arcs and the geometric alignment between a foreground deflector and background sources.

\subsubsection{Sample Validation}
To explain the source of the newly classified 139 grade A/B candidates (130 new candidates only in this paper, 9 candidates identified in SLDE F and this paper) relative to SLDE A, an important factor to consider is the more inclusive filtering criterion applied in pre-selecting sources, which accounts for 52 new A/B lenses. If the same filtering from SLDE A is applied to these 52 candidates, 12 are removed by requiring no \textit{Gaia} cross-match (which are generally sources with $\IE \lesssim 19$), then 38 are removed by the magnitude cut ($\IE \le 22.5$), and finally \texttt{SPURIOUS\_PROB} $>$ 0.05 removed the last 2 candidates. Of these 12 new lenses collected by relaxing the \textit{Gaia} cross-match criterion, 9 were also recovered in the work of SLDE F, which specifically examined sources previously removed by the \textit{Gaia} cross-match filter applied in SLDE A, and found 68 grade A and B lens candidates relative to SLDE A. Taking SLDE F in consideration, the number of new A/B candidates in this work is adjusted to 130. The rest of the new candidates not removed by the filtering criterion ($139-52=87$ candidates) is mostly dominated by lower confidence grade B lens candidates, and can be a result of variance in human classification and the model's unique selection function.


An additional discrepancy between this work and SLDE A concerns the candidates examined in both studies that received significantly different scores. Specifically, this includes 4 candidates classified as Grade A in this work, where two were poorly ranked by the models in SLDE C and did not reach the visual inspection stage, while the other two were visually inspected but rejected by citizen scientists in Space Warps, a galaxy classification project on Zooniverse platform used in SLDE A. These four lenses are shown in Fig.~\ref{fig:prev_in_lines}. This divergence may be attributed to differences in visualisation schemes, where variations in intensity and colour scaling could lead to interpretation discrepancies or potentially overly optimistic classification. 


\subsubsection{Analysis of false positives and model robustness}

\begin{figure}[htbp!]
    \centering
    \includegraphics[width=\linewidth]{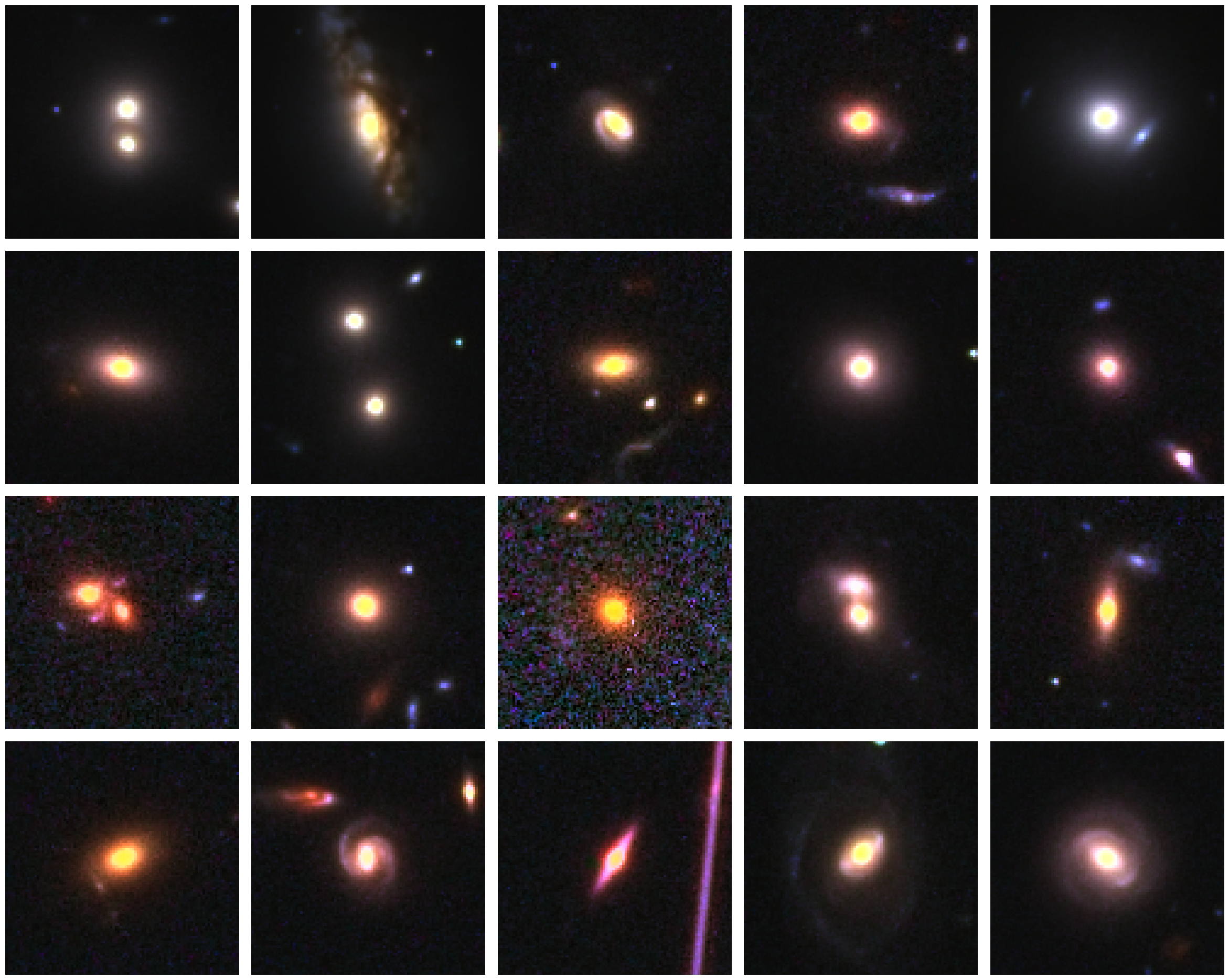}
    \caption{20 false positives within the top 4000 classifications of the model. Within these we can see an example with a diffraction spike, several elliptical and spiral galaxies, with many containing evidence of recent mergers or an ongoing merger in the form of a tidal stream, as well as cut-outs containing multiple objects.}
    \label{fig:final_false_positives}
\end{figure}

Among the objects incorrectly identified as lenses by the model, several recurring morphological types emerge -- namely, diffraction spikes, spiral galaxies (particularly barred spirals), isolated red elliptical galaxies, and galaxies with nearby companions. In many cases, these false positives can be attributed to structural similarities with genuine strong lenses. For example, spiral galaxies often exhibit a bright central bulge and a circular or ring-like morphology, features that may resemble lensing arcs. A representative panel of arbitrary false positives is shown in Fig.~\ref{fig:final_false_positives}. Nevertheless, in the majority of cases, the model successfully prioritises clearly identifiable lenses over more ambiguous or misleading candidates.

These findings suggest that the model’s susceptibility to false positives arises primarily from structural degeneracies, rather than random misclassification. While certain morphologies -- such as barred spirals or ellipticals with nearby companions -- can mimic lens-like features, the low prevalence of such cases among top-ranked candidates indicates that the model retains discriminative power even in morphologically ambiguous regimes. This behaviour reflects a degree of robustness and highlights the importance of incorporating domain-specific priors or multi-band information in future iterations to further suppress false detections, elaborated in Sect. \ref{sec:strandlim}.

\subsection{Characteristics of Discovered Lenses}

\begin{figure}[htbp]
    \centering
    \includegraphics[width=\linewidth]{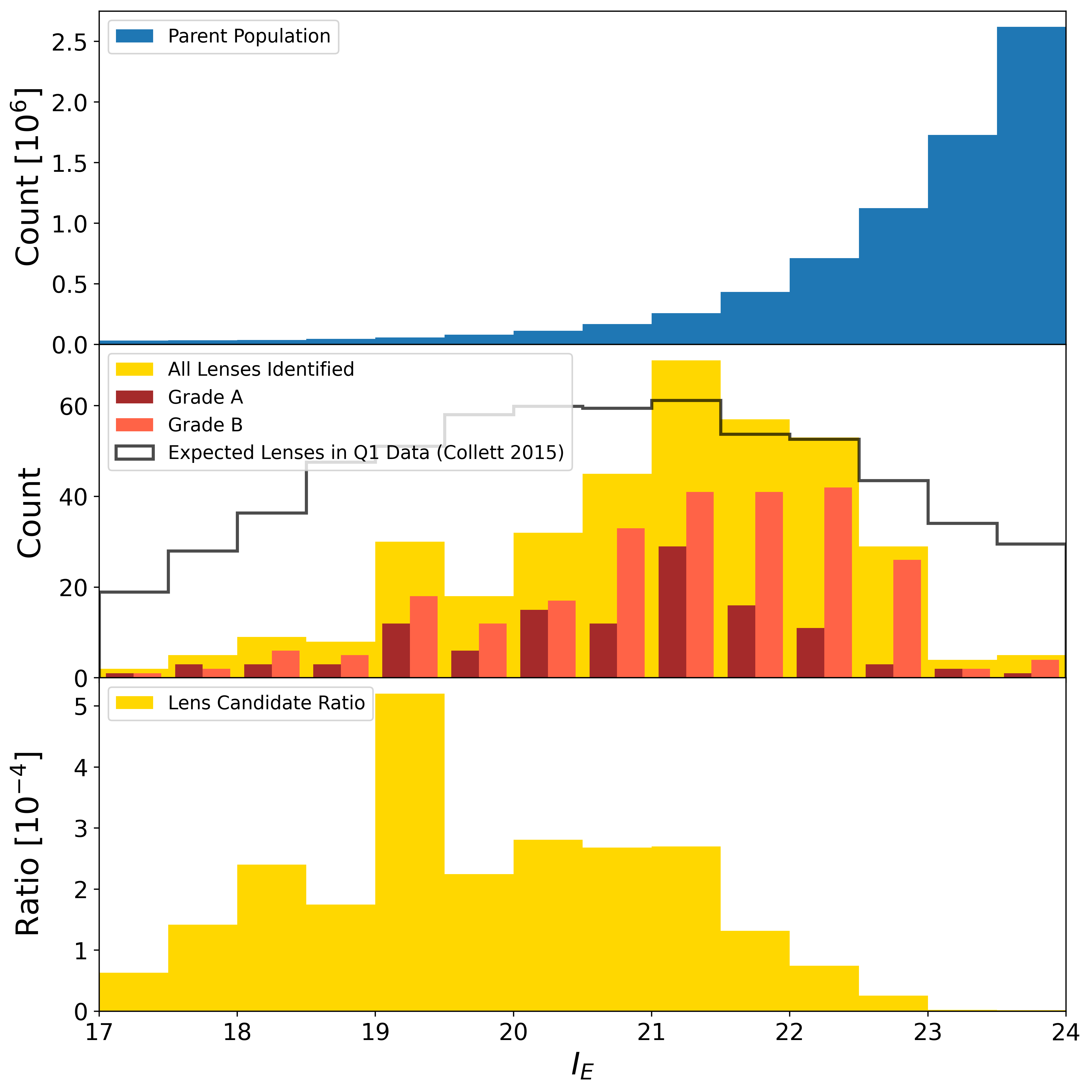}
    \caption{Histogram of object counts as a function of detection magnitude. \textit{Top:} Blue bars represent the full set of inspected candidates with $\IE \le 24.0$; \textit{Middle:} dark and light colour corresponds to the grade A and grade B lens candidates respectively, and the background bars are the sum of the two. For reference, we also show the forecast strong lens magnitude distribution from \cite{Collett_2015} normalised to \Euclid Q1 area. \textit{Bottom:} The ratio of lens candidates inside the parent population as a function of magnitude.}
    \label{fig:vis_mag_hist}
\end{figure}

\begin{figure}[htbp]
    \centering
    \includegraphics[width=\linewidth]{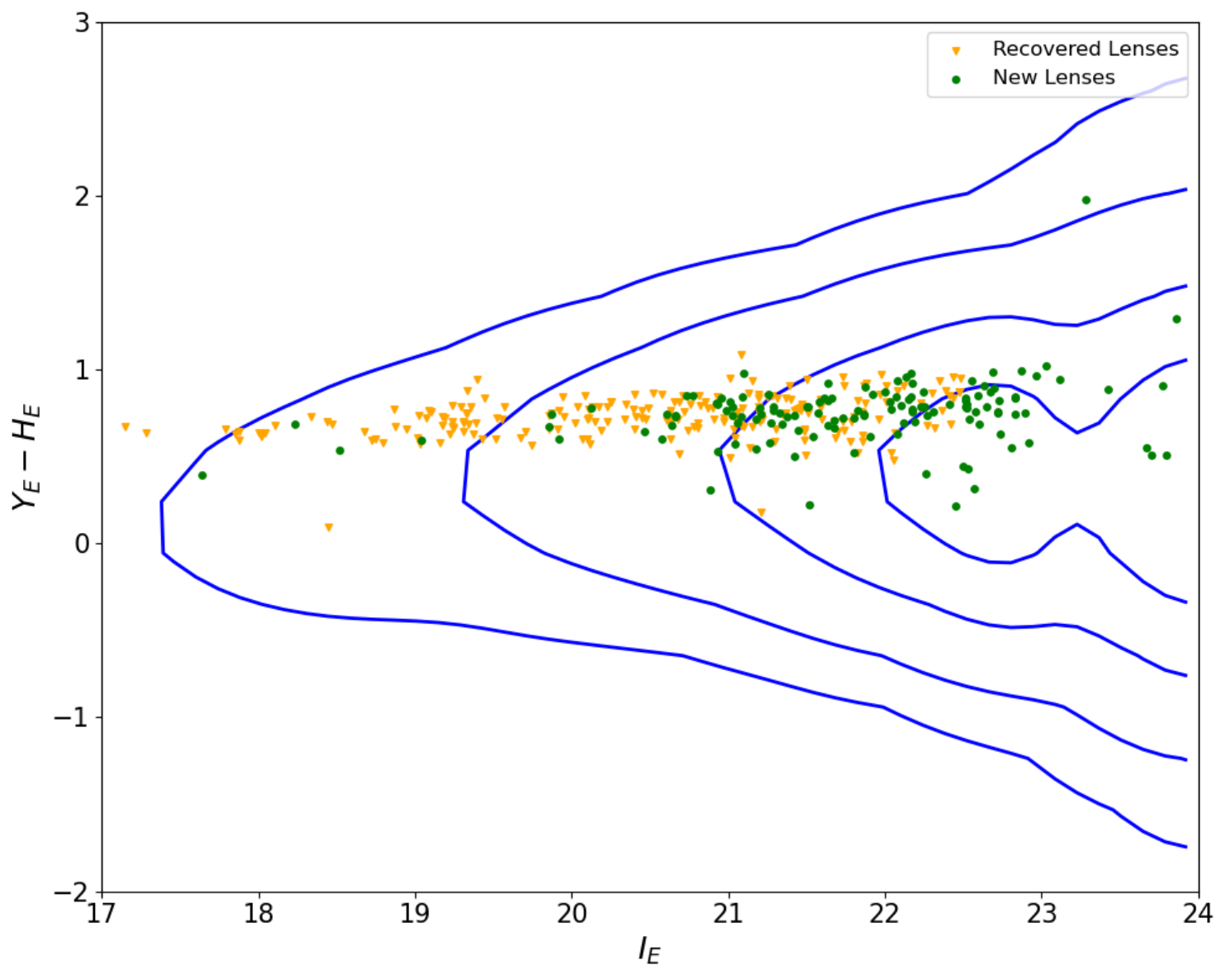}
    \caption{$\YE-\HE$ colour vs.\ \IE magnitude distribution for model-discovered lenses and the parent population. Orange triangles represent the lenses previously found in SLDE A, while green circles represent lenses not found in SLDE A. The blue contours represent the distribution of the parent population, with contour lines at $0.5\sigma$, $\sigma$, $1.5\sigma$, and $2\sigma$ confidence levels.}
    \label{fig:vis_y-h_plot}
\end{figure}

One of the key distinguishing features of the lens candidates analysed is their distribution in \IE, inferred from \texttt{FLUX\_DETECTION\_TOTAL}. As shown in Fig.~\ref{fig:vis_mag_hist}, the high-confidence lens candidates within $17\le \IE \le 24$, have a median of 21.2, a mean of 21.1, and an interquartile range (IQR) of 20.3-22.0. The minimum and maximum observed values are 17.1 and 23.9, respectively. To further characterise the photometric properties of the sample, we examined the $\YE-\HE$ colour as a function of \IE for the classified lenses, in comparison with the full parent population (Fig.~\ref{fig:vis_y-h_plot}). We find that lenses consistently exhibit higher $\YE-\HE$ values, indicating redder colours relative to the general galaxy population. This trend is consistent with expectations that lensing galaxies are typically massive early-type systems. To statistically confirm these findings, we performed two Kolmogorov--Smirnov (KS) tests on the \IE and $\YE-\HE$ distributions, yielding $p$-values of $3.5 \times 10^{-64}$ and $8.1 \times 10^{-118}$, respectively. These extremely low $p$-values strongly reject the null hypothesis that the lensing galaxies are randomly drawn from the full galaxy population observed in the imaging data.

\section{Discussion}
\label{sec:strandlim}

\subsection{Strengths and limitations of developmental pipeline} 

The greatest strength of the model development pipeline we present in this paper is its efficiency, requiring only limited involvement of human resources as such involvement is a bottleneck in large-scale astronomical surveys \citep{fortson2011galaxyzoomorphologicalclassification}. During the final two iterations of the model, after the key methods and functions were fully implemented, each iteration took approximately one week for two people. Each cycle showed noticeable improvements in model performance (Fig.~\ref{fig:evalmetprogress}). 

This developmental pipeline is well suited to building a machine-learning model within a reasonable time frame when staffing for tasks such as image simulation, data processing, or human labelling is limited. Its efficiency rests on prioritising the removal of false positives from model outputs, thereby avoiding the need to inspect large numbers of objects by eye. With each fine-tuning step, the false-positive rate among the top candidates falls, so more true positives rise into the review set or the highest-ranked predictions. Previously identified true positives remain in this set throughout. In the early stages, false positives were common and true positives scarce, so the primary bottleneck was assembling a robust sample of true positives. By design, the pipeline allows a small initial set of true positives to expand rapidly -- often within one or two iterations -- into a larger dataset that can be augmented and used to train the final model.

One limitation of our current approach is the potential for reduced recall in our final model. Due to our emphasis on precision and efficiency (Fig.~\ref{fig:lens_frac_dist}), as well as the narrower scope of our human labelling (top 4000 here vs. top 10\,000 in SLDE A), we likely miss considerable number of genuine lenses that appear lower in the rankings. While this deficiency can be mitigated by inspecting a larger range of candidates, a more fundamental issue arises from the iterative "bootstrapping" nature of the pipeline. Because the model learns from the lenses it finds, it risks overfitting to the specific characteristics of early detections -- such as specific colours or brightness levels -- at the expense of diversity. For instance, we observed that the model consistently assigned lower scores to lenses with atypical "whiter" colouring. This sensitivity confirms that our pipeline does not sample the sky uniformly, but rather has a distinct selection bias that must be characterised before any physical conclusions can be drawn. Addressing this issue will require further investigation in future work, for example by incorporating a broader variety of lens types, such as edge-on systems and low-redshift lenses, into the training set to improve the robustness and diversity of the model.

\subsection{Selection Function}
The population of the lenses detected in this work is a product of the true lens population and the network's selection effect as a function of lensing configurations. Consequently, the distribution of candidates cannot be directly interpreted as the underlying astrophysical population without correction. For example, comparing our detections to the forecasts of \cite{Collett_2015} reveals a selection bias that peaked at $I_{\rm E} \sim 21.5$. Understanding the selection function is thus crucial for enabling inference on population level \citep{Sonnenfeld_2022}. We plan to determine this function by feeding simulated lenses into our network while keeping the network parameters frozen. These test images will be synthesised by compositing simulated lensed arcs with real galaxy cutouts drawn from Q1 data. The lensing configurations will be drawn from a uniform prior spanning the lens parameter space, rather than from an astrophysical prior, allowing us to comprehensively map the network's decision boundaries. The resulting probability map will provide the correction weights necessary to reconstruct the parent population of strong lenses. Full details of the selection function and the associated lens modelling pipeline will be presented in our follow up work.

\subsection{Comparison with traditional approaches}

A common strategy to address the small number of confirmed lens candidates is to generate simulated images of lenses to compose the training sets for the model \citep{Metcalf_2019, Schuldt2023HOLISMOKESIX,Q1-SP052,AcevedoBarroso2026UNIONS}. This is a particularly useful strategy for projects that rely on early-stage mission data, where there are few or virtually no images of lenses in the style of the target data. However, in simulating astronomical data there are many things to consider in order to create sufficiently realistic samples, such as morphological diversity, noise, PSF, and imaging artefacts; failing to adequately address these factors can significantly limit model performance (\citealt{Bottrell_2019, huppenkothen2023constructingimpactfulmachinelearning, Pearce-Casey24,Q1-SP053}). As a result, the trained model will be limited by any differences found in the simulated data and the actual population of inference. In contrast to traditional methods, our model minimised reliance on simulated data and primarily utilised real data that the model found, which allowed us to focus on model development. The drawback of this element of our approach, however, is that model performance may be capped by the small volume of data, which would be insufficient to train a maximally accurate model. Quantitative measurement the selection function of the model will be necessary to fully characterise this insufficiency. 

Additionally, our approach utilised a much faster human-rating process. Since efficiency was prioritised, we opted not to use citizen science platforms such as Space Warps and Galaxy Judges until evaluation of the final model was reached. Instead, for each iteration, the top 4000 images produced by the model underwent visual inspection by the authors with priority given to unambiguous lenses that could be readily identified by their distinct characteristics (beyond this range, lens-finding again becomes difficult for manual inspection as highlighted in Fig.~\ref{fig:lens_frac_dist}). While this approach significantly reduced the time required to generate a labelled training set to be typically under one week per iteration, compared to 2--3 weeks using Galaxy Judges, it inherently risks the possibility of missing some less distinct lenses. As a result, the training sets in each iteration may under-represent less distinct lenses, potentially limiting our model’s scope of inference. Despite this, we believe that this was an effective solution for our goals. Our final model was able to make predictions with an exceptionally high purity in the top 1000 (427/1000 classified as Grade A/B lenses\footnote{Lenses that were classified as Grade A/B by both this work and SLDE A/B/F, if available.}) and recover most of the lenses found using more human resource-intensive approaches such as SLDE A. These results suggest that our efficiency-focused, iterative approach is a viable solution to future lens-finding initiatives, especially when human resources are relatively limited.

\subsection{Implication for future work and \Euclid data releases}

The iterative model training approach we have presented here can be easily scaled to meet the demands of large datasets, including the upcoming \Euclid DR1 data release (\citealt{EuclidSkyOverview}). Both the model and the developmental approach could be applied effectively to larger data volumes without modification. While the time and number of candidates to sort through would inevitably increase as data increase, this is also expected to result in a proportionally larger set of lenses found. For example, assuming one labeller could quickly sort through 4000 candidates within a week, such that inspection of tens of thousands by a larger groups such as the \Euclid SL SWG is feasible, and over the six year long planned \Euclid mission, the discovery of up to 170\,000 lenses \citep{Collett_2015} with this procedure realistic. Thus, the core advantage of this approach, mainly the efficiency in development under limited human resources, remains  preserved even when scaled to the full scope of the \Euclid Wide Survey.

\begin{figure}
    \centering
    \includegraphics[width=\linewidth]{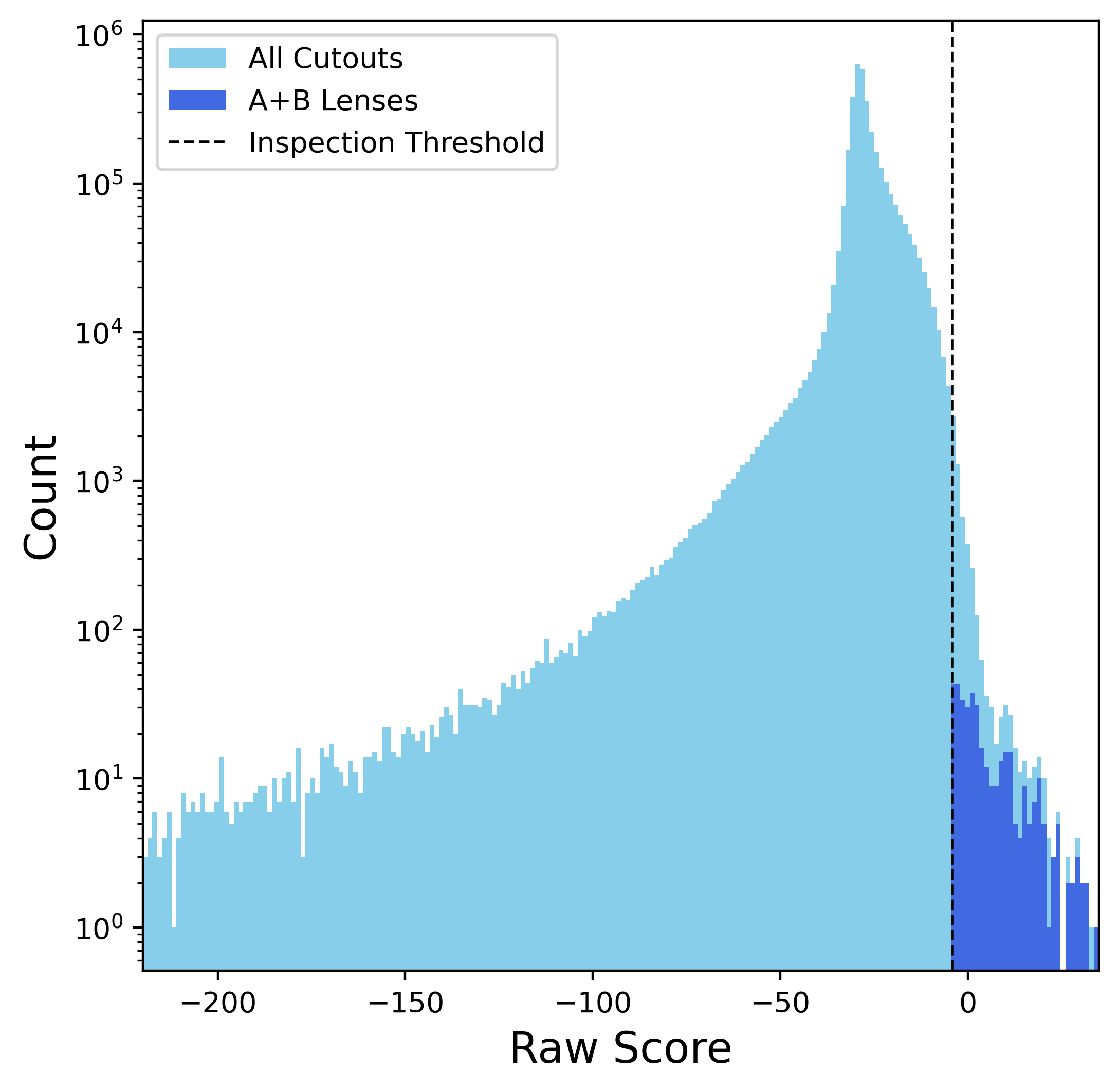}
    \caption{"Raw score'' distribution of all cut-outs and the identified lenses. The dashed line represents the inspection threshold of the top 4000 "raw scores''; all cutouts to the left of the dashed line were not inspected.}
    \label{fig:prediction_dist}
\end{figure}

Finally, it is important to note that although we inspected only the top 4000 lenses, Fig.~\ref{fig:prediction_dist} indicates that there are likely significantly more lenses beyond our range of inspection. The main challenge to address when looking beyond this range is the diminishing proportion of lenses, which returns to the original "needle-in-a-haystack'' problem. Future work can be performed to inspect beyond the current range of the top 4000 rated lenses that were the focus of this paper. Furthermore, as increasing samples of lenses are gathered from other \Euclid lens search outcomes to train on, the model performance
should continue to improve through the iterative and human-in-the-loop pipeline. In the meantime, it is crucial to adjust estimates of the proportion of lenses expected from a sky search such as \Euclid taking into account the model capabilities and identification limitations. 

\section{\label{conclusion}Conclusions and future work}

Here we have presented an end-to-end, automated pipeline for discovering strong gravitational lenses in the \Euclid Q1 imaging spanning over a total imaging area of 63.1\,deg$^{2}$ in three deep fields. Starting from the VIS instrument catalogues, we applied strict source and artefact vetting, mainly point-source identification to remove stars, magnitude thresholding, pixel-level defect and systematic effects filters, and built $96\times96$ pixel VIS+NISP colour cutouts using a VIS-anchored luminance scheme that preserves VIS morphology while retaining NISP colour contrast. Rather than rely on fully simulated training datasets, we bootstrapped from a VIS-only seed classifier to harvest real \Euclid positives and to curate a balanced, morphology-aware negative set that accounts for contamination from spiral arms, unlensed ellipticals, edge-on galaxies, and diffraction spikes. Among six lightweight CNN models, a modified VGG16 that included \texttt{GlobalAveragePooling} with 256/128 dense layers and a sigmoid with trainable last nine layers returned the best precision at the top of the ranked list. Iterative fine-tuning grew the training set from 27 seed lenses to a balanced colour set of $\sim$30.7\,k images (with 15\,343 positive and 15\,343 negative, respectively, with additional hard-negative mining and targeted diversification for underrepresented morphologies).

Applying the final model to Q1 data, the team arranged visual inspection of the top 4000 ranked candidates through the Galaxy Judges project on Zooniverse with exactly 10 independent votes per object. This procedure was aimed at repeating the same analysis scheme as the one that was used previously to identify strong lensing systems in Q1 imaging data \citep{Q1-SP048}. These rankings resulted in 441 grade~A/B systems, with 302 lenses overlapping with the previously reported candidates and an additional 130 Grade A/B candidates uncovered in this work. VGG16 model scores and human grades are monotonically correlated, with the lens yield strongly concentrated at the highest scores. Independently, of the 905 high-confidence lens candidates in this work and SLDE A/B/F studies in combination and containing off-centred ones, 740 (81.8\%) lie within this model's top 20\,000 predictions. Photometrically, recovered lenses span $\IE \approx$ 17–24 (median 21.2) and exhibit redder $\YE - \HE$ colour than the parent population, consistent with massive early-type deflectors. Operationally, each training iteration required roughly a week of effort by a small team, involving manual visual inspection and model retraining, because the pipeline steadily suppresses recurrent impostors so that true lenses rise into the top-ranked, human-reviewed subset.

The main contribution of this work is a practical, scalable workflow for rapidly adapting a dedicated CNN to \Euclid survey images using iterative self-training on real Q1 cutouts, paired with a fast expert-vetting loop. This design produces a high-purity ranking on the target data domain, enables week-timescale iteration during early releases, and yields new lens candidates that complement existing Q1 discovery efforts. Confirming these systems as genuine lensing events remains challenging. The standard approach relies on spectroscopic observations to determine the redshifts of both the source and the lens, combined with detailed lens modelling. However, this method requires substantial observing time at ground-based facilities. Even multi-fibre spectrographs do not fully alleviate this limitation, because the lenses are widely distributed on the sky and the intrinsic lensing rate is low. In the future, \Euclid grism data will provide a valuable complement, at least for the subset of systems whose lens or source redshifts lie within the grism sensitivity window.

The presented work will be expanded in the future to include (i) a measurement of the survey-realistic selection function by injecting physically motivated simulated lenses into real \Euclid backgrounds (native PSF, noise, and artefacts) and reporting calibrated precision–recall at fixed candidate budgets; (ii) a possible extension beyond the top 4000 via uncertainty-aware active learning (e.g., entropy/ensemble disagreement) to raise recall without inflating human load; (iii) a test with multi-scale/attention architectures and adding fast post-hoc lens-model vetters to prune residual impostors; and (vii) adaptation to Rubin and \textit{Roman} \citep{Gezari2022R2D2} with minimal additional labels, accompanied by a harmonised public candidates table (thumbnails, scores, and key metadata) for follow-up studies from the ground and space observatories.

The principal limitations of the presented pipeline are completeness for faint or atypical morphologies and sensitivity to early training seeds and display stretch; these effects can bias recall at fixed magnitude depth used for human classification. Moreover, our current evaluation is tied to a top-$N$ candidate budget, and a fully calibrated selection function capturing the completeness and purity versus lens/source properties remains to be established. Naturally, the selection function is a crucial ingredient in converting this or any other galaxy strong lens sample for inferences related to either cosmological parameters through statistics related to lensing or astrophysical properties of either the foreground lenses or background sources. Despite these caveats, the approach is immediately scalable to larger \Euclid releases and portable to Rubin and \textit{Roman}, and the code structure is suitable for public reuse.

\begin{acknowledgements}
  \AckEC 
   \AckQone
  XX, RC, TL and AC acknowledge the support from NASA ROSES Grant 12-EUCLID11-0004. SS has received funding from the European Union’s Horizon 2022 research and innovation programme under the Marie Skłodowska-Curie grant agreement No 101105167--FASTIDIoUS.

\end{acknowledgements}

\bibliographystyle{aa}
\bibliography{Euclid,Q1,external_bib}

\begin{appendix}
\onecolumn

\section{Deterministic Data Augmentation Scheme\label{apd:imgaug}}
\label{app:augmentations}

We applied a fixed set of 67 deterministic image augmentations implemented using the
\texttt{imgaug} library. Each transformation was applied independently to the training
images. The augmentations consist of geometric operations (horizontal and vertical flips,
scaling), photometric adjustments (brightness, contrast, saturation), and combinations
thereof.

\subsection{Geometric Transformations}

\begin{itemize}
    \item G1: Horizontal flip (Fliplr)
    \item G2: Vertical flip (Flipud)
    \item G3: Horizontal + vertical flip
    \item G4: Affine scaling (scale = 1.1)
    \item G5: Affine scaling (scale = 1.2)
\end{itemize}

\subsection{Contrast Adjustments}

\begin{itemize}
    \item C1: LinearContrast 1.2
    \item C2: LinearContrast 1.1
    \item C3: LinearContrast 0.9
    \item C4: LinearContrast 0.8
\end{itemize}

\subsection{Brightness Adjustments}

\begin{itemize}
    \item B1: MultiplyBrightness 1.1
    \item B2: MultiplyBrightness 0.9
\end{itemize}

\subsection{Saturation Adjustments}

\begin{itemize}
    \item S1: MultiplySaturation 1.1
    \item S2: MultiplySaturation 0.9
\end{itemize}

\subsection{Combined Transformations}

The remaining augmentations consist of deterministic combinations of flips with
brightness, contrast, and saturation modifications.

\begin{itemize}
    \item M1: Fliplr + LinearContrast 1.2
    \item M2: Flipud + LinearContrast 1.2
    \item M3: Fliplr + LinearContrast 1.1
    \item M4: Flipud + LinearContrast 1.1
    \item M5: Fliplr + Flipud + LinearContrast 1.2
    \item M6: Fliplr + Flipud + LinearContrast 1.1

    \item M7: Fliplr + MultiplyBrightness 1.1
    \item M8: Flipud + MultiplyBrightness 0.9
    \item M9: Fliplr + Flipud + MultiplyBrightness 0.9
    \item M10: Fliplr + Flipud + MultiplyBrightness 1.1

    \item M11: Fliplr + MultiplyBrightness 1.1 + LinearContrast 1.1
    \item M12: Flipud + MultiplyBrightness 0.9 + LinearContrast 1.1
    \item M13: Fliplr + Flipud + MultiplyBrightness 0.9 + LinearContrast 1.1
    \item M14: Fliplr + Flipud + MultiplyBrightness 1.1 + LinearContrast 1.1

    \item M15: Fliplr + MultiplyBrightness 1.1 + LinearContrast 1.2
    \item M16: Flipud + MultiplyBrightness 0.9 + LinearContrast 1.2
    \item M17: Fliplr + Flipud + MultiplyBrightness 0.9 + LinearContrast 1.2
    \item M18: Fliplr + Flipud + MultiplyBrightness 1.1 + LinearContrast 1.2

    \item M19: Fliplr + MultiplyBrightness 1.1 + LinearContrast 0.8
    \item M20: Flipud + MultiplyBrightness 0.9 + LinearContrast 0.8
    \item M21: Fliplr + Flipud + MultiplyBrightness 0.9 + LinearContrast 0.8
    \item M22: Fliplr + Flipud + MultiplyBrightness 1.1 + LinearContrast 0.8

    \item M23: Fliplr + MultiplyBrightness 1.1 + LinearContrast 0.9
    \item M24: Flipud + MultiplyBrightness 0.9 + LinearContrast 0.9
    \item M25: Fliplr + Flipud + MultiplyBrightness 0.9 + LinearContrast 0.9
    \item M26: Fliplr + Flipud + MultiplyBrightness 1.1 + LinearContrast 0.9

    \item M27: Fliplr + MultiplySaturation 1.1
    \item M28: Flipud + MultiplySaturation 0.9
    \item M29: Fliplr + Flipud + MultiplySaturation 0.9
    \item M30: Fliplr + Flipud + MultiplySaturation 1.1

    \item M31: Fliplr + MultiplySaturation 1.1 + MultiplyBrightness 1.1
    \item M32: Flipud + MultiplySaturation 0.9 + MultiplyBrightness 1.1
    \item M33: Fliplr + Flipud + MultiplySaturation 0.9 + MultiplyBrightness 1.1
    \item M34: Fliplr + Flipud + MultiplySaturation 1.1 + MultiplyBrightness 1.1

    \item M35: Fliplr + MultiplySaturation 1.1 + MultiplyBrightness 0.9
    \item M36: Flipud + MultiplySaturation 0.9 + MultiplyBrightness 0.9
    \item M37: Fliplr + Flipud + MultiplySaturation 0.9 + MultiplyBrightness 0.9
    \item M38: Fliplr + Flipud + MultiplyBrightness 1.1 + LinearContrast 0.9

    \item M39: Fliplr + MultiplySaturation 1.1 + LinearContrast 1.1
    \item M40: Flipud + MultiplySaturation 0.9 + LinearContrast 1.1
    \item M41: Fliplr + Flipud + MultiplySaturation 0.9 + LinearContrast 1.1
    \item M42: Fliplr + Flipud + MultiplySaturation 1.1 + LinearContrast 1.1

    \item M43: Fliplr + MultiplySaturation 1.1 + LinearContrast 1.2
    \item M44: Flipud + MultiplySaturation 0.9 + LinearContrast 1.2
    \item M45: Fliplr + Flipud + MultiplySaturation 0.9 + LinearContrast 1.2
    \item M46: Fliplr + Flipud + MultiplySaturation 1.1 + LinearContrast 1.2

    \item M47: Fliplr + MultiplySaturation 1.1 + LinearContrast 0.8
    \item M48: Flipud + MultiplySaturation 0.9 + LinearContrast 0.8
    \item M49: Fliplr + Flipud + MultiplySaturation 0.9 + LinearContrast 0.8
    \item M50: Fliplr + Flipud + MultiplySaturation 1.1 + LinearContrast 0.8

    \item M51: Fliplr + MultiplySaturation 1.1 + LinearContrast 0.9
    \item M52: Flipud + MultiplySaturation 0.9 + LinearContrast 0.9
    \item M53: Fliplr + Flipud + MultiplySaturation 0.9 + LinearContrast 0.9
    \item M54: Fliplr + Flipud + MultiplySaturation 1.1 + LinearContrast 0.9
\end{itemize}

\section{Images with grading discrepancy\label{apdx:A}}

\begin{figure}[htbp!]
    \centering
    \includegraphics[width=0.95\textwidth]{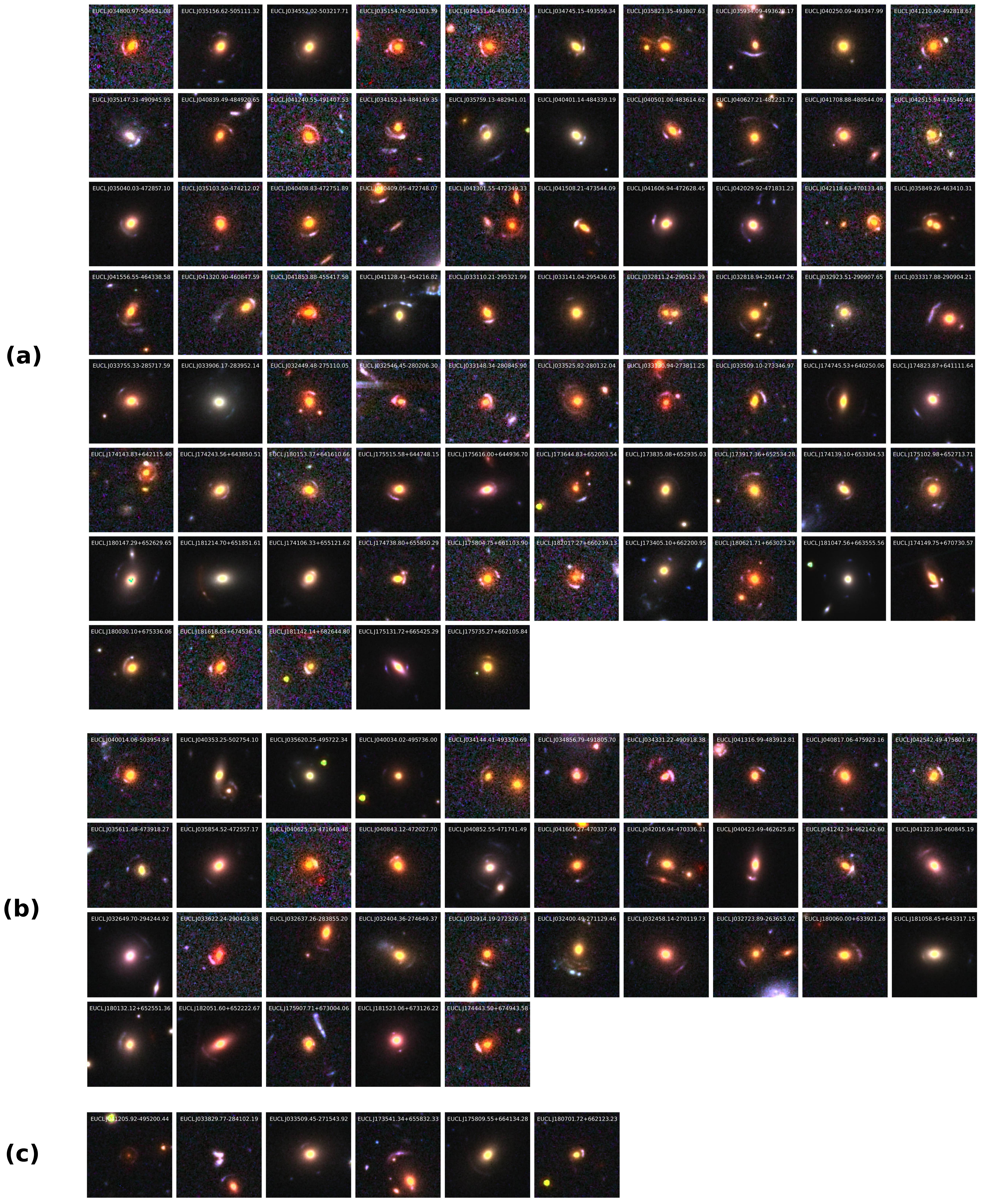}
    \caption{Centred lens candidates that received lower grades in this work compared with SLDE A. Panel (a) displays objects that received Grade A in SLDE A but were graded as Grade B in this work. Panel (b) displays objects that received Grade B in SLDE A but were graded as Grade C in this work. Panel (c) displays objects that received Grade A in SLDE A but were graded as Grade C in this work.}
    \label{fig:app-a-to-b}
\end{figure}

In this Appendix, in Fig.~\ref{fig:app-a-to-b}, we highlight the strong lens candidates that differed in rating from \cite{Q1-SP048}.

\begin{figure}[htbp!]
    \centering
    \includegraphics[width=0.95\textwidth]{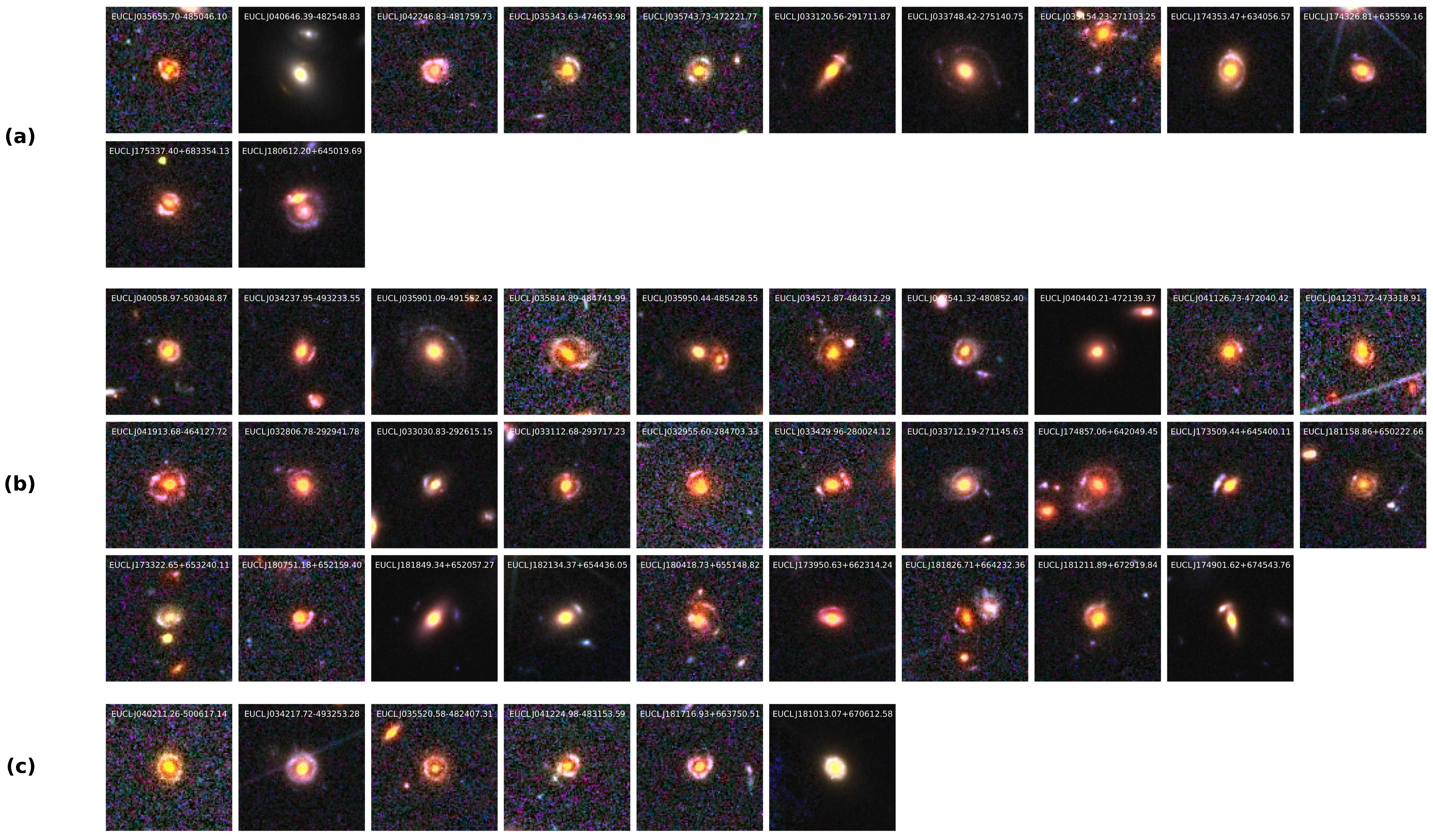}
    \caption{Centred lens candidates that received higher grades in this work compared with SLDE A. Panel (a) displays objects that received Grade A in this work but were graded as Grade B in SLDE A. Panel (b) displays objects that received Grade B in this work but were graded as C in SLDE A. Panel (c) displays objects that received Grade A in this work but were graded as C in SLDE A.}
    \label{fig:app-mismatch-walmsley-lower}
\end{figure}

\section{100 hand-drawn positive training set\label{apdx:B}}

\begin{figure}[htbp!]
    \includegraphics[width=\linewidth]{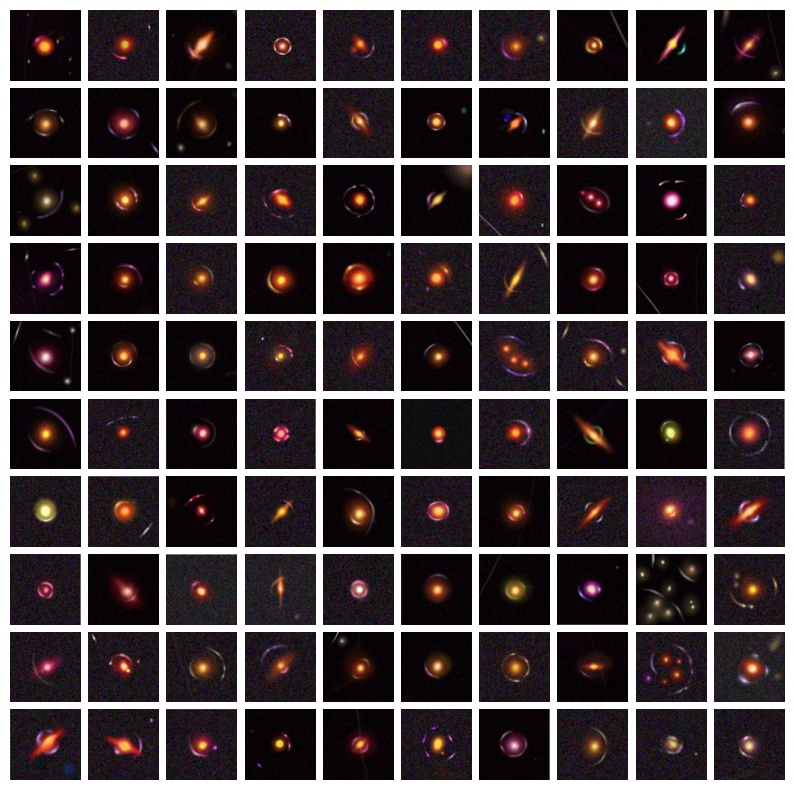}
    \caption{A panel of the 100 experimental hand-drawn lens simulations used to enhance the diversity of the positive training dataset in the second round of the iterative training process. These images were created using Procreate and were based on visual features commonly exhibited among found lenses. These features include the red--orange colour of the lensing galaxy, purple colour of the background galaxy, various observed lensing configurations, etc.}
    \label{fig:app-art-lens}
\end{figure}

In this Appendix, in Fig.~\ref{fig:app-art-lens}, we highlight the 100 experimental hand-drawn lens simulations used to enhance the diversity of the positive training dataset in the second round of the iterative training process.

\section{Plots for all classified A/B Lenses within the top 4000 predictions\label{apdx:master_plots}}

\begin{figure}[htbp!]
    \centering

    \includegraphics[width=\linewidth]{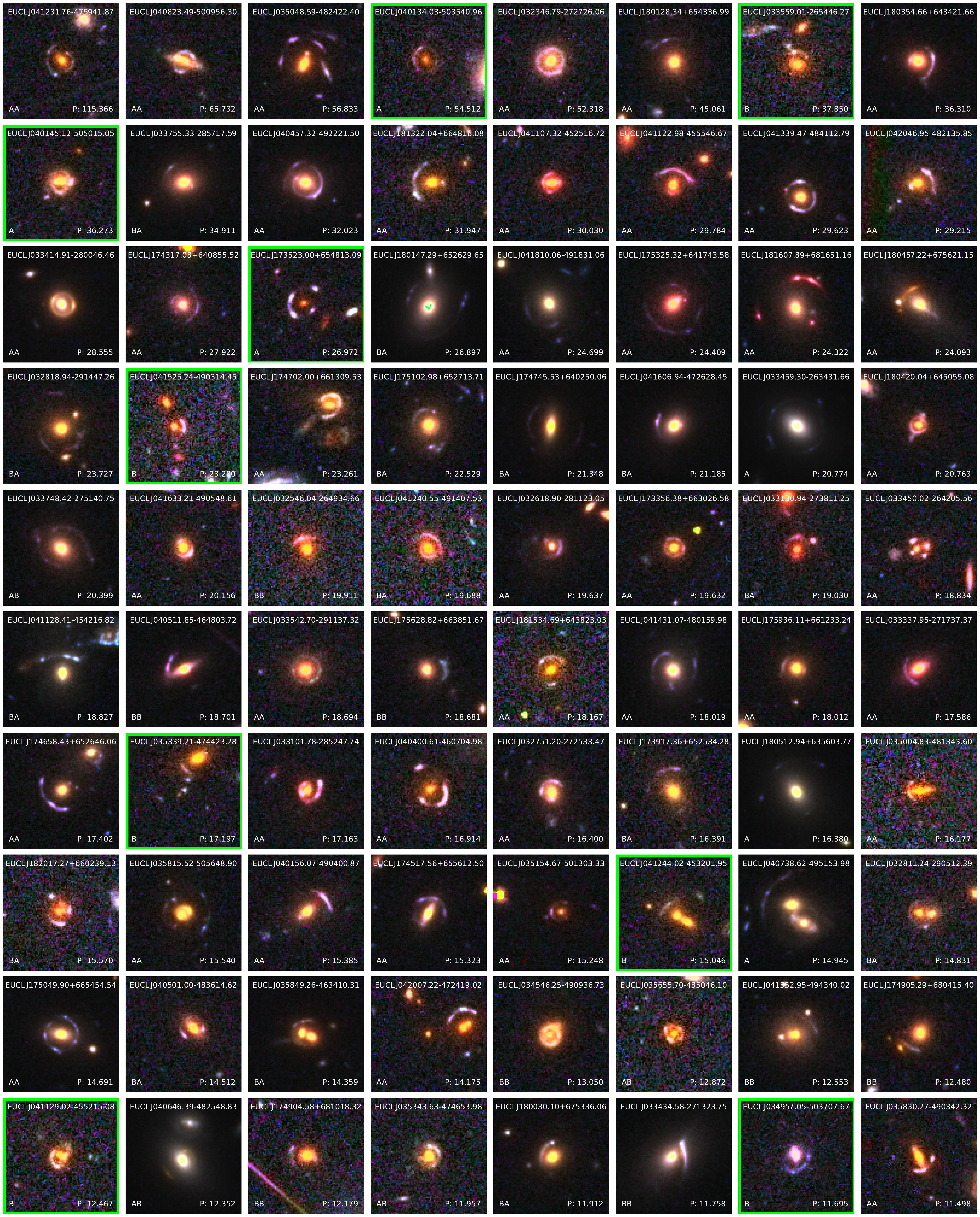}
    \caption{(1--80) 367 classified grade A/B lens candidates. New lens candidates not included in \cite{Q1-SP048} are outlined in green. Images displayed are slightly enhanced with an arcsinh function; however, this function was only applied for display -- the model itself was applied to linearly stretched images in training and inference.}
    \label{fig:0-80 total}
\end{figure}

\begin{figure}[htbp!]
    \centering
    \includegraphics[width=\linewidth]{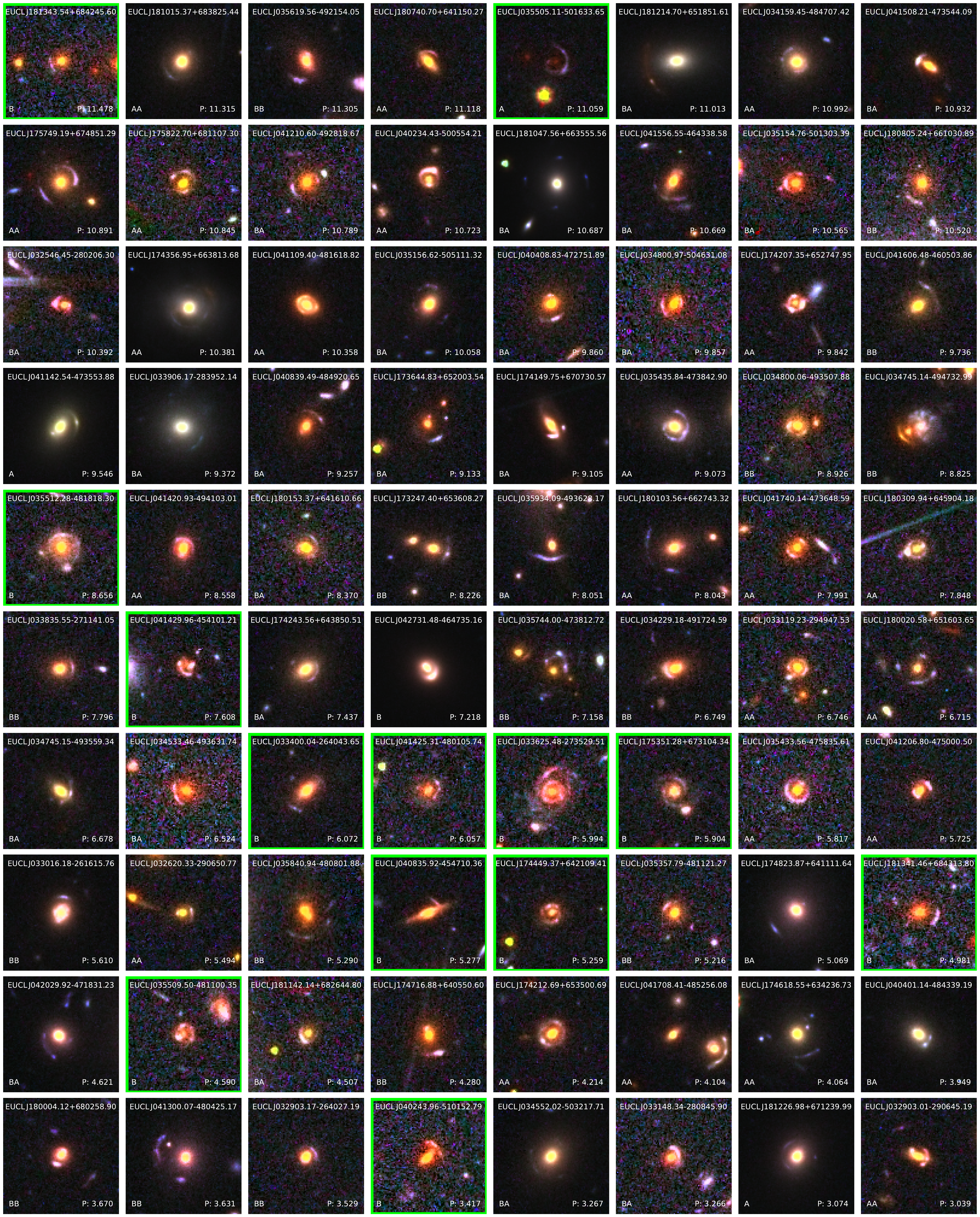}
    \caption{(81--160) 367 classified grade A/B lens candidates. New lens candidates not included in SLDE A are outlined in green. Images displayed are slightly enhanced with an arcsinh function; however, this function was only applied for display -- the model itself is applied to linearly stretched images in training and inference.}
    \label{fig:80-160 total}
\end{figure}

\begin{figure}[htbp!]
    \centering
    \includegraphics[width=\linewidth]{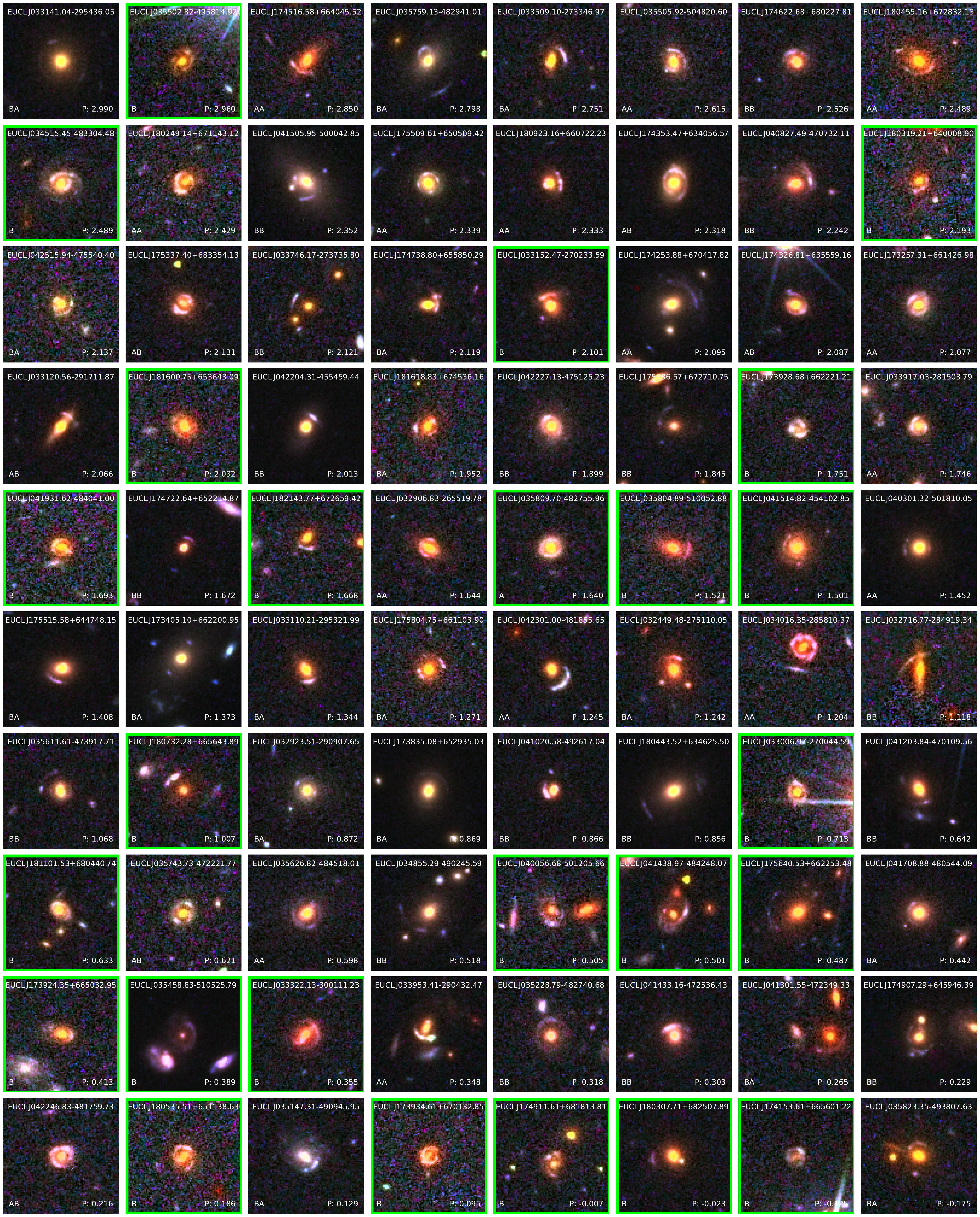}
    \caption{(161--240) 367 classified grade A/B lens candidates. New lens candidates not included in SLDE A are outlined in green. Images displayed are slightly enhanced with an arcsinh function; however, this function was only applied for display -- the model itself was applied to linearly stretched images in training and inference.}
    \label{fig:160-240 total}
\end{figure}

\begin{figure}[htbp!]
    \centering
    \includegraphics[width=\linewidth]{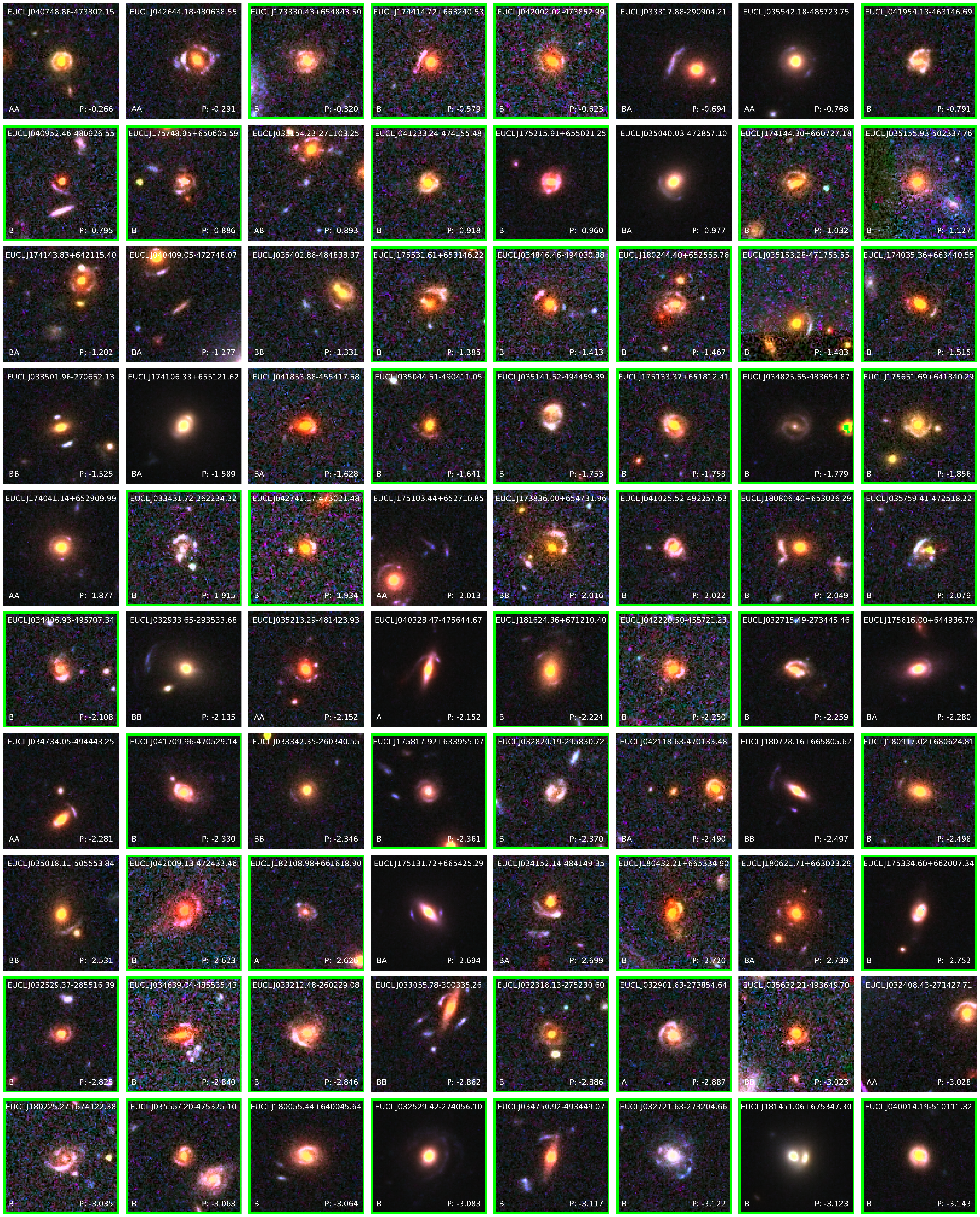}
    \caption{(241--320) 367 classified grade A/B lens candidates. New lens candidates not included in SLDE A are outlined in green. Images displayed are slightly enhanced with an arcsinh function; however, this function was only applied for display -- the model itself was applied to linearly stretched images in training and inference.}
    \label{fig:240-320 total}
\end{figure}

\begin{figure}[htbp!]
    \centering
    \includegraphics[width=\linewidth]{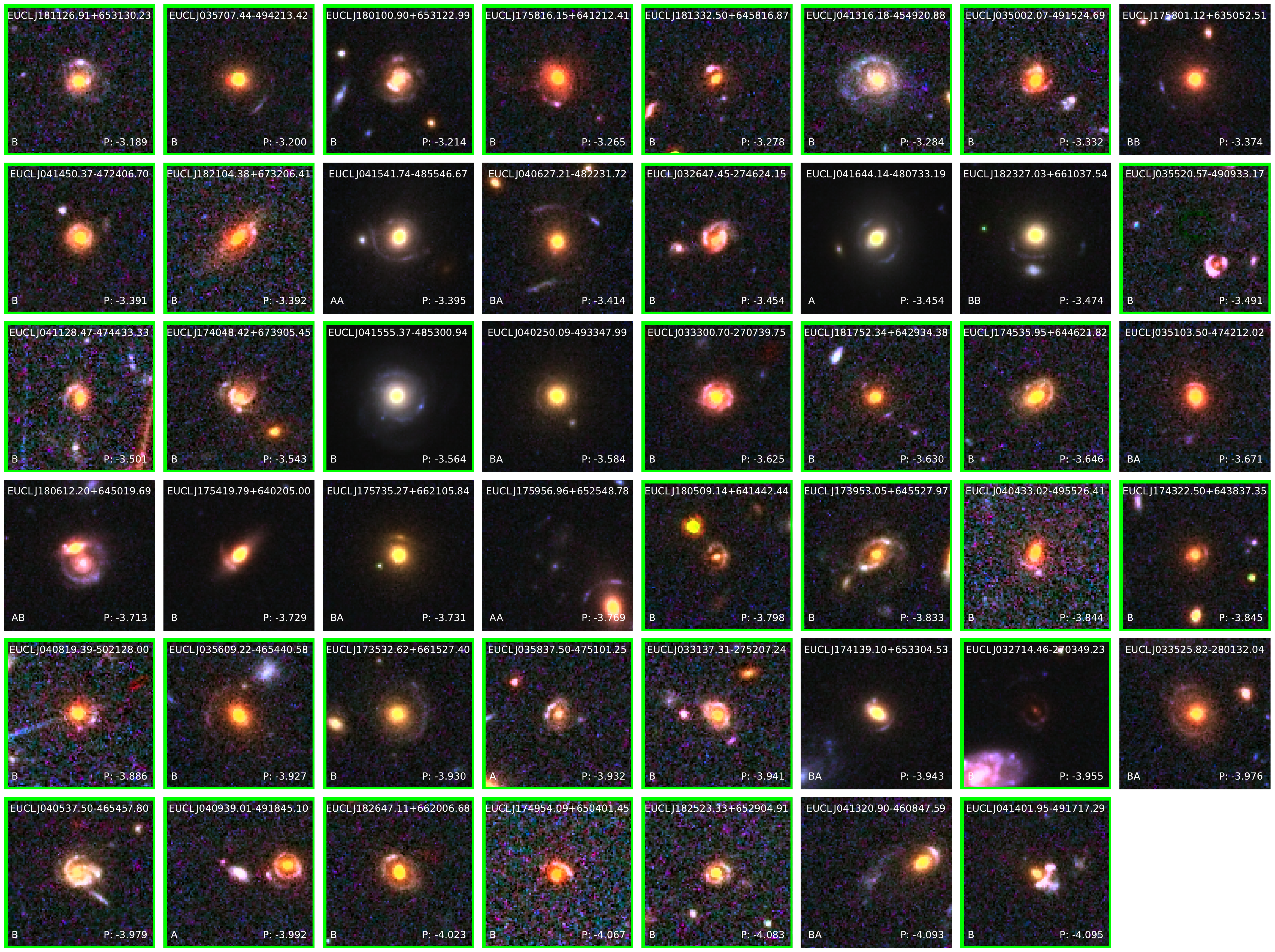}
    \caption{(321--367) 367 classified grade A/B lens candidates. New lens candidates not included in SLDE A are outlined in green. Images displayed are slightly enhanced with an arcsinh function; however, this function was only applied for display -- the model itself was applied to linearly stretched images in training and inference.}
    \label{fig:320-367 total}
\end{figure}

In this Appendix, in Figs.~\ref{fig:0-80 total}, \ref{fig:80-160 total}, \ref{fig:160-240 total}, \ref{fig:240-320 total}, and \ref{fig:320-367 total} we show all 367 strong lens candidates that were classified as grade A/B lenses. Newly identified lenses are marked with green outlines. The other lenses, which were previously identified, are selected from having union grades between AA, AB, BA, and BB. For each union grade, it consists of grade received in this work and grade given in previous research. Note that in text we refer to a total A/B lens candidate sample of 441. The 441 includes 74 additional candidates that received 
AC, BC, AX and BX union grades. These candidates are shown in Fig.~\ref{fig:app-a-to-b}.

\clearpage
\begin{landscape}
\section{Master table displaying RA; DEC; \IE, \YE, \JE, and \HE Magnitudes; Raw prediction output from model; and citations\label{apdx:master_table}}

In this Appendix, in Table~\ref{tab:master_table}, ranked by raw scores, we tabulate coordinates, \Euclid magnitudes, classification grades and model raw scores of grade A/B lens candidates.

\begin{table}[htbp!]
    \centering
    \caption{A few example entries of the master table of classified lenses depicted in \ref{apdx:master_plots}. Currently depicting the top 18 lenses from Fig. \ref{fig:0-80 total}. The properties displayed are (1): the object id/name for each candidate (2): right ascension (3): declination (4): \IE band magnitude, using \texttt{FLUX\_DETECTION\_TOTAL} (5) \YE band magnitude, found using \texttt{FLUX\_Y\_2FWHM\_APER} (6): \JE band magnitude, found using \texttt{FLUX\_J\_2FWHM\_APER} (7): \HE band magnitude, found using \texttt{FLUX\_H\_2FWHM\_APER} (8) The processed grades of human experts classification. If there are two letters, then the object was previously identified and classified, with grade represented by the second letter (9): The "raw score'' output from the model in this work (10): citations and any additional notes. Two undisplayed properties are included in the full electronically available table in additional resources: whether the lens is new; and an alternative name for each object composed of the mosaic tile and cutout id. Under notes, L25: \cite{Q1-SP053}, W25: \cite{Q1-SP048}, R25: \cite{Q1-SP052}, and E25: \cite{Q1-SP099}. R25 and E25 not shown here.}
    \label{tab:master_table}
\begin{tabularx}{\linewidth}{
                    >{\small\raggedright\arraybackslash}p{4.5cm} 
                    S[table-format=3.6, round-precision=6, group-digits=false] 
                    S[table-format=3.6, round-precision=6, group-digits=false] 
                    S[table-format=2.2, round-precision=2, group-digits=false] 
                    S[table-format=2.2, round-precision=2, group-digits=false] 
                    S[table-format=2.2, round-precision=2, group-digits=false] 
                    S[table-format=2.2, round-precision=2, group-digits=false] 
                    c 
                    S[table-format=3.2, round-precision=2] 
                    X} 
    \hline
    \noalign{\smallskip}
    \textbf{OBJ NAME} &
    \textbf{RA} &
    \textbf{Dec} &
    \textbf{\IE} &
    \textbf{\YE} &
    \textbf{\JE} &
    \textbf{\HE} &
    \textbf{GRADE} &
    \textbf{RAW SCORE} &
    \textbf{NOTES} \\
    \noalign{\smallskip}
    (1) &
    \multicolumn{1}{c}{(2)} &
    \multicolumn{1}{c}{(3)} &
    \multicolumn{1}{c}{(4)} &
    \multicolumn{1}{c}{(5)} &
    \multicolumn{1}{c}{(6)} &
    \multicolumn{1}{c}{(7)} &
    (8) &
    \multicolumn{1}{c}{(9)} &
    (10) \\
\end{tabularx}
\end{table}
\end{landscape}

\end{appendix}

\label{LastPage}
\end{document}